\documentclass[a4paper,12pt,two-side]{article}

\usepackage[english
]{babel}
\usepackage[T1]{fontenc}
\usepackage{hyperref}

\usepackage{dsfont}
\usepackage{amsfonts}
\usepackage{amsmath}
\usepackage{amssymb}
\usepackage{amsthm}
\usepackage{eucal}
\usepackage{mathrsfs}
\usepackage[all]{xy}
\usepackage{units}

\usepackage{fancyhdr}

\usepackage[dvips]{graphicx}
\usepackage{color}

\newtheorem{teo}{\textsc{Theorem}}[section]
\newtheorem{defi}[teo]{\textsc{Definition}}
\newtheorem{lem}[teo]{\textsc{Lemma}}
\newtheorem{propos}[teo]{\textsc{Proposition}}
\newtheorem{concl}[teo]{\textsc{Conclusion}}

\newtheorem*{teo*}{\textsc{Theorem}}
\newtheorem*{defi*}{\textsc{Definition}}
\newtheorem*{corol*}{\textsc{Corollary}}
\newtheorem{concl*}{\textsc{Conclusion}}

\theoremstyle{definition}
\newtheorem*{ex*}{\textsc{Example}}
\newtheorem{rk}[teo]{\textsc{Remark}}
\newtheorem*{rk*}{\textsc{Remark}}
\newtheorem*{qst*}{\textsc{Question}}

\newcommand{\Proof}{\begin{proof}[\textsc{\bf{Proof}}]}
\newcommand{\CVD}{\end{proof}}

\newcommand{\R}{\mathbb R}
\newcommand{\N}{\mathbb N}
\newcommand{\C}{\mathbb C}                           

\newcommand{\Z}{\mathbb Z}
\newcommand{\T}{\mathbb T}

\newcommand{\sss}[1]{\CMcal{#1}}
\newcommand{\ssss}[1]{\mathcal{#1}}                  
\newcommand{\bbb}[1]{\mathscr{#1}}
\newcommand{\rrr}[1]{\mathfrak{#1}}
\newcommand{\num}[1]{\mathds {#1}}
\newcommand{\ttb}[1]{\boldsymbol{\mathsf{#1}}}
\newcommand{\ttt}[1]{\mathsf{#1}}

\newcommand{\bra}[1]{\langle #1|}
\newcommand{\ket}[1]{|#1\rangle}
\newcommand{\ketbra}[2]{|#1\rangle\langle#2|}

\newcommand{\expo}[1]{\mbox{e}^{#1}}                 

\newcommand{\ncint}{\mathrel{{\ooalign{$\int$\cr\kern+.07em\raise.15ex\hbox{$\pmb{\scriptstyle-}$}\cr}}}}           
\newcommand{\ncpartial}{\mathrel{{\ooalign{$\partial$\cr\kern+.29em\raise.79ex\hbox{$\pmb{\scriptstyle-}$}\cr}}}}

\newcommand{\virg}[1]{\lq\lq#1\rq\rq}                
\newcommand{\ie}{{\sl i.\,e. }}

\newcommand{\Hi}{\sss{H}}
\newcommand{\B}{\mathcal{B}}

\newcommand{\epsi}{\varepsilon}
\newcommand{\moy}{\, \sharp \,}

\newcommand{\eg}{{\sl e.g.\/ }}

\def\({\left(}
\def\){\right)}
\def\<{\left\langle}
\def\>{\right\rangle}

\DeclareMathOperator{\Ran}{Ran} 
\DeclareMathOperator{\Op}{Op}

\title{\vspace{-10mm}\textbf{\Huge{Effective models for conductance\\ in magnetic fields:}\vspace{1mm}\\
\Large{ derivation of Harper and Hofstadter models}}}
\author{\vspace{2mm}\\ \Large{$\text{G. De Nittis}^*$ and $\text{G. Panati}^{**}$
}\\
\vspace{5mm}\\
\normalsize{$^\ast$ SISSA Scuola Internazionale Superiore di Studi Avanzati, Trieste, Italy}\\
\footnotesize{\texttt{denittis@sissa.it}}\vspace{5mm}
\\
\normalsize{$^{**}$ Dipartimento di Matematica, Universit\`{a} di Roma \virg{La Sapienza}}, Roma, Italy \\
\footnotesize{\texttt{panati@mat.uniroma1.it}}\\
}
\date{27th July 2010}

\begin{document}
\maketitle              

\vspace{-5mm}

\begin{abstract}
Some relevant transport properties of solids do not depend only on
the spectrum of the electronic Hamiltonian, but on finer properties
 preserved only by unitary equivalence, the most striking example
 being the  conductance.  When interested in such properties,
 and aiming to a simpler model, it is mandatory to check that the simpler effective Hamiltonian is
 approximately unitarily equivalent to the original one, in the appropriate
 asymptotic regime.
In this paper, we consider the Hamiltonian of an electron in a
$2$-dimensional periodic potential (\eg generated by the ionic cores
of a crystalline solid) under the influence of a uniform transverse
magnetic field. We prove that such Hamiltonian  is approximately
unitarily equivalent to a Hofstadter-like (resp. Harper-like)
Hamiltonian, in the limit of weak (resp. strong) magnetic field. The
result concerning the case of weak magnetic field holds true in any
dimension. Finally, in the limit of strong uniform magnetic field,
we show that an additional periodic magnetic potential induces a
non-trivial coupling of the Landau bands.
\end{abstract}



\newpage

{\scriptsize\tableofcontents}


\section{Introduction}\label{sec_int}
Schr\"odinger operators with periodic potentials and magnetic fields
have been fascinating physicist and mathematicians for the last
decades. Due to the competition between the crystal length scale and
the magnetic length scale, these operators reveals striking features
as fractal spectrum {\cite{Geyler}}, anomalous localization {\cite{
Ostrovsky}}, quantization of the transverse conductance
\cite{TKNN,sim_let2,bellissard}, anomalous thermodynamic phase
diagrams \cite{avron1,osad}.

\goodbreak
\medskip

Due to its relevance for the {Quantum Hall Effect} (QHE) \cite{morandi,Graf}, we focus
on the two-dimensional \emph{Bloch-Landau Hamiltonian}
\begin{equation}\label{Hamilt BL}
\ttb{H}_\text{BL}:=\dfrac{1}{2m}\left(-i\hslash\nabla_r-\iota_q\frac{|q|B}{2c}e_\bot\wedge
r\right)^2+\ttt{V}_\Gamma\left(r\right),
\end{equation}
acting in the Hilbert space $\sss{H}_\text{phy}= L^2(\R^2,d^2r)$,
$r=(r_1,r_2)\in\R^2$. Here $c$ is the speed of  light, $h:=2\pi
\hslash$ is the {Planck constant},  $m$ is the {mass} and $q$ the
{charge} (positive if $\iota_q = 1$ or negative if $\iota_q = -1$)
of the charge carrier, $B$ is the {strength} of the external uniform
time-independent magnetic field, $e_\bot=(0,0,1)$ is a unit vector
orthogonal to the sample, and $\ttt{V}_\Gamma$ is a periodic
potential describing the interaction of the carrier with the ionic
cores of the crystal. For the sake of a simpler notation, in this
introduction we assume that the periodicity lattice $\Gamma$ is
simply $\Z^2$.

\medskip

While extremely interesting, a direct analysis of the fine
properties of the operator $\ttb{H}_\text{BL}$ is a formidable task.
Thus the need to study simpler effective models which capture the
main features of (\ref{Hamilt BL}) in suitable physical regimes, as
for example in the limit of weak (resp. strong) magnetic field. The
relevant dimensionless parameter appearing in the problem is
$h_B:=\nicefrac{\Phi_0}{Z\Phi_B} \propto B^{-1}$, where $\Phi_0
=\nicefrac{hc}{e}$ is the \emph{magnetic flux quantum}, $\Phi_B=
\Omega_\Gamma B$ is the flux of the external magnetic field through
the unit cell of the periodicity lattice $\Gamma$ (whose area is
$\Omega_\Gamma$) and $Z =\nicefrac{|q|}{e}$ is the \emph{magnitude}
of the charge $q$ of the carrier in units of $e$ (the
\emph{positron} charge). It is also useful to introduce the reduced
constant $\hslash_B:=\nicefrac{h_B}{2\pi}$.

\medskip

In the limit of weak magnetic field, $\hslash_B \to \infty $, one
expects that the relevant features are captured by the well-known
\emph{Peierls' substitution} \cite{peierls,harper,Hof}, thus
yielding to consider, for each Bloch band
{$\sss{E}_\ast=\sss{E}_\ast(k_1,k_2)$} of interest, the following
effective model: in the Hilbert space $L^2(\num{T}^2, d^2k)$, $k$
being the Bloch momentum and
$\num{T}^2\simeq[0,2\pi)\times[0,2\pi)$,  one considers the
Hamiltonian operator

\begin{equation}\label{Hamilt_Hofstadter}
 H_{\text{Hof}}\ \psi\ = {\sss{E}_\ast}\left(k-\left(\frac{\iota_q}{\hslash_B}\right)\frac{1}{2} \, e_\bot\wedge i\nabla_k\right)\ \psi, \qquad
\qquad \psi \in L^2(\T^2, dk).
\end{equation}

\noindent In physicists' words, the above Hamiltonian corresponds to
replace the variables $k_1$ and $k_2$ in $\sss{E}_n$ with the
symmetric operators (\emph{magnetic momenta})
\begin{equation}\label{eq03}
\bbb{K}_1:=k_1+\frac{i}{2}\left({\frac{\iota_q}{\hslash_B}}\right) \dfrac{\partial}{\partial
k_2},\qquad \qquad \bbb{K}_2:=k_2- \frac{i}{2}\left({\frac{\iota_q}{\hslash_B}}\right)
\dfrac{\partial}{\partial k_1}.
\end{equation}
Since $[\bbb{K}_1, \bbb{K}_2] \neq 0$ the latter prescription is
formal and (\ref{Hamilt_Hofstadter}) must be defined by an
appropriate variant of the Weyl quantization. The rigorous
justification of the Peierls' substitution and the definition and
the derivation of the Hamiltonian \eqref{Hamilt_Hofstadter} are the
content of Section~\ref{Sec_Hofs}.

\medskip

 \noindent The use of the
operator (\ref{Hamilt_Hofstadter}) traces back to the pioneering
works of R. Peierls \cite{peierls} and P. G. Harper \cite{harper},
and its spectrum was extensively studied by D. Hofstadter in the
celebrated paper \cite{Hof}, where he specialized to the case
${\sss{E}_\ast}(k_1, k_2) = 2 \cos k_1 + 2 \cos k_2$. In view of
that, we call \textbf{Hofstadter-like Hamiltonian} any operator in
the form \eqref{Hamilt_Hofstadter}, while the name Hofstadter
Hamiltonian is used only for the special case above. However, this
nomenclature is far to be unique. For instance M. A. Shubin in
\cite{shub} names \emph{discrete magnetic Laplacian} the same
operator (up to a Fourier transform).

\medskip

As for the case of a strong magnetic field, $\hslash_B \to 0$, the
periodic potential can be considered a small perturbation of the
Landau Hamiltonian, which provides the leading order approximation
of $\ttb{H}_\text{BL}$. To the next order of accuracy in $\hslash_B$, to each
Landau level there corresponds an effective Hamiltonian, acting in
$L^2(\R,dx)$, given (up to a suitable rescaling of the energy scale) by
\begin{equation}\label{Hamilt_Harper}
 H_\text{Har}\ \psi =  \ttt{V}_\Gamma\left( - i (\iota_q\hslash_B) \frac{\partial}{\partial x}, x\right) \, \psi, \qquad
 \qquad \psi \in L^2(\R, dx),
\end{equation}
where the r.h.s refers to the ordinary $(\iota_q\hslash_B)$-Weyl quantization
of the $\Z^2$-periodic
function
 $\ttt{V}_\Gamma: \R^2 \to \R$.
We refer to Section \ref{Sec_Har} for the definition of the
effective Hamiltonian \eqref{Hamilt_Harper}. We call
\textbf{Harper-like Hamiltonian} any operator of the form
\eqref{Hamilt_Harper}, using the name Harper Hamiltonian for the
special case $\ttt{V}_\Gamma(p,x) = 2 \cos(2\pi p) + 2 \cos(2\pi x)
$.

\begin{rk}[nomenclature and historical overview]
In a remarkable series of papers \cite{h_sI,h_sII,h_sIII} B. Helffer
and J. Sj\"{o}strand studied the relation between the spectrum of
the operator (\ref{Hamilt_Harper}) and the spectra of a
one-parameter family of {one-dimensional} operators on $\ell^2(\Z)$
defined by
\begin{equation}\label{eq_alm_mat}
(h_{\beta}u)_n:=u_{n-1}+u_{n+1}+2\cos(2\pi\theta n+\beta)u_n,
\end{equation}
where $\theta\in\R$ is a fixed number (\emph{deformation parameter})
and $\beta\in[0,2\pi)$ is the parameter of the family. In the work
of the French authors the operator defined by \eqref{eq_alm_mat} is
called  \emph{Harper operator} (and indeed it was introduced by
Harper in \cite{harper}). However, in the last three decades, the
operator \eqref{eq_alm_mat} has been extensively studied by many
authors (see \cite{Last1,Last2} for an updated review) with the name
of \emph{almost-Mathieu operator}. To avoid confusion and make the
nomenclature clear, we chose to adhere to the most recent
convention, using the name \emph{almost-Mathieu operator} for
\eqref{eq_alm_mat}. We thus decided to give credits to Harper's work
by associating his name to the operator (\ref{Hamilt_Harper}).\hfill $\blacklozenge\lozenge$
\end{rk}

The regime of strong magnetic field was originally investigated by
A. Rauh \cite{rau1,rau2}. However the  correct effective model, the
operator (\ref{Hamilt_Harper}), was derived firstly by M. Wilkinson
\cite{wilk} and then, rigorously, by J. Bellissard in an algebraic
context \cite{bell-2D} and by  B. Helffer and J. Sj\"{o}strand in
\cite{hel-sjo}, inspired by the latter paper. In both these papers
the case of weak magnetic field is also considered. In particular,
in \cite{hel-sjo} it is proven that $H_\text{Har}$ (resp.
$H_\text{Hof}$) has, locally on the energy axis, the same spectrum
and the same density of states of $\ttb{H}_\text{BL}$, as $\hslash_B
\to 0$ (resp. $\hslash_B \to \infty$).

\bigskip

Beyond the spectrum and the density of states, there are other
mathematical properties of $\ttb{H}_\text{BL}$ which reveal
interesting physics, as for example the orbital magnetization
\cite{avr_gat,CTVR2005,CTVR2006} or the transverse (Hall)
conductance\footnote{Some authors refer to the  \virg{Hall
conductivity} rather
 than to the \virg{Hall conductance}. Indeed, in two space dimensions the
 conductivity (microscopic quantity)  coincides exactly with the conductance
 (macroscopic quantity), so the two concepts are synonymous. In this sense
  the quantization of the Hall conductivity is
 a macroscopic quantum phenomenon.}. These properties are not
invariant under a loose equivalence relation as isospectrality, then
it is important to show that $H_\text{Har}$ (resp. $H_\text{Hof}$)
is approximately \emph{unitarily equivalent} to $\ttb{H}_\text{BL}$
in the appropriate limit. This is the main goal of this paper. The
problem is not purely academic, since there exist examples of two
isospectral Schr\"odinger operators which are however \emph{not}
unitarily equivalent and which exhibit different values of the
transverse conductance. In particular, in [DFP10b] we prove that the
Hofstadter operator and the Harper operator are isospectral but not
unitarily equivalent. One concludes that, in the study of the
conductance, it is not enough to prove that the effective model is
isospectral to the
original Hamiltonian. \\

We thus introduce the stronger notion of \textbf{unitarily effective
model}, referring to the concept of almost-invariant subspace
introduced by G. Nenciu \cite{nen2} and to the related notion of
effective Hamiltonian \cite{PST1,PST2,stefan_book}, which we shortly
review. Let us focus on the regime of weak (resp. strong) magnetic
field and define $\varepsilon = \varepsilon(B) =
\nicefrac{1}{\hslash_B}$ (resp. $\varepsilon(B) = {\hslash_B}$) so
that $\varepsilon \to 0$ in the relevant limit. Let
$\Pi_\varepsilon$ be  an orthogonal projector in $\Hi_\text{phy}$
such that, for any $N \in \N, N \leq N_0$ there exist a constant
$C_N$ such that
\begin{equation}\label{Pi commutator}
\| [\ttb{H}_\text{BL}; \, \Pi_{\varepsilon}] \| \leq C_N \, \epsi^N
\end{equation}
for $\epsi$ sufficiently small. Then $\Ran \Pi_\varepsilon$ is
called an \textbf{almost-invariant subspace} \cite{nen2,stefan_book}
at accuracy $N_0$, since it follows by a Duhammel's argument that
$$
\| (1 - \Pi_\varepsilon) \,\, \expo{- i s \, {\ttb{H}_\text{BL}}}
\,\, \Pi_\varepsilon \| \leq C_N \, \varepsilon^{N} \, |s|
$$
for every $s \in \R$, $N \leq N_0$. Granted the existence of such a
subspace, we call \textbf{(unitarily) effective Hamiltonian} a
self-adjoint operator $H_\text{eff}$ acting on a Hilbert space
$\Hi_\text{ref}$, such that there exists a unitary $U_{\varepsilon}:
\Ran \Pi_{\epsi} \to \sss{H}_\text{ref}$ such that for any $N \in
\N, N \leq N_0$, one has
\begin{equation}\label{Effective Hamiltonian}
\| \left( \, \Pi_\epsi \, \ttb{H}_\text{BL}   - U_{\varepsilon}^{-1}
H_\text{eff} \, U_{\varepsilon} \right) \Pi_{\epsi} \| \leq
{C_N}\!\!\!'  \,\,\, \epsi^N .
\end{equation}
The estimates (\ref{Pi commutator}) and (\ref{Effective
Hamiltonian}) imply that
\begin{equation}\label{unit equiv}
\|  \left(\expo{ -i s \, \ttb{H}_\text{BL}}  - U_\varepsilon^{-1} \,
\expo{- i s \, H_{\text{eff}}} \, U_{\varepsilon} \right) \,
\Pi_\varepsilon \| \leq {C_N}\!\!\!''  \,\, \epsi^N |s|.
\end{equation}
When the macroscopic time-scale $t =\epsi s$ is physically relevant,
the estimate above is simply rescaled. The triple $(
\sss{H}_\text{ref}, U_{\epsi}, H_{\text{eff}})$ is, by definition, a
unitarily effective model for $\ttb{H}_\text{BL}$. \noindent To our
purposes, it is important to notice that the asymptotic unitary
equivalence in (\ref{Effective Hamiltonian}) assures that the
\emph{topological invariants} related with the spectral projections
of $\Pi_\varepsilon \, \ttb{H}^\varepsilon_\text{BL} \,
\Pi_\varepsilon$ ($K$-theory, Chern numbers, \ldots) are equal to
those of $ H_\text{eff}$, for $\varepsilon$ sufficiently small
[DP09].

In this paper we prove that in the limit $\hslash_B \to \infty$ the
Hofstadter-like Hamiltonian (\ref{Hamilt_Hofstadter}) provides a
unitarily effective model for $\ttb{H}_\text{BL}$ with accuracy
$N_0=1$, and we exhibit an iterative algorithm to construct an
effective model at any order of accuracy $N_0 \in \N$ (Theorem
\ref{teo2}). As for the limit $\hslash_B\to\infty$, up to a
rescaling of the energy, the non-trivial leading order (accuracy
$N_0=1$) for the effective Hamiltonian is given by the Haper-like
Hamiltonian (\ref{Hamilt_Harper}). We also exhibit the effective
Hamiltonian with accuracy $N_0=2$, i.e. up to errors of order
$\sss{O}(\varepsilon^2)$ (Theorem  \ref{teo1}). Moreover, due to the
robustness of the adiabatic techniques, we can generalize the simple
model described by \eqref{Hamilt BL} to include other potentials,
see \eqref{eq2}, including in particular a periodic vector potential
$\ttb{A}_{\Gamma}$.  This terms produces interesting consequences
especially in the Harper regime (see Section \ref{Sec_Har_per}) and
it could play a relevant role in the theory of orbital
magnetization. This kind of generalization is new with respect to
both \cite{bell-2D} and \cite{hel-sjo}.

Our proof is based on the observation that both the Hofstadter and
the Harper regime are space-adiabatic limits, and can be treated in
the framework of \emph{space-adiabatic perturbation theory}\, (SAPT)
\cite{PST1,PST2}, see also \cite{stefan_book}. As for the Hofstadter
regime, the proof follows ideas similar to the ones in \cite{PST2}.
Our generalization allows however to consider a constant magnetic
field (while in \cite{PST2} the vector potential is assumed in
$C^{\infty}_\text{b}(\R^d)$) and to include a periodic vector
potential. Moreover the proof extends the one in \cite{PST2}, in
view of the use of the special symbol classes defined in Section
\ref{appA2}. On the contrary, from the discussion of the Harper
regime $\hslash_{B} \to 0$ some new mathematical problems emerge.
Then, although the \virg{philosophy} of the proof of Theorem
\ref{teo1} is  of SAPT-type, the technical part is new as it will be
explained in Section \ref{Sec_Har}. Notice that the regime of weak
magnetic field can also be conveniently approached by using the
\emph{magnetic Weyl quantization} \cite{MP04, MPR05,IMP05, IMP09}, a
viewpoint which is investigated in \cite{DL10}.

For the sake of completness, we summarize some salient aspects of
the SAPT. We refer to specific references (e.g. \cite{stefan_book})
for a complete exposition. Let $H$ be the Hamiltonian of a generic
physical system which acts on the total (or physical) Hilbert space
$\sss{H}$. For the SAPT to be applicable, three important
ingredients needs: (i) a distinction between \emph{fast} and
\emph{slow degrees of freedom} which is mathematically expressed by
a unitary decomposition of the physical space $\sss{H}$ into a
product space $\sss{H}_\text{s}\otimes\sss{H}_\text{f}$ (or, more
generally, a direct integral), the first factor being the
\emph{space of the slow degrees of freedom} and the second  the
\emph{space of the fast degrees of freedom}; $\sss{H}_\text{s} \cong
L^2(\sss{M})$ for suitable measure space $\sss{M}$ is also required;
(ii) a dimensionless \emph{adiabatic parameter} $\varepsilon\ll1$
that quantifies the separation of scales between the fast and slow
degrees of freedom and which measures how far are the slow degrees
of freedom to be \virg{classical} in terms of some process of
quantization; (iii) a \emph{relevant part of the spectrum} for the
fast dynamics which remains separated from the rest of the spectrum
under the perturbation caused by the slow degrees of freedom.

\medskip

A numerical simulation of the spectrum of the Hofstadter operator,
as a function of the parameter $\varepsilon =
\nicefrac{1}{\hslash_B}$, leads to a fascinating picture known as
Hofstadter butterfly \cite{Hof}. Since the spectrum has zero measure
as a subset of the square, the physically relevant object is its
complement, the resolvent set. It has been pointed out by D. Osadchy
and J. Avron \cite{osad} that the open connected regions of the
resolvent set (\emph{islands}) can be associated to different
thermodynamic phases (at zero temperature) of a gas of non
interacting fermions in a periodic potential, with $\varepsilon
\propto B$ and the chemical potential as thermodynamic coordinates.
The different phases are labeled by an integer (\emph{topological
quantum number}), interpreted as  the value of the transverse
conductance of the system in units of $\nicefrac{e^2}{h}$ in the
limit of weak magnetic field. The latter integers are conveniently
visualized by different colors, thus leading to the \emph{colored
Hofstadter butterfly} \cite{osad,avron1}. With this language in
mind, the main result of this paper can be reformulated by saying
that the Hofstadter-like and Harper-like Hamiltonians are
\virg{colour-preserving effective models} for the original
Bloch-Landau Hamiltonian. Thus they describe, though in a distorted
and approximated way, some aspects of the thermodynamics of the
original system.

\bigskip
\goodbreak

\textbf{Acknowledgments.} It is a pleasure to thank Y. Avron, J.
Bellissard, G. Dell'Antonio, M. M\u{a}ntoiu, D. Masoero, H. Spohn
and S. Teufel for many useful discussions, and F. Faure for
stimulating comments on a previous version of the paper. We are
grateful to B. Hellfer for suggesting  useful references. Financial
support by the {{INdAM-GNFM} project \emph{Giovane ricercatore
2009}} is gratefully acknowledged.


\section{Description of the model}\label{Sec_mod}

\subsubsection*{A generalized  Bloch-Landau Hamiltonian}
The Hamiltonian \eqref{Hamilt BL} describes the dynamics of  particle  with mass $m$ and  charge $q$ which interacts with the ionic structure of a two dimensional crystal and with an external orthogonal uniform magnetic field. A more general model is provided by the operator
\begin{equation}\label{eq2}
\ttb{H}_\text{BL}:=\dfrac{1}{2m}\left[-i\hslash\nabla_r-\frac{q}{c}
\ttb{A}_\Gamma\left(r\right)-\frac{q}{c}\ttb{A}\left(r\right)\right]^2
+\ttt{V}_\Gamma\left(r\right)+q\ {\Phi} \left(r\right)
\end{equation}
still called \emph{Bloch-Landau Hamiltonian} and, with an abuse of notation, still denoted with the same symbol used in \eqref{Hamilt BL}. The vector-valued
function
  $\ttb{A}:=(\ttt{A}_1,\ttt{A}_2)$ is a \emph{vector potential} corresponding to an (orthogonal)
  \emph{external magnetic field} $\ttb{B}=\nabla_r\wedge\ttb{A}=\left(\partial_{1}\ttt{A}_2-\partial_{2}\ttt{A}_1\right)
  \ e_\bot $, $\Phi$ is a \emph{scalar potential} corresponding to a (parallel) \emph{external
electric field} $\ttb{E}=-\nabla_r\Phi$ and $\ttb{A}_\Gamma$ and
$\ttt{V}_\Gamma$ are \emph{internal periodic potentials} which
describe the electromagnetic interaction with the ionic cores of the
crystal lattice. The external vector potential is assumed to have
the following structure
\begin{equation}\label{eq3}
\ttb{A}\left(r\right)=\ttb{A}_0\left(r\right)+\ttb{A}_B\left(r\right),
\end{equation}
where  $\ttb{A}_0$ is a bounded function and $\ttb{A}_B$ describes a
uniform orthogonal magnetic field of strenght $B$, i.e. in the
symmetric gauge
\begin{align} \label{eq4a}
&\ttb{A}_B\left(r\right)=\dfrac{B}{2}\ e_{\bot}\wedge
r=\left(-\frac{B}{2}r_2,\frac{B}{2}r_1\right),&& \nabla_r\wedge
\ttb{A}_B=B\ e_\bot,&& \nabla_r\cdot \ttb{A}_B=0.
\end{align}

\noindent The evolution of the system is prescribed by the
Schr\"odinger equation
\begin{equation}\label{eq1}
i\hslash\frac{d}{ds}\psi(r,s)=\ttb{H}_\text{BL}\ \psi(r,s),
\end{equation}
where $s$ corresponds to the \emph{microscopical time-scale}.

\subsubsection*{Mathematical description of the crystal structure}
The periodicity of the crystal is described by a \emph{two
dimensional lattice} $\Gamma\subset\R^2$ (namely  a discrete
subgroup of maximal dimension of the Abelian group $(\R^2,+)$), thus
$\Gamma\simeq\Z^2$. Let $\{a,b\}\subset\R^2$ be two generators of
$\Gamma$, i.e.
$$
\Gamma=\{ \gamma\in\R^2\ :\ \gamma=n_1a+n_2b,\ \ \
 n_1,n_2\in\Z\}.
$$
The \emph{fundamental} or \emph{Voronoi cell} of $\Gamma$ is $
M_\Gamma:=\{ r\in\R^2\ |\ r=l_1\ a+ l_2\ b,\ \ \
 l_1,l_2\in[0,1]\}$
and its area is given by $\Omega_\Gamma=|a\wedge b|$. We fix the
orientation of the lattice in such a way that
$\Omega_\Gamma=(a_1b_2-a_2b_1)>0$. We say that a function
$f_\Gamma:\R^2\to\C$ is $\Gamma$-\emph{periodic} if $ f_\Gamma(r+
\gamma)=f_\Gamma(r)$ for all $\gamma \in\Gamma$ and all $r\in\R^2$.
The electrostatic and magnetostatic crystal potentials $V_\Gamma$
and $A_\Gamma$ are assumed to be $\Gamma$-periodic according to the
previous definition.

An important notion is that of \emph{dual lattice} $\Gamma^\ast$ which is  the set of the vectors  $\gamma^\ast\in\R^2$ such that $\gamma^\ast\cdot \gamma\in2\pi\Z$ for all $\gamma\in\Gamma$.
Let $\{a^\ast,b^\ast\}\subset\R^2$ be defined by the relations $a^\ast\cdot a=b^\ast\cdot b=1$ and $a^\ast\cdot b=b^\ast\cdot a=0$; these vectors are the generators of the lattice $\Gamma^\ast$, i.e.
$$
\Gamma^\ast=\{ \gamma^\ast\in\R^2\ :\ \gamma^\ast=m_1\ 2\pi a^\ast+m_2\ 2\pi b^\ast,\ \ \
 m_1,m_2\in\Z\}.
$$
The \emph{Brillouin zone} $
M_{\Gamma^\ast}:=\{  k\in\R^2\ |\ k=k_1\ a^\ast+ k_2\ b^\ast,\ \ \
 k_1,k_2\in[0,2\pi]\}
$ is the fundamental cell of the dual lattice $\Gamma^\ast$.
The explicit expressions for the dual generators $\{a^\ast,b^\ast\}$ in terms of the basis $\{a,b\}$
is
\begin{equation}\label{eq5}
a^\ast=\frac{e_\bot\wedge b}{|a\wedge b|}=\dfrac{1}{\Omega_\Gamma}(b_2,-b_1),\ \ \ \ \ \ \ b^\ast=-\frac{e_\bot\wedge a}{|a\wedge b|}=\dfrac{1}{\Omega_\Gamma}(-a_2,a_1).
\end{equation}
It follows from \eqref{eq5}  that the  surface of the Brillouin zone is  $\Omega_{\Gamma^\ast}=(2\pi)^2|a^\ast\wedge b^\ast|=\nicefrac{(2\pi)^2}{\Omega_\Gamma}$.

Given a $\Gamma$-periodic function $f_\Gamma$, we denote its Fourier decomposition as
\begin{equation}\label{eq6}
f_\Gamma(r)=\sum_{\gamma^\ast\in\Gamma^\ast}f(\gamma^\ast)\ \expo{i\gamma^\ast\cdot r}=\sum_{{m_1,m_2}\in\Z}f_{m_1,m_2}\ \expo{i2\pi (m_1\ a^\ast+ m_2\ b^\ast)\cdot r}.
\end{equation}
 A \emph{$\Z^2$-periodic} function  $f:\R^2\to\C$ is a function periodic with respect to an orthonormal lattice, namely such that $f(x_1+1,x_2)=f(x_1,x_2+1)=f(x_1,x_2)$ for all $x_1,x_2\in\R$. If
one replaces the two real variables by $x_1:=a^\ast\cdot r$ and
$x_2:={b}^\ast\cdot r$ one has that $ f_\Gamma(r):=f({a}^\ast\cdot
r,{b}^\ast\cdot r) $ is  $\Gamma$-periodic in $r$. Every
$\Gamma$-periodic function can be obtained in this way.
\subsubsection*{Assumptions on the regularity of the potentials and self-adjointness}
Let us denote by $C^n_{\text{{\upshape b}}}(\R^2,\R)$ the space of
real-valued $n$-times differentiable functions (smooth functions if
$n=\infty$) with continuous and bounded derivatives up to order $n$.
Concerning the internal potentials $\ttb{A}_\Gamma$ and $\ttt{V}_\Gamma$ we need to assume that:\\
\\
{\bf \textsc{Assumption} ($\text{A}_\text{s}$) [internal potentials, strong form].} {\it The $\Gamma$-periodic potential $\ttt{V}_\Gamma$ and the two components of the $\Gamma$-periodic vector potential $\ttb{A}_\Gamma$ are functions of class $C^\infty_{\text{{\upshape b}}}(\R^2,\R)$.}\\
\\
Sometime will be enough to consider a weaker version of this assumption, namely:\\
\\
{\bf \textsc{Assumption} ($\text{A}_\text{w}$) [internal potentials, weak form].}
{\it The two components of the $\Gamma$-periodic vector potential $\ttb{A}_\Gamma$  are in
$C^1_{\text{{\upshape b}}}(\R^2,\R)$. The $\Gamma$-periodic potential  $\ttt{V}_\Gamma$
verifies the condition $\int_{M_\Gamma}|\ttt{V}_\Gamma(r)|^2\ d^2r<+\infty$.}\\
\\
Assumption ($\text{A}_\text{w}$) implies that $\ttt{V}_\Gamma$  is
\emph{uniformly locally} $L^2$  and this implies also that
$\ttt{V}_\Gamma$ is infinitesimally bounded with respect to $-\Delta_r$
(see Theorem XIII.96 in \cite{red-sim4}).
Concerning the external potentials $\ttb{A}$ and $\Phi$, we need to assume that:\\
\\
 {\bf \textsc{Assumption} (B) [external potentials].} {\it The scalar potential $\Phi$ is of
 class $C^\infty_{\text{{\upshape b}}}(\R^2,\R)$.
The vector potential $\ttb{A}$ consists of a linear term $\ttb{A}_B$
of the form (\ref{eq4a}) plus a bounded term
 $\ttb{A}_0$ which  is of class $C^\infty_{\text{{\upshape b}}}(\R^2,\R)$.}\\
 \\
When the external potentials $\ttb{A}$ and $\Phi$ vanish, the
Bloch-Landau Hamiltonian \eqref{eq2} reduces to the \emph{periodic
Hamiltonian} (or \emph{Bloch Hamiltonian})
\begin{equation}\label{eq588}
\ttb{H}_\text{per}:=\dfrac{1}{2m}\left[-i\hslash\nabla_r-\frac{q}{c}\ttb{A}_\Gamma(r)\right]^2+\ttt{V}_\Gamma(r).
\end{equation}
The domains of self-adjointness of $\ttb{H}_\text{{\upshape BL}}$
and $\ttb{H}_\text{{\upshape per}}$ are described in the following
proposition. Its proof, together with some basic notion about the
\emph{Sobolev space} $\ssss{H}^2(\R^2)$ and  the \emph{magnetic-Sobolev
space} $\ssss{H}_{\text{\upshape M}}^2(\R^2)$, is postponed to
Section \ref{appA1}.

\begin{propos}\label{prop1}
Let {\upshape Assumptions ($\text{A}_\text{w}$)} and {\upshape(B)}
hold true. Then both\  $\ttb{H}_\text{{\upshape BL}}$ and \
$\ttb{H}_\text{{\upshape per}}$ are  essentially self-adjoint
operators on $ L^2(\R^2,d^2r)$ with common domain of essential
self-adjointness the space of smooth functions with compact support
$C^\infty_{\text{{\upshape c}}}(\R^2,\C)$. Moreover the domain of
self-adjointness of \ $\ttb{H}_\text{{\upshape per}}$ is
 $\ssss{H}^2(\R^2)$ while the  domain of self-adjointness of \ $\ttb{H}_\text{{\upshape BL}}$ is  $\ssss{H}_{\text{\upshape M}}^2(\R^2)$.
\end{propos}

\section{Space-adiabatic theory for the Hofstadter regime}\label{Sec_Hofs}

\subsection{Adiabatic parameter for weak magnetic fields}
\label{Sec_Hofs_par}

The SAPT for a Bloch electron developed in \cite{PST2} is based on
the existence  of a separation between the \emph{microscopic space
scale} fixed by the lattice spacing  $\ell:=\sqrt{\Omega_\Gamma}$,
and a \emph{macroscopic space scale} fixed by the scale of variation
of the \virg{slowly varying} external potentials. The existence of
such a separation of scales is expressed by introducing a
dimensionless parameter $\varepsilon\ll1$ (\emph{adiabatic
parameter}) to control the scale of variation of  the vector
potential and the scalar potential $\Phi$ appearing in \eqref{eq2},
namely by setting $\ttb{A}=\ttb{A}(\varepsilon r)$ and
$\Phi=\Phi(\varepsilon r)$. In particular  the external magnetic and
electric fields are weak compared to the fields generated by the
ionic cores.

It is useful to rewrite the ($\varepsilon$-dependent) Hamiltonian
\eqref{eq2} in a dimensionless form. The microscopic unit of length
being $\ell$, we introduce the dimensionless position vector
$x:=\nicefrac{r}{\ell}$ and the dimensionless gradient
${\nabla}_{x}=\ell{\nabla}_r$. Moreover, since the vector potential
has the dimension of a length times a magnetic field, then
$A(\varepsilon x):=\nicefrac{\varepsilon}{\ell B}\
\ttb{A}(\varepsilon \ell x)$ is a dimensionless function, with $B$ a
dimensional constant which fixes the order of magnitude of the
magnetic field due to the external vector potential $\ttb{A}$.
Similarly for $\ttb{A}_\Gamma$ (with $\varepsilon=1$). Factoring out
the dimensional constants one finds
\begin{equation}\label{eq9}
H_\text{BL}:=\dfrac{1}{\sss{E}_0}\ttb{H}_\text{BL}=\dfrac{1}{2}\left[-i{\nabla}_x-\underbrace{\frac{q\Omega_\Gamma
B_\Gamma}{c\hslash}}_{=:\hslash_\Gamma^{-1}}{A}_\Gamma\left({x}\right)-\iota_q\underbrace{\frac{|q|\Omega_\Gamma
B}{c\hslash}}_{=\hslash_B^{-1}}\frac{1}{\varepsilon}{A}\left(\varepsilon
{x}\right)\right]^2+V_\Gamma\left({x}\right)+\phi \left(\varepsilon
{x}\right),
\end{equation}
where $\sss{E}_0:=\nicefrac{\hslash^2}{m\Omega_\Gamma}$ is the
\emph{natural unit of the energy} fixed by the problem,
$V_\Gamma(x):=\nicefrac{1}{\sss{E}_0}\ \ttt{V}_{\Gamma}(\ell x)$ and
$\phi(\varepsilon x):=\nicefrac{q}{\sss{E}_0}\ \Phi(\varepsilon\ell
x)$ are both dimensionless quantities. The constant $\hslash_\Gamma$
will play no particular role in the rest of this paper, so it is
reabsorbed into the definition of the dimensionless vector potential
${A}_\Gamma$, i.e. formally $\hslash_\Gamma=1$.

Comparing the dimensional Hamiltonian \eqref{eq9} with the original
Hamiltonian \eqref{eq2} , or observing that the strenght of the
magnetic field goes to zero (at least linearly) with $\varepsilon$,
it is physically reasonable to estimate  $\varepsilon
\hslash_B\propto1$. This is rigorously true in the case in which the
external electromagnetic field is uniform.

The external force due to ${A}$ and $\phi$ are of order of
$\varepsilon$ and therefore have to act over a time of order
$\varepsilon^{-1}$ to produce a finite change, which defines the
macroscopic time-scale. The  \emph{macroscopic dimensionless (slow)
time-scale} is fixed by  $t:=\varepsilon\frac{\sss{E}_0}{\hslash}s$
where $s$ is the dimensional microscopic \emph{(fast)} time-scale.
With this change of scale the Schr\"odinger equation \eqref{eq1}
reads
\begin{equation}\label{eq12}
i\varepsilon\frac{d}{dt}\psi=H_\text{BL}\ \psi
\end{equation}
with $H_\text{BL}$  given by equation \eqref{eq9}.
\begin{rk}\label{rk01}
Observe that from the definition of the dimensionless periodic
potential $A_\Gamma$ and $V_\Gamma$ it follows that they are
periodic with respect to the transform $x\mapsto x+
\nicefrac{\gamma}{\ell}$. This means that $A_\Gamma$ and $V_\Gamma$
are periodic with respect to a \virg{normalized} lattice whose
fundamental cell has surface 1. \hspace{\stretch{1}}\hfill $\blacklozenge\lozenge$
\end{rk}
\subsection{Separation of scales: the Bloch-Floquet transform}\label{sec_BF}
To make explicit the presence of the linear term of the external vector potential, we can rewrite the \eqref{eq9} as follows
\begin{equation}\label{eq56}
H_\text{BL}=\dfrac{1}{2}\left[-i{\nabla}-{A}_\Gamma\left({x}\right)-{A}_0\left(\varepsilon{x}\right)-\iota_q\ \frac{1}{2}{e}_\bot\wedge\varepsilon{x}\right]^2+V_\Gamma\left({x}\right)+\phi
\left(\varepsilon{x}\right),
\end{equation}
where the  adiabatic parameter $\varepsilon$
expresses the separation between the macroscopic length-scale,
defined by the external potentials, and the microscopic
length-scale, defined by the internal $\Gamma$-periodic potentials.
The separation between slow and fast degrees of
freedom can be expressed decomposing the physical Hilbert space
$\sss{H}_\text{phy} = L^2(\R^2,d^2x)$ into a product of two Hilbert
spaces or, more generally, into a direct integral. To this end, we
use the Bloch-Floquet transform \cite{kuc}. As in \cite{PST2} we
define the \emph{(modified) Bloch-Floquet transform} $\sss{Z}$ of a
function $\psi\in\sss{S}(\R^2)$  to be
\begin{equation}\label{eq58}
(\sss{Z}\psi)({k},{\theta}):=\sum_{{\gamma}\in\Gamma}\expo{-i({\theta}+{\gamma})\cdot{k}}\psi({\theta}+{\gamma}),\ \ \ \ \ \ \ \ ({k},{\theta})\in\R^2\times\R^2.
\end{equation}
Directly from the definition one can check the following periodicity properties:
\begin{align}
& (\sss{Z}\psi)({k},{\theta}+{\gamma})=(\sss{Z}\psi)({k},{\theta})&&\ \ \ \ \  \ \forall\ {\gamma}\in\Gamma\label{eq59a}\\
&(\sss{Z}\psi)({k}+{\gamma}^\ast,{\theta})=\expo{-i{\theta}\cdot{\gamma}^\ast}(\sss{Z}\psi)({k},{\theta})& &\ \ \ \ \ \ \forall\ {\gamma}^\ast\in\Gamma^\ast.\label{eq59b}
\end{align}
Equation \eqref{eq59a} shows that for any fixed ${k}\in\R^2$,
$(\sss{Z}\psi)({k},\cdot)$ is a $\Gamma$-periodic function and can
be seen as an element of
$\sss{H}_\text{f}:=L^2(\num{V},d^2{\theta})$ with
$\num{V}:=\R^2/\Gamma$  a two-dimensional slant torus (\emph{Voronoi
torus}). The torus $\num{V}$ coincides with the  the fundamental
cell $M_\Gamma$ endowed with the identification of the opposite
edges and $d^2{\theta}$ denotes the (normalized)  measure induced on
$\num{V}$ by the identification with $M_\Gamma$.
 The Hilbert space $\sss{H}_\text{f}$ is the \emph{space of fast degrees of freedom},
corresponding to the microscopic scale. Equation \eqref{eq59b}
involves a unitary representation
$\tau:\Gamma^\ast\longrightarrow\bbb{U}(\sss{H}_\text{f})$ of the
group of the (dual) lattice translations $\Gamma^\ast$ on the
Hilbert space $\sss{H}_\text{f}$. For every
${\gamma}^\ast\in\Gamma^\ast$ the unitary operator
$\tau({\gamma}^\ast)$ is the multiplication with
$\expo{i{\theta}\cdot{\gamma}^\ast}$. It will be convenient to
introduce the Hilbert space
\begin{equation}\label{eq60}
 \sss{H}_\tau:=\left\{\psi\in L^2_\text{loc}\left(\R^2,d^2\underline{k},\sss{H}_\text{f}\right)\ :\ \psi({k}-{\gamma}^\ast,\cdot)=\tau({\gamma}^\ast)\ \psi({k},\cdot)\right\}
\end{equation}
equipped with the inner product $
\langle\psi;\varphi\rangle_{\sss{H}_\tau}:=\int_{M_{\Gamma^\ast}}(\psi({k});\varphi({k}))_{\sss{H}_\text{f}}\
d^2\underline{k}$ where $d^2\underline{k}:=\frac{d^2{k}}{(2\pi)^2}$
is the normalized measure. There is a natural isomorphism from
$\sss{H}_\tau$ to $L^2\left(M_{\Gamma^\ast},d^2\underline{k}, \,
\sss{H}_\text{f}\right)$ given by restriction from $\R^2$ to
$M_{\Gamma^\ast}$, and with inverse given by $\tau$-covariant
continuation, as suggested by \eqref{eq59b}. The Bloch-Floquet
transform \eqref{eq58} extends to a unitary map
\begin{equation}\label{eq61}
\sss{Z}:\sss{H}_\text{phy}\longrightarrow\sss{H}_\tau\simeq
L^2\left(M_{\Gamma^\ast},d^2\underline{k},\sss{H}_\text{f}\right)\simeq
L^2(M_{\Gamma^\ast},d^2\underline{k})\otimes \sss{H}_\text{f}.
\end{equation}
The Hilbert space $L^2(M_{\Gamma^\ast},d^2\underline{k})$ can be
seen as the \emph{space of slow degrees of freedom} and in this
sense the transform $\sss{Z}$ produces a decomposition of the
physical Hilbert space according to the existence of fast and slow
degrees of freedom.

We need to discuss how differential and multiplication operators
behave under $\sss{Z}$. Let ${Q}=(Q_1,Q_2)$ be the multiplication by
${x}=(x_1,x_2)$ defined on its maximal domain and
 ${P}=(P_1,P_2)=-i{\nabla}_x$ with domain the
 Sobolev space $\ssss{H}^1(\R^2)$, then from \eqref{eq58} it follows:
\begin{equation}\label{eq62}
\sss{Z}\ {P}\ {\sss{Z}}^{-1}={k}\otimes\num{1}_{\sss{H}_\text{f}}+\num{1}_{L^2(M_{\Gamma^\ast})}\otimes -i{\nabla}_{{\theta}},\ \ \ \ \ \ \ \ \ \ \ \
\sss{Z}\ {Q}\ {\sss{Z}}^{-1}=i{\nabla}_{{k}}^\tau
 \end{equation}
where $-i\nabla_{{\theta}}$ acts on the domain $\ssss{H}^1(\num{V})$
while the domain of the differential operator $i{\nabla}_{{k}}^\tau$
is the space $\sss{H}_\tau\cap
\ssss{H}^1_\text{loc}\left(\R^2,\sss{H}_\text{f}\right)$, namely it
consists of  vector-valued distributions which are in
$\ssss{H}^1\left(M_{\Gamma^\ast},\sss{H}_\text{f}\right)$ and
satisfy the ${\theta}$-dependent boundary condition associated with
\eqref{eq59b}. The central feature of the Bloch-Floquet transform
is, however, that multiplication operators corresponding to
$\Gamma$-periodic functions like ${A}_\Gamma$ or $V_\Gamma$  are
mapped into multiplication operators corresponding to the same
function, i.e.
\begin{equation}\label{eq63}
\sss{Z}\ {A}_\Gamma({x})\ {\sss{Z}}^{-1}=\num{1}_{L^2(M_{\Gamma^\ast})}\otimes {A}_\Gamma({\theta})\ \ \ \ \ \ \ \ \ \ \sss{Z}\ V_\Gamma({x})\ {\sss{Z}}^{-1}=\num{1}_{L^2(M_{\Gamma^\ast})}\otimes V_\Gamma({\theta}).
\end{equation}
Let ${H}^\sss{Z}:=\sss{Z}\ H_\text{BL}\ {\sss{Z}}^{-1}$ be the
 Bloch-Floquet transform of the Bloch-Landau Hamiltonian \eqref{eq56}. According to the relations \eqref{eq62} and \eqref{eq63} one obtains
 from (\ref{eq56}) that
\begin{align}\label{eq68}
 H^\sss{Z}=\dfrac{1}{2}\left[-i{\nabla}_{{\theta}}+{k}-{A}_\Gamma\left({\theta}\right)-{A}_0\left(i\varepsilon{\nabla}_{{k}}^\tau\right)-\iota_q\dfrac{1}{2}{e}_\bot \wedge(i\varepsilon{\nabla}_{{k}}^\tau)\right]^2+V_\Gamma\left({\theta}\right)+\phi\left(i\varepsilon{\nabla}_{{k}}^\tau\right)
\end{align}
with domain of self-adjointness $\sss{Z}\ssss{H}^2_\text{M}(\R^2)\subset\sss{H}_\tau$, i.e.  the image under $\sss{Z}$ of the second magnetic-Sobolev space.
\subsection{The periodic Hamiltonian and the gap condition}\label{sec_rel_pat}
When $\varepsilon=0$ the  Bloch-Landau Hamiltonian \eqref{eq56}
reduces to the periodic Hamiltonian
\begin{equation}\label{eq56'}
H_\text{per}=\dfrac{1}{2}\left[-i{\nabla}_x-{A}_\Gamma\left({x}\right)\right]^2+V_\Gamma\left({x}\right).
\end{equation}
According to \eqref{eq68} the Bloch-Floquet transform maps
$H_\text{per}$ into a fibered operator. In other words, denoting
$H_\text{per}^\sss{Z}:=\sss{Z}\ H_\text{per}\ {\sss{Z}}^{-1}$, one
has $
 H_\text{per}^\sss{Z}=\int_{M_{\Gamma^\ast}}^\oplus H_\text{per}({k})\ d^2\underline{k}$
where, for each ${k}\in M_{\Gamma^\ast}$
\begin{equation}\label{eq644}
 H_\text{per}({k})=\dfrac{1}{2}\left[-i\nabla_{{\theta}}+{k}-{A}_\Gamma({\theta})\right]^2+V_\Gamma({\theta}).
\end{equation}
The operator $H_\text{per}({k})$ acts on $\sss{H}_\text{f}=
L^2(\num{V},d^2{\theta})$ with self-adjointness domain
$\sss{D}:=\ssss{H}^2(\num{V})$ (the second Sobolev space)
independent of ${k}\in M_{\Gamma^\ast}$. Moreover it is easy to
check that the Bloch-Floquet transform induces the following
property of periodicity, called \emph{$\tau$-equivariance}:
 \begin{equation}\label{eq_tau_equi_per}
H_\text{per}([k]-{\gamma}^\ast)=\tau({\gamma}^\ast)\ H_\text{per}([k])\ \tau({\gamma}^\ast)^{-1}\in\Gamma^\ast\ \ \ \ \ \ \ \ \ \ \ \forall \gamma^\ast\in\Gamma^\ast.
\end{equation}
 where  the notation $k:=[k]- \gamma^\ast$ denotes the a.e.-unique decomposition of $k\in\R^2$ as a sum of $[k]\in M_{\Gamma^\ast}$  and $\gamma^\ast\in\Gamma^\ast$.

\begin{rk}[Analiticity]\label{rk_analyt}
For any $k\in\R^2$, let $I(k)$ be the unitary operator acting on
$\sss{H}_\text{f}$ as the multiplication by $\expo{-i \theta\cdot
k}$. Obviously
$I(k)=I([k]-\gamma^\ast)=I([k])\tau(\gamma^\ast)^{-1}$. A simple
computation shows that
\begin{equation}\label{eq_analitic}
 H_\text{per}({k})=I(k)\ H_\text{per}(0)\ I(k)^{-1}
\end{equation}
where the equality holds on the fixed domain of self-adjointness
$\sss{D}=\ssss{H}^2(\num{V})$. The $\tau$-equivariance property
\eqref{eq_tau_equi_per} follows immediately from
\eqref{eq_analitic}. Moreover from \eqref{eq_analitic} is evident
that $H_\text{per}({k})$ defines an \emph{analytic family (of type
A) in the sense of Kato} (see \cite{red-sim4} Chapter XII). Finally
a short computation shows
$$
(\partial_{k_j}H_\text{per})(k)=-iI(k)\ [\theta_j;H_\text{per}(0)]\ I(k)^{-1}=I(k)\ \left(-i{\nabla}_\theta-{A}_\Gamma\left({\theta}\right)\right)_j\ I(k)^{-1}
$$
and $(\partial^2_{k_j}H_\text{per})(k)=\num{1}_\sss{D}$, $(\partial^2_{k_1,k_2}H_\text{per})(k)=0$ on the domain $\sss{D}$.
\hfill $\blacklozenge\lozenge$
\end{rk}

The spectrum of $H_\text{per}$, which coincides with the spectrum of
$H_\text{per}^\sss{Z}$, is given by the union of all the spectra of
$H_\text{per}({k})$. The following classical results hold true:
\begin{propos}\label{prop2}
Let  $V_\Gamma$ and ${A}_\Gamma$  satisfy {\upshape Assumption
($\text{A}_\text{w}$)}, then:
\begin{enumerate}
\item[{\upshape (i)}] for all ${k}\in \R^2$ the operator  $H_\text{{\upshape per}}({k})$ defined by \eqref{eq56'} is self-adjoint with domain $\sss{D}=\ssss{H}^2(\num{V})$ and is  bounded below;
\item[{\upshape (ii)}] $H_\text{{\upshape per}}({k})$  has compact resolvent and its spectrum  is purely discrete with eigenvalues $\sss{E}_n({k})\to+\infty$ as $n\to+\infty$;
\item[{\upshape (iii)}] let the eigenvalues be arranged in increasing order and repeated according to their multiplicity for any ${k}\in M_{\Gamma^\ast}$, i.e.  $\sss{E}_1({k})\leqslant \sss{E}_2({k})\leqslant \sss{E}_3({k})\leqslant\ldots$ then $\sss{E}_n({k})$ is a continuous $\Gamma^\ast$-periodic function of ${k}$.
\end{enumerate}
\end{propos}
The above result differs from the standard theory of periodic
Schr\"{o}dinger operators just for the presence of a periodic vector
potential $A_\Gamma$. Since we were no able to find a suitable
reference in the literature, we sketch its proof in  Appendix
\ref{appA1}.

\medskip

We  call $\sss{E}_n(\cdot)$ the  $n^\text{th}$ \emph{Bloch band} or \emph{energy band}.
The corresponding normalized eigenstates $\{\varphi_n({k})\}_{n\in\N}\subset \sss{D}$ are called {\it Bloch functions} and form, for any ${k}\in M_{\Gamma^\ast}$, an orthonormal basis of $\sss{H}_\text{f}$.  Notice that, with this choice of the labelling, $\sss{E}_n(\cdot)$ and $\varphi_n(\cdot)$ are continuous in $k$, but generally they are not smooth functions  if \emph{eigenvalue crossings} are present.

We say that a family of Bloch bands
$\{\sss{E}_n(\cdot)\}_{n\in\sss{I}}$, with
$\sss{I}:=[I_+,I_-]\cap\N$, is \emph{ isolated} if
\begin{equation}\label{eq64}
\inf_{{k}\in M_{\Gamma^\ast}}\ \text{{\upshape dist}}\left(\bigcup_{n\in\sss{I}}\{\sss{E}_n({k})\},\bigcup_{j\notin\sss{I}}\{\sss{E}_j({k})\}\right)=C_g>0.
\end{equation}
The existence of an isolated part of the spectrum is a necessary ingredient for an adiabatic theory.
We introduce the following:\\
\\
{\bf \textsc{Assumption} (C) [constant gap condition].} {\it The spectrum of $H_\text{{\upshape per}}$ admits a family of Bloch bands $\{\sss{E}_n(\cdot)\}_{n\in\sss{I}}$ which is isolated in the sense of \eqref{eq64}.}\\
\\
Let $P_\sss{I}({k})$ be the spectral projector of
$H_\text{per}({k})$ corresponding to the family of eigenvalues
$\{\sss{E}_n({k})\}_{n\in\sss{I}}$, then
$P_\sss{I}^\sss{Z}:=\int_{M_{\Gamma^\ast}}^\oplus P_\sss{I}({k})\
d^2\underline{k}$ is the projector on the  isolated family of Bloch
bands labeled by $\sss{I}$. In terms of Bloch functions (using the
Dirac notation), one has that
$P_\sss{I}(\cdot)=\sum_{n\in\sss{I}}\ketbra{\varphi_n(\cdot)}{\varphi_n(\cdot)}$.
However, in general, $\varphi_n(\cdot)$ are not smooth functions of
${k}$ at eigenvalue crossing, while $P_\sss{I}(\cdot)$ is a smooth
function of ${k}$ because of the gap condition. Moreover, from the
periodicity of $H_\text{per}(\cdot)$, one argues
$P_\sss{I}([k]-{\gamma}^\ast)=\tau({\gamma}^\ast)\ P_\sss{I}([k])\
\tau({\gamma}^\ast)^{-1}$.
In general the smoothness of $P_\sss{I}(\cdot)$ is not enough to assure
 the existence of family of orthonormal basis for the subspaces
 $\text{{\upshape Ran}}P_\sss{I}(\cdot)$ which varies smoothly
 (or only continuously) with respect to $k\in M_{\Gamma^\ast}$. Then we need the following assumption.\\
\\
{\bf \textsc{Assumption} (D) [smooth frame].} {\it  Let
$\{\sss{E}_n(\cdot)\}_{n\in\sss{I}}$ be a family of Bloch bands
($|\sss{I}|=m>1$). We assume that there exists an orthonormal basis
$\{\psi_n(\cdot)\}_{j=1}^m$ of $\text{{\upshape
Ran}}P_\sss{I}(\cdot)$ whose elements are smooth and (left)
$\tau$-covariant with respect to ${k}$, i.e.
$\psi_j(\cdot-{\gamma}^\ast)=\tau({\gamma}^\ast)\psi_j(\cdot)$ for
all $j=1,\ldots,m$ and $\gamma^\ast\in\Gamma^\ast$.
}\\
\\
Note that it is not required that $\psi_j({k})$  is an eigenfunction
of $H_\text{per}({k})$. However, in the special but important case
in which the family of bands consist of a single isolated $m$-fold
degenerate eigenvalue, i.e. $\sss{E}_n({k})=\sss{E}_\ast({k})$ for
every $n=1,\ldots,m$, then the Assumption (D)  is equivalent to the
existence of an orthonormal basis consisting of smooth and
$\tau$-covariant Bloch functions.

\begin{rk}[Time-reversal symmetry breaking] As far as low dimensional models
are concerned ($ d \leq 3$), Theorem 1 in \cite{Pan} assures that
Assumption (D) is true  whenever the Hamiltonian $H_\text{per}$ is
invariant with respect to the \emph{time-reversal symmetry}, which
is implemented in the Schr\"{o}dinger representation by the complex conjugation operator.
However, the term $A_\Gamma\neq0$ in $H_\text{per}$ generically
breaks the time reversal symmetry. Therefore, to consider also the
effects due to a periodic vector potential, we need to assume the
existence of a smooth family of frames. Anyway is opinion of the
authors that the result in \cite{Pan} can be extended to the case of
a periodic vector potential, at least assuming that
$A_\Gamma $ is small in a suitable sense.\hfill
$\blacklozenge\lozenge$
\end{rk}

Let ${k}_0$ be a fixed point in $M_{\Gamma^\ast}$ and define the
projection $\pi_\text{r}:=P_\sss{I}({k}_0)$. If the Assumption (C)
holds true then $\text{dim}\ \pi_\text{r}=\text{dim}\
P_\sss{I}({k})$ for all ${k}\in\R^2$. Let $\{\chi_n\}_{j=1}^m$ be an
orthonormal basis for $\text{Ran}\ \pi_\text{r}$ and define a
unitary map
\begin{equation}\label{Unitary}
   u_0({k}):=\widetilde{u}_0({k})+{u}_0^\bot({k}), \qquad \qquad
   \text{with}\qquad \qquad\widetilde{u}_0({k}):=\sum_{1 \leq j \leq \ell} \ketbra{\chi_j}{\psi_j({k})},
\end{equation}
which maps  $\text{Ran}\ P_\sss{I}({k})$ in $\text{Ran}\
\pi_\text{r}$. The definition of this unitary is not unique because
the freedom in the choice of the frame and of the orthogonal
complement ${u}_0^\bot({k})$. From the definition and the
$\tau$-covariance of $\psi_j(\cdot)$ one has that $ u_0({k})\
P_\sss{I}({k})\ u_0({k})^{-1}=\pi_\text{r}$ and
$u_0([k]-{\gamma}^\ast)=u_0([k])\ \tau({\gamma}^\ast)^{-1}$
(\emph{right $\tau$-covariance}).
\subsection{$\tau$-equivariant and special $\tau$-equivariant symbol classes}\label{appA2}
Proposition \ref{prop2} shows that for all $k\in\R^2$, the operator
$H_\text{per}(k)$ defines an unbounded self-adjoint operator on the
Hilbert space $\sss{H}_\text{f}$ with dense domain
$\sss{D}:=\ssss{H}^2(\num{V})$. However the domain $\sss{D}$ can be
considered itself as a Hilbert space with respect to the Sobolev
norm
$\|\cdot\|_{\sss{D}}:=\|(\num{1}_{\sss{H}_\text{f}}-\Delta_{{\theta}})\cdot\|_{\sss{H}_\text{f}}$
and so $H_\text{per}(k)$ can be seen as a bounded linear operator
from $\sss{D}$ to $\sss{H}_\text{f}$, i.e. as an element of the
Banach space $\bbb{B}(\sss{D},\sss{H}_\text{f})$. The map $\R^2\ni
k\mapsto H_\text{per}(k)\in \bbb{B}(\sss{D},\sss{H}_\text{f})$ is  a
special example of a \emph{operator-valued symbol}. For a summary
about the theory of the Weyl quantization of vector-valued symbols,
we refer to Appendices A and B in \cite{stefan_book}. In what
follows we will need the following definition.
\begin{defi}[H\"{o}rmander symbol classes]\label{hor_sym} A {\upshape symbol} is any map $F$ from the (cotangent) space $\R^2\times\R^2$ to the  Banach space $\bbb{B}(\sss{D},\sss{H}_\text{{\upshape f}})$, i.e. $\R^2\times\R^2\ni (k,\eta)\mapsto F(k,\eta)\in \bbb{B}(\sss{D},\sss{H}_\text{{\upshape f}})$. A function $w:\R^2\times\R^2\to[0,+\infty)$ is said to be an {\upshape order function} if there exists constants $C_0>0$ and $N_0>0$ such that
\begin{equation}\label{order function}
w(k,\eta)\leqslant
C_0\left(1+|k-k'|^2+|\eta-\eta'|^2\right)^{\frac{N_0}{2}}w(k',\eta')
\end{equation}
for every $(k,\eta),(k',\eta')\in\R^2\times\R^2$. A symbol $F\in
C^\infty(\R^2\times\R^2,\bbb{B}(\sss{D},\sss{H}_\text{{\upshape
f}}))$ is an element of the {\upshape (H\"{o}rmander) symbol class}
$S^w(\bbb{B}(\sss{D},\sss{H}_\text{{\upshape f}}))$ with order
function $w$, if for every $\alpha,\beta\in\N^2$ there exists a
constant $C_{\alpha,\beta}>0$ such that
$\|(\partial_k^\alpha\partial_\eta^\beta
F)(k,\eta)\|_{\bbb{B}(\sss{D},\sss{H}_\text{{\upshape f}})}\leqslant
\, C_{\alpha,\beta} \, w(k,\eta)$ for every
$(k,\eta)\in\R^2\times\R^2$.
\end{defi}
According to the previous definition, the vector-valued map
$H_\text{per}(\cdot)$ defines a H\"{o}rmander symbol constant in the
$\eta$-variables and with order function $v(k,\eta):=1+|k|^2$ (see
the proof of Proposition \ref{prop3} below). However, as showed by
equation \eqref{eq_tau_equi_per}, the symbol $H_\text{per}(\cdot)$
satisfies an extra condition of periodicity.

\begin{defi}[$\tau$-equivariant symbols]
Let $\Gamma^\ast$ be a two dimensional lattice (the dual lattice
defined in Section \ref{Sec_mod} for our aims) and
$\tau:\Gamma^\ast\to\bbb{U}(\sss{H}_\text{{\upshape f}})$ the
unitary representation defined in Section \ref{sec_BF}. Denote by
$\widetilde{\tau}:=\left.\tau\right|_{\sss{D}}$ the bounded-operator
\footnote{Clearly $\tau(\gamma^\ast)$ acts as an invertible bounded
operator on the space $\sss{D}$, but it is no longer unitary with
respect to the Sobolev-norm defined on $\sss{D}$.} representation of
$\Gamma^\ast$ in $\sss{D}$. A symbol $F\in
S^w(\bbb{B}(\sss{D},\sss{H}_\text{{\upshape f}}))$ is said to be
{\upshape $\tau$-equivariant} if
$$
F(k-\gamma^\ast,\eta)=\tau(\gamma^\ast)\ F(k,\eta)\
\widetilde{\tau}(\gamma^\ast)^{-1}\ \ \ \ \ \ \ \ \ \forall\
\gamma^\ast\in\Gamma^\ast, k \in \R^2.
$$
The space of $\tau$-equivariant symbols is denoted as
$S^w_\tau(\bbb{B}(\sss{D},\sss{H}_\text{{\upshape f}}))$.
\end{defi}

For the purposes of this work, it is convenient to focus on special
classes of symbols. By considering the \emph{kinetic momentum
function} $
\R^2\times\R^2\ni(k,\eta)\stackrel{\kappa}{\longmapsto}\kappa(k,\eta):=k-A(\eta)\in
\R^2, $ with $A$ fulfilling Assumption (B), one defines the
\emph{minimal coupling map} by
\begin{equation}
(k,\eta)\stackrel{\jmath_\kappa}{\longmapsto}\jmath_\kappa(k,\eta):=(\kappa(k,\eta),\eta)\in\R^2\times\R^2.
\end{equation}

\begin{defi}[Special $\tau$-equivariant symbols]
Let $w$ be an order function, in the sense of (\ref{order
function}). We define
$$
S^w_{\kappa;\tau}(\bbb{B}(\sss{D},\sss{H}_\text{{\upshape
f}})):=\{\widetilde{F} = F\circ\jmath_\kappa\ :\ F\in
S^w_\tau(\bbb{B}(\sss{D},\sss{H}_\text{{\upshape f}}))\}.
$$
\end{defi}

We refer to
$S^w_{\kappa;\tau}(\bbb{B}(\sss{D},\sss{H}_\text{{\upshape f}}))$ as
the class of \emph{special $\tau$-equivariant symbols}. The
following result shows that special symbols can be considered as
genuine $\tau$-equivariant symbols with respect to a modified order
function. The key ingredient is the linear growth of the kinetic
momentum.

\begin{lem}\label{lem_app_A2}
With the above  notations
$S^w_{\kappa;\tau}(\bbb{B}(\sss{D},\sss{H}_\text{{\upshape
f}}))\subset S^{w'}_\tau(\bbb{B}(\sss{D},\sss{H}_\text{{\upshape
f}}))$ where $w':=w\circ \jmath_\kappa$.
\end{lem}
\Proof If $F\in S^w_\tau(\bbb{B}(\sss{D},\sss{H}_\text{{\upshape
f}}))$ then also $F\circ\jmath_\kappa$ is $\tau$-equivariant, indeed
$\kappa(k-\gamma^\ast,\eta)=\kappa(k,\eta)-\gamma^\ast$ and
$(F\circ\jmath_\kappa)(k-\gamma^\ast,\eta)=\tau(\gamma^\ast)\
(F\circ\jmath_\kappa)(k,\eta)\ \widetilde{\tau}(\gamma^\ast)^{-1}$.
Since $\jmath_\kappa$ is a smooth function, then also the
composition $F\circ\jmath_\kappa$ is a smooth function. Observing
that $(F\circ\jmath_\kappa)(k,\eta)=F(k-A(\eta),\eta)$ it follows
that
\begin{align*}
&(\partial_{k_j}(F\circ\jmath_\kappa))(k,\eta)=((\partial_{k_j}F)\circ\jmath_\kappa)(k,\eta)\\
&(\partial_{\eta_j}(F\circ\jmath_\kappa))(k,\eta)=((\partial_{\eta_j}F)\circ\jmath_\kappa)(k,\eta)+\sum_{i=1}^2(\partial_{\eta_j}\kappa_i)(k,\eta)\
((\partial_{k_i}F)\circ\jmath_\kappa)(k,\eta)
\end{align*}
where $\partial_{\eta_j}\kappa_i$ are bounded functions because
Assumption (B). From the first equation it follows that
$$
\|\partial_{k_j}(F\circ\jmath_\kappa)(k,\eta)\|_{\bbb{B}(\sss{D},\sss{H}_\text{f})}\leqslant
C_{j,0}\ (w\circ \jmath_\kappa)(k,\eta)\ \ \ \ \ \ \ j=1,2
$$
for suitable positive constants $C_{j,0}$. Similarly the second
equation implies
$$
\|\partial_{\eta_j}(F\circ\jmath_\kappa)(k,\eta)\|_{\bbb{B}(\sss{D},\sss{H}_\text{f})}\leqslant [C_{0,j}+K(C_{1,0}+C_{2,0})] (w\circ \jmath_\kappa)(k,\eta).
$$
where $K>0$ is a bound for the functions $\partial_{\eta_j}\kappa_i$. By an inductive argument on the number of the derivatives one can proof that the derivatives of $F\circ\jmath_\kappa$ are bounded by the function $w':=w\circ \jmath_\kappa$. To complete the proof we need to show that $w'$ is an order function according to Definition \ref{hor_sym}. This follows by a simple computation using the fact that $\kappa$ has a linear growth in $k$ and $\eta$.
\CVD

In view of Lemma \ref{lem_app_A2}, all the results of Appendix B of
\cite{stefan_book} hold true for symbols in
$S^w_{\kappa;\tau}(\bbb{B}(\sss{D},\sss{H}_\text{f}))$ and in
particular  the quantization of a symbol in
$S^w_{\kappa;\tau}(\bbb{B}(\sss{D},\sss{H}_\text{f}))$ preserves the
$\tau$-equivariance  and that the pointwise product or the Moyal
product of two symbols of order $w_1$ and $w_2$ produce a symbol of
order $w_1w_2$ (see \cite{stefan_book}, Propositions B.3 and B.4).
\begin{rk}[Notation]\label{not_k}
In what follows we will use the short notation $F(\kappa;\eta):=(F\circ\jmath_\kappa)(k,\eta)$ to denote the special symbol
$F\circ\jmath_\kappa\in S^w_{\kappa;\tau}(\bbb{B}(\sss{D},\sss{H}_\text{{\upshape f}}))$ related to the $\tau$-equivariant symbols $F\in S^w_\tau(\bbb{B}(\sss{D},\sss{H}_\text{{\upshape f}}))$. We emphasize on the use of the semicolon \virg{;} instead the comma \virg{,} and of the symbol of the kinetic momentum $\kappa$ instead the quasi-momentum $k$.\\
\phantom{a}\hfill $\blacklozenge\lozenge$
\end{rk}

\subsection{Semiclassics: quantization of equivariant symbols}\label{sec_quan_pair}

As explained in Section \ref{sec_BF}, the Bloch-Floquet transform
$\sss{Z}$ provides the separation between the fast degrees of
freedom, associated to the Hilbert space
$\sss{H}_\text{f}=L^2(\num{V},d^2{\theta})$, and the slow degrees of
freedom, associated to the Hilbert
$L^2(M_{\Gamma^\ast},d^2\underline{k})$. A fruitful point of view is
to consider the slow degrees of freedom \virg{classical} with
respect to the \virg{quantum} fast degrees of freedom.
Mathematically, this is achived by recognizing that the Hamiltonian
$H^\sss{Z}$ defined in \eqref{eq68} is the  \emph{Weyl quantization}
of an operator-valued \virg{semiclassical} symbol over the classical
phase space $\R^2\times\R^2$. As explained rigorously in the
Appendices A and B of \cite{stefan_book}, the quantization procedure
maps an operator-valued symbol
$F:\R^2\times\R^2\to\bbb{B}(\sss{D},\sss{H}_\text{{\upshape f}})$
into a linear operator
$\text{Op}_\varepsilon(F):\sss{S}(\R^2,\sss{D})\to\sss{S}(\R^2,\sss{H}_\text{f})$,
where $\sss{S}(\R^2,\sss{H})$ denotes the space of $\sss{H}$-valued
Schwartz functions. The quantization procedure concerns only the
slow degrees of freedom and at a formal level can be identified with
the prescription
\begin{equation}\label{eq100}
k\longmapsto\text{Op}_\varepsilon(k):=\text{multiplication by}\
k\otimes\num{1}_\sss{D};\ \ \ \ \ \ \
\eta\longmapsto\text{Op}_\varepsilon(\eta):=i\varepsilon\nabla_k\otimes\num{1}_\sss{D}.
\end{equation}
 Let us consider the operator-valued symbol ${H}_0:\R^2\times\R^2\to\bbb{B}(\sss{D},\sss{H}_\text{{\upshape f}})$ defined by
\begin{align}\label{eq69}
{H}_0({k},\eta)&:=
\dfrac{1}{2}\left[-i{\nabla}_{{\theta}}+{k}-{A}_\Gamma\left({\theta}\right)-{A}_0({\eta})-\iota_q\dfrac{1}{2}{e}_\bot\wedge
\eta\right]^2+V_\Gamma\left({\theta}\right)+\phi(\eta).
\end{align}
The symbol ${H}_0$ does not depend on $\varepsilon$ and in view of
Proposition \ref{prop2} it defines an unbounded operator on
$\sss{H}_\text{f}$  with domain of self-adjointness
$\sss{D}=\ssss{H}^2(\num{V})$ for all choice of
$(k,\eta)\in\R^2\times\R^2$. According to the notation of Section
\ref{appA2} and comparing \eqref{eq69} with \eqref{eq644} we can
write
\begin{equation}\label{eq6969}
{H}_0(k,\eta)=
H_\text{per}\left({\kappa}({k},{\eta})\right)+\phi({\eta})=(H_\phi\circ \jmath_\kappa)(k,\eta).
\end{equation}
where $H_\phi(k,\eta):=H_\text{per}(k)+\phi(\eta)$. As suggested by equation \eqref{eq_tau_equi_per},  $H_\phi$ is a
$\tau$-equivariant symbol. Thus the symbol ${H}_0$ is
$\tau$-equivariant with respect to the kinetic momentum $\kappa$.
The following result establishes the exact symbol class for ${H}_0$.

\begin{propos}\label{prop3}
If {\upshape Assumption ($\text{A}_\text{w}$)} and {\upshape (B)}
hold true then ${H}_0\in
S^v_{\kappa;\tau}(\bbb{B}(\sss{D};\sss{H}_\text{{\upshape f}}))$
with order function $v(k,\eta):=1+|k|^2$.
\end{propos}
\Proof
Using the result of Lemma \ref{lem_app_A2}, we only need to show that $H_\phi\in S^v(\bbb{B}(\sss{D},\sss{H}_\text{f}))$. The later claim is easy to verify, indeed the derivative in $\eta$ are bounded functions, the second derivative in $k$ is a constant and the derivatives of higher order in $k$ are zero. Then we need only to check the growth of the first derivative in $k$. A simple computation shows that $$\|(\partial_{k_j}H_\phi)(k,\eta)\|_{\bbb{B}(\sss{D},\sss{H}_\text{f})}=\|(\partial_{k_j}H_{\text{per}})(k)\ (\num{1}_{\sss{H}_\text{f}}-\Delta_\theta)^{-1}\|_{\bbb{B}(\sss{H}_\text{f})}$$ and since $\partial_{k_j}H_{\text{per}}$
is  $\tau$-equivariant
(see Remark \ref{rk_analyt}), then
$$\|(\partial_{k_j}H_\phi)(k,\eta)\|_{\bbb{B}(\sss{D},\sss{H}_\text{f})}=\|(\partial_{k_j}H_{\text{per}})([k])\tau(\gamma^\ast)^{-1}(\num{1}_{\sss{H}_\text{f}}-\Delta_\theta)^{-1}\|_{\bbb{B}(\sss{H}_\text{f})}.$$ Observing that $\tau(\gamma^\ast)$ is the multiplication by $\expo{i\theta\cdot\gamma^\ast}$ in $\sss{H}_\text{f}$ and by a simple computation that $(\partial_{k_j}H_{\text{per}})([k])\tau(\gamma^\ast)^{-1}=\tau(\gamma^\ast)^{-1}[-2\gamma^\ast_j+(\partial_{k_j}H_{\text{per}})([k])]$ one has
$$
\|(\partial_{k_j}H_\phi)(k,\eta)\|_{\bbb{B}(\sss{D},\sss{H}_\text{f})}\leqslant C_1|\gamma^\ast_j|+\|(\partial_{k_j}H_{\text{per}})([k])\|_{\bbb{B}(\sss{D},\sss{H}_\text{f})}\leqslant C_1(|k|+C_3)+C_2
$$
where $C_1=2\|(\num{1}_{\sss{H}_\text{f}}-\Delta_\theta)^{-1}\|_{\bbb{B}(\sss{H}_\text{f})}$, $C_2:=\max_{k\in M_{\Gamma^\ast}}\|(\partial_{k_j}H_{\text{per}})([k])\|_{\bbb{B}(\sss{D},\sss{H}_\text{f})}$ and $|\gamma^\ast_j|\leqslant|\gamma^\ast|=|k-[k]|\leqslant |k|+C_3$ with $C_3:=\max_{k\in M_{\Gamma^\ast}}|k|$. The claim follows observing that $1+|k|\leqslant 2(1+|k|^2)$.
\CVD

\begin{figure}[htbp]
\begin{center}
\fbox{
\includegraphics[width=12cm]{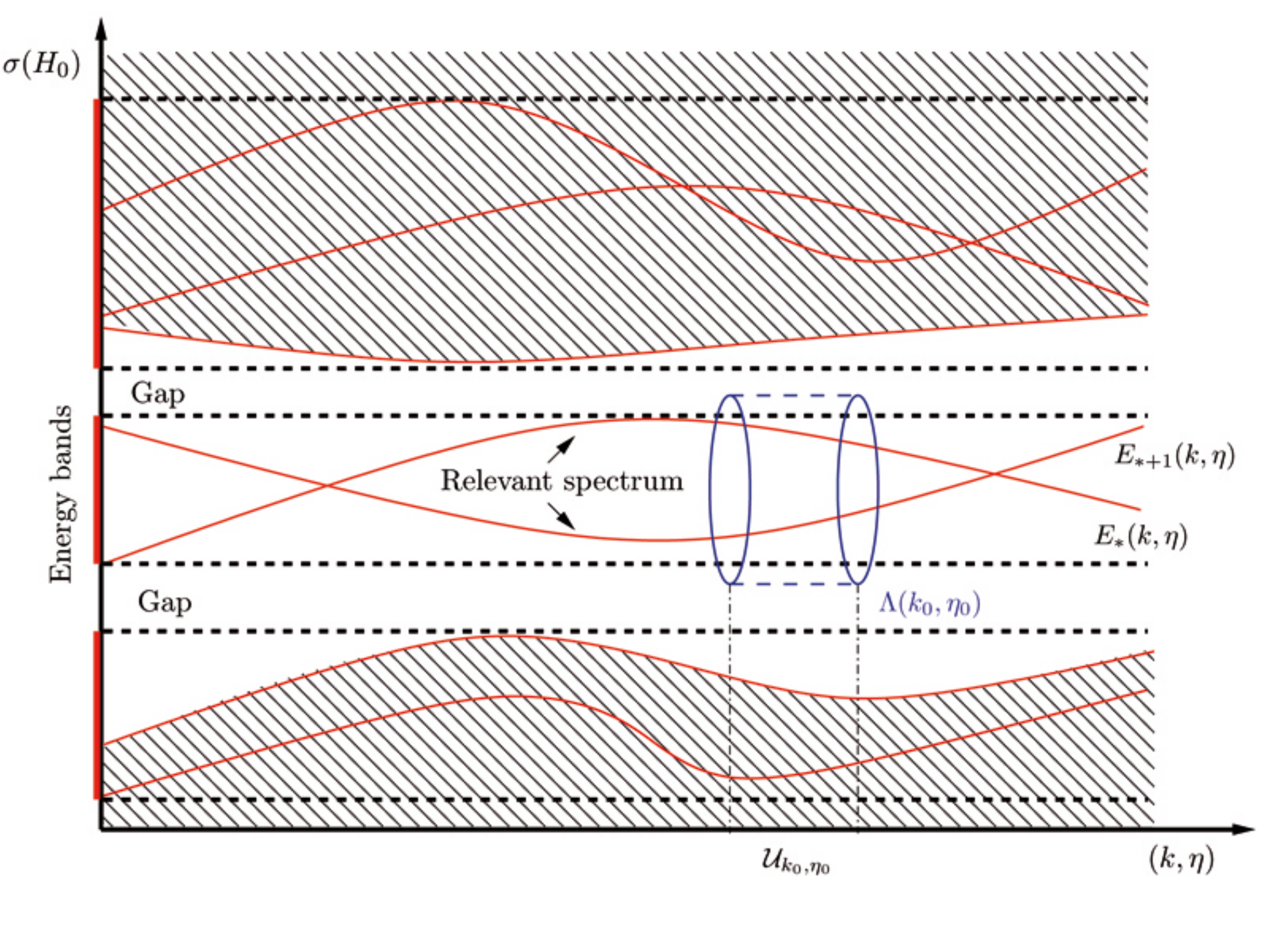}
}
\end{center}
\caption{{\footnotesize Structure of the spectrum of $H_0(k,\eta)$.
The picture shows schematically a \virg{relevant part of the
spectrum},  consisting of two energy bands $\{E_{\ast},
E_{\ast+1}\}$, with
$E_{\ast+j}(k,\eta)=\sss{E}_{\ast+j}(\kappa(k,\eta))+\phi(\eta)$.
Notice that we assume only a local gap condition, as stated in
(\ref{eq72}), while in the picture a stronger condition is
satisfied: a gap exists when projecting the relevant bands on the
vertical axis.}} \label{fig01}
\end{figure}

Equation \eqref{eq6969} provides information
about the dependence on $k$ and $\eta$ of the spectrum of $H_0$. The
$n^{\text{th}}$ eigenvalue $E_n(k,\eta)$ of the operator
$H_0(k,\eta)$ is related to the $n^{\text{th}}$ eigenvalue
$\sss{E}_n(k)$ of the periodic Hamiltonian $H_\text{per}(k)$ by the
relation $E_n({k},\eta)=\sss{E}_n({\kappa}({k},\eta))+\phi(\eta)$.
The  function $E_n:\R^2\times\R^2\to\R$
 is still $\Gamma^\ast$-periodic in ${k}$ but only oscillating with bounded variation in $\eta$.
Assumption (C) for the family of Bloch  bands $\{\sss{E}_n(\cdot)\}_{n\in\sss{I}}$ immediately implies that
\begin{equation}\label{eq72}
\inf_{({k},\eta)\in M_{\Gamma^\ast}\times\R^2}\ \text{{\upshape dist}}\left(\bigcup_{n\in\sss{I}}\{E_n({k},\eta)\},\bigcup_{j\notin\sss{I}}\{E_j(k,\eta)\}\right)=C_g>0.
\end{equation}
This is the relevant part of the spectrum of $H_0$ which we are interested in.

According to the general theory (see Appendices A and B of
\cite{stefan_book}), one has that
\begin{equation}\label{eq74}
\text{Op}_\varepsilon(H_0)=\dfrac{1}{2}\left[-i{\nabla}_{{\theta}}+{k}-{A}_\Gamma({\theta})-{A}_0\left(i\varepsilon{\nabla}_{{k}}\right)
-\iota_q\dfrac{1}{2}{e}_\bot\wedge(i\varepsilon{\nabla}_{{k}})\right]^2+V_\Gamma({\theta})+\phi\left(i\varepsilon{\nabla}_{{k}}\right)
\end{equation}
defines a linear operator from $\sss{S}(\R^2,\sss{D})$ in
$\sss{S}(\R^2,\sss{H}_\text{f})$  and by duality it extends to a continuous
mapping $\text{Op}_\varepsilon(H_0):\sss{S}'(\R^2,\sss{D})\to
\sss{S}'(\R^2,\sss{H}_\text{f})$ (with an abuse of notation we use
the same symbol for the extended operator). The $\tau$-equivariance
assures that
$\text{Op}_\varepsilon(H_0)\varphi([k]-\gamma^\ast)=\tau(\gamma^\ast)\text{Op}_\varepsilon(H_0)\varphi([k])$
(see \cite{stefan_book} Proposition B.3). Since
$\text{Op}_\varepsilon(H_0)$ preserves $\tau$-equivariance it can
then be restricted to an operator on the domain
$\sss{Z}\ssss{H}^2_\text{M}(\R^2)\subset \sss{S}'(\R^2,\sss{D})$
which is the domain of self-adjointness of $H^\sss{Z}$, according to
\eqref{eq68}. To conclude that $\text{Op}_\varepsilon(H_0)$,
restricted to $\sss{Z}\ssss{H}^2_\text{M}(\R^2)$, agrees with
$H^\sss{Z}$ it is enough to recall that $i\nabla_{{k}}^\tau$ is
defined as $i\nabla_{{k}}$ restricted to its natural domain
$\ssss{H}^1\left(\R^2,\sss{D}\right)\cap\sss{H}_\tau$ and to use the
spectral calculus. These arguments justify the following:
\begin{propos}
The Hamiltonian $H^\sss{Z}$, defined by \eqref{eq68}, agrees on its domain of definition with the Weyl quantization of the operator-valued symbol $H_0$ defined by \eqref{eq69}.
\end{propos}
With a little abuse of notation, we refer to this result by writing
$H^\sss{Z}=\text{Op}_\varepsilon(H_0)$.

\subsection{Main result: effective dynamics for weak magnetic fields}\label{Sec_Hofs_ham}
Let $A_{\epsi}$ and $B_{\epsi}$ be  $\varepsilon$-dependent
(possibly unbounded) linear operators in $\Hi$. We write $A_{\epsi}
= B_{\epsi} + \sss{O}_0(\epsi^{\infty})$ if: for any $N \in \N$
there exist a positive constant $C_N$ such that
\begin{equation}\label{Approx}
    \|A_{\epsi} - B_{\epsi} \|_{\B(\Hi)} \leq C_N \, \epsi^{N}
\end{equation}
for every $\epsi \in [0, \epsi_0)$. Notice that, though the
operators are unbounded, the difference is required to be a bounded
operator.

We refer to Appendices A and B of \cite{stefan_book} for the basic
terminology concerning pseudodifferential operators, and in
particular as for the notions of \emph{principal symbol},
\emph{asymptotic expansion}, \emph{resummation}, \emph{Moyal
product}.

\bigskip

\begin{teo}\label{teo2}
Let {\upshape Assumptions ($\text{A}_\text{w}$)},
{\upshape (B)}, {\upshape (C)} and {\upshape (D)} be satisfied and
let $\{E_n(\cdot)\}_{n\in\sss{I}}$ (with $|\sss{I}|=m$) be an
isolated family of energy bands for $H_0$ satisfying condition
\eqref{eq72}. Then:

\medskip

\noindent \textbf{1. Almost-invariant subspace:} there exist an
orthogonal projection $\Pi_\epsi \in\bbb{B}(\sss{H}_\tau)$, with
$\Pi_\epsi = \text{{\upshape
Op}}_\varepsilon(\pi)+\sss{O}_0(\varepsilon^\infty)$ and the symbol
$\pi({k},\eta)\asymp\sum_{j=0}^\infty \varepsilon^j\ \pi_j(k,\eta)$
having principal part $\pi_0({k},\eta)= P_\sss{I}({k}-{A}(\eta))$,
so that
$$
[{H}^\sss{Z};\Pi_\varepsilon]=\sss{O}_0(\varepsilon^\infty).
$$
In particular for any 
$N \in \N$ there exist a
$C_N$ such that
\begin{equation}\label{eq_hof_stima1}
\| (1 - \Pi_\varepsilon) \,\, \expo{ - i \frac{t}{\varepsilon}
{H}^\sss{Z}} \,\, \Pi_\varepsilon \| \leq  C_N\ \varepsilon^N\
|t|
\end{equation}
for $\epsi$ sufficiently small, $t \in \R$.

\medskip

\noindent \textbf{2. Effective dynamics:} let $\sss{H}_\text{\upshape ref}=
L^2({M_{\Gamma^\ast}},d^2\underline{k})\otimes\sss{H}_\text{{\upshape
f}}$,  $\pi_\text{{\upshape r}}$ as defined above (\ref{Unitary}) and $
\Pi_{\rm r} = \num{1}_{L^2({M_{\Gamma^\ast}})} \otimes \pi_\text{{\upshape r}} \in
\bbb{B}(\sss{H}_\text{\upshape ref})$. Then there exist a unitary operator
$$
U_\varepsilon:\sss{H}_\tau\to\sss{H}_\text{{\upshape ref}}
$$ such that
\renewcommand{\labelenumi}{{\rm(\roman{enumi})}}
\begin{enumerate}
    \item $ U_\varepsilon =\Op_\varepsilon(u)+\sss{O}_0(\varepsilon^\infty)$,
    where the symbol $u\asymp\sum_{j=0}^{\infty}\varepsilon^j u_j$
    has principal part $u_0$ given by (\ref{Unitary}) with  $k$ replaced by $\kappa(k,\eta)$;
    \item $\Pi_\text{{\upshape r}}=U_\varepsilon \ \Pi_\varepsilon\ {U_\varepsilon}^{-1};$
    \item posing $\sss{K} := \Pi_\text{\upshape r} \sss{H}_\text{\upshape ref}$,  one has
    $$
    U_\varepsilon \, \Pi_\varepsilon \, {H}^\sss{Z} \, \Pi_\varepsilon \, U_\varepsilon^{-1} =  {H}^\varepsilon_\text{{\upshape
eff}}+\sss{O}_0(\varepsilon^\infty) \in\bbb{B}(\sss{K})
    $$
    with {${H}_{\text{\upshape eff}}^\varepsilon = \Op_\epsi(h)$} and $h$ a
    resummation of the formal symbol $u\, \moy\, \pi \,\moy \, H_0 \, \moy \, \pi
    \moy u^{-1}$ (thus algorithmically computable at any finite order).
Moreover,
\begin{equation}\label{eq_hof_stima2}
\|  (\expo{ -i \frac{t}{\varepsilon} {H}^\sss{Z}}  -
U_\varepsilon^{-1} \, \expo{ - i \frac{t}{\varepsilon}
H^\varepsilon_{\text{\upshape eff}}} \, U_{\varepsilon}) \,
\Pi_\varepsilon \| \leq {C_N}\!\!' \,\, \ \varepsilon^N\
({\varepsilon}+|t|).
\end{equation}
\end{enumerate}
\end{teo}

\begin{rk} The previous theorem  and the following proof generalize straightforwardly to any dimension $d \in \N$. We prefer to state it only
in the case $d =2$ in view of the application to the QHE and of the
comparison with the results in Section 3, the latter being valid
only for $d =2$.  \hfill $\blacklozenge\lozenge$
\end{rk}


\goodbreak

\subsection*{Proof of Theorem \ref{teo2}}
\label{Sec_proof Hof}

\subsubsection*{Step 1. {Almost-invariant subspace}}

The proof of the existence of the super-adiabatic projection is very
close to the proof of Proposition 1 of \cite{PST2}, so we only
sketch the strategy and emphasize the main differences with respect
to that proof.

\medskip

First of all, one constructs a formal symbol $\pi \asymp\sum_{j=0}^\infty \varepsilon^j \,
\pi_j$ (the \emph{Moyal
projection}) such that: {\upshape(i)} $\pi \moy \pi \asymp
\pi$; {\upshape(ii)} $\pi^\dag=\pi$;
{\upshape(iii)} ${H}_0 \moy \pi \asymp \pi \moy
{H}_0$ where $\asymp$ denotes the asymptotic equivalence of formal series.

\smallskip

\noindent The symbol $\pi$ is constructed recursively at any order
$j \in \N$ starting from $\pi_0$ and $H_0$. One firstly show the
uniqueness of $\pi$ (see Lemma 2.3. in \cite{PST1}). The uniqueness
allows us to construct $\pi$ locally, i.e. in a neighborhood of some
point $z_0:=(k_0,\eta_0)\in\R^2\times\R^2$. From the continuity of
the map $k\mapsto H_\text{per}(k)$ and the condition \eqref{eq64} it
follows that there exists a neighborhood $\sss{U}_{k_0}$ of $k_0$
such that for every $k\in\sss{U}_{k_0}$ the set
$\{\sss{E}_n(k)\}_{n\in\sss{I}}$ can be enclosed by a
positively-oriented circle $\Sigma(k_0)\subset\C$ independent of
$k$. Moreover it is possible to choose $\Sigma(k_0)$ in such a way
that: it is symmetric with respect to the real axis;
$\text{dist}(\Sigma(k_0),\sigma(H_\text{per}(k)))\geqslant\frac{1}{4}C_g$
for all $k\in\sss{U}_{k_0}$; $\text{Radius}(\Sigma(k_0))\leqslant
C_r$ is bounded by a constant $C_r$ independent of $k_0$;
$\Sigma(k_0-\gamma^\ast)=\Sigma(k_0)$ for all
$\gamma^\ast\in\Gamma^\ast$.

\noindent With the notation of Section \ref{appA2} we have $H_0 =
H_\phi \circ\jmath_\kappa$ with
$H_\phi(k,\eta):=H_\text{per}(k)+\phi(\eta)$. Let
$\widetilde{\Lambda}(k_0,\eta_0):=\Sigma(k_0)+\phi(\eta_0)$ denote
the translation of the circle $\Sigma(k_0)$ by $\phi(\eta_0)\in \R$
and pose $\Lambda:=\widetilde{\Lambda}\circ\jmath_\kappa$.  From the
smoothness of $\phi$ it follows that there exists a neighborhood
$\sss{U}_{z_0}\subset\R^2\times\R^2$ of $z_0$ such that
$\text{dist}(\Lambda(z_0),\sigma(H_0(z)))\geqslant\frac{1}{4}C_g$
for all  $z\in \sss{U}_{z_0}$. Moreover $\Lambda(z_0)$ is symmetric
with respect to the real axis, has radius bounded by $C_r$ and is
$\Gamma^\ast$-periodic in the variable $\kappa = k-A(\eta)$ {(see
Figure \ref{fig01})}.

We proceed by using the \emph{Riesz formula}, namely by posing
$$
\pi_j(z):=\frac{i}{2\pi}\oint_{\Lambda(z_0)}d\lambda\ R_j(\lambda,z)
 \qquad \text{on }\quad \sss{U}_{z_0}
$$
where $R_j(\lambda,\cdot)$ denotes the $j$-th term in the
\emph{Moyal resolvent} $R(\lambda,\cdot)
=\sum_{j=0}^{\infty}\varepsilon^j R_j(\lambda,\cdot)$ (also known as
the \emph{parametrix}), defined by the request that
$$
(H_0(\cdot)- \lambda \num{1}_\sss{D})\moy R(\lambda,\cdot) =
\num{1}_{\sss{H}_\text{f}},\quad\quad   R(\lambda,\cdot) \moy
(H_0(\cdot)- \lambda \num{1}_\sss{D})=\num{1}_\sss{D}\qquad \text{on
}\quad \sss{U}_{z_0}.$$ Each term $R_j$ is computed by a recursive
procedure starting from
$R_0(\lambda,\cdot):=(H_0(\cdot)-\lambda\num{1}_\sss{D})^{-1}$, as
illustrated in {\cite{ger-mar-sjo}. Following \cite[equations (30)
and (31)]{PST2} one obtains that
\begin{equation}\label{R induction}
R_j(\lambda,z)=-R_0(\lambda,z)\ L_j(\lambda,z)
\end{equation}
where $L_j$ is the $(j-1)$-th order obstruction for $R_0$ to be the
Moyal resolvent, i.e.
\begin{equation}\label{L_obstruction}
(H_0(\cdot)- \lambda \num{1}_\sss{D})\moy  \left(\sum_{n=0}^{j-1}\varepsilon^n\
R_j(\lambda,\cdot)\right)=\num{1}_{\sss{H}_\text{f}}+\varepsilon^j\
L_j(\cdot)+\sss{O}(\varepsilon^{j+1}).
\end{equation}
At the first order
$L_1=-\frac{i}{2}\left\{H_0,R_0\right\}_{k,\eta}$, with $\{\cdot,
\cdot \}_{k,\eta}$  the Poisson brackets.

\medskip

\noindent The technical (and crucial) part of the proof is to show
that
$$
\pi_j\in S^v_{\kappa;\tau}\left(\bbb{B}(\sss{H}_\text{{\upshape
f}},\sss{D})\right)\cap
S^1_{\kappa;\tau}\left(\bbb{B}(\sss{H}_\text{{\upshape f}})\right)
$$
for all $j \in \N$, with $v({k},\eta):=(1+|{k}|^2)$.  By means of
the recursive construction each $R_j(\lambda,\cdot)$ inherits the
special $\tau$-equivariance from the principal symbol
$R_0(\lambda,\cdot)
=((H_\phi\circ\jmath_\kappa)(\cdot)-\lambda\num{1}_\sss{D})^{-1}$.
The special periodicity in $\kappa$ of the domain of integration
$\Lambda(\cdot)$ which appears in the Riesz formula assures also the
special $\tau$-equivariance of each
 $\pi_j(\cdot)$.

\noindent Since
$\|(\partial^{\alpha}_z\pi_j)(z)\|_\flat\leqslant2\pi C_r\
\sup_{\lambda\in
\Lambda(z_0)}\|\partial^{\alpha}_z(R_j)(\lambda,z)\|_\flat$ ($\flat$
means either $\bbb{B}(\sss{H}_\text{f})$ or
$\bbb{B}(\sss{H}_\text{f};\sss{D})$, $\alpha\in\N^{4}$ is a
multiindex and
$\partial^{\alpha}_z:=\partial^{\alpha_1}_{k_1}\partial^{\alpha_2}_{k_2}\partial^{\alpha_3}_{\eta_1}\partial^{\alpha_4}_{\eta_2}$),
we need only to prove that $R_j(\lambda,\cdot)\in
S^v_{\kappa;\tau}\left(\bbb{B}(\sss{H}_\text{{\upshape
f}},\sss{D})\right)\cap
S^1_{\kappa;\tau}\left(\bbb{B}(\sss{H}_\text{{\upshape f}})\right)$
uniformly in $\lambda$. This is the delicate point of the proof.

First of all, from the definition of $\Lambda(z_0)$ it follows that
$$\|R_0(\lambda,z)\|_{\bbb{B}(\sss{H}_\text{f})}=[\text{dist}(\lambda,\sigma(H_0(z)))]^{-1}\leqslant\nicefrac{4}{C_g}$$
uniformly in $\lambda$. Let $\sigma \in \N^4$, with $|\sigma |= 1$.
One observes that
$\partial^\sigma_zR_0(\lambda,z)=-R_0(\lambda,z)N^\sigma_z(\lambda,z)$
with $N^\alpha_z(\lambda,z):=\partial^\alpha_z H_0(z)\
R_0(\lambda,z)$. From the  relation
$$
\partial^\sigma_{z}N^{\alpha}_{z}=N^{\alpha+\sigma}_z-N^\alpha_zN^\sigma_z
$$
and an inductive argument, it follows the \emph{chain rule}
$$
\partial^\alpha_zR_0=R_0\
\sum\omega_{\beta_1\ldots\beta_{|\alpha|}}N^{\beta_1}_z\ldots
N^{\beta_{|\alpha|}}_z
$$
where $\beta_1,\ldots,\beta_{|\alpha|}\in\N^4$,
$|\alpha|:=\alpha_1+\ldots+\alpha_4$,
$\omega_{\beta_1\ldots\beta_{|\alpha|}}=\pm1$ is a suitable sign
function and the sum runs over all the combinations of multiindices
such that $\beta_1+\ldots+\beta_{|\alpha|}=\alpha$ with the
convention $N^{0}_z=\num{1}$. The chain rule implies that $R_0\in
S^1_{\kappa;\tau}\left(\bbb{B}(\sss{H}_\text{{\upshape f}})\right)$
provided that $$ \|N^{\alpha}_{z}\|_{\bbb{B}(\sss{H}_\text{{\upshape
f}})}=\|\partial^\alpha_zH_0\ R_0\|_{\bbb{B}(\sss{H}_\text{{\upshape
f}})}\leqslant C_{\alpha} \qquad \mbox{uniformly in $\lambda$}.
$$
The latter condition is true since
$\|(\partial^\alpha_zH_0)(k,\eta)\
R_0(\lambda,k,\eta)\|_{\bbb{B}(\sss{H}_\text{{\upshape
f}})}\leqslant (g\circ\jmath_\kappa)(k,\eta)$, for a suitable $g(k,\eta)$,
$\Gamma^\ast$-periodic in $k$ and bounded in $\eta$; the latter claim can be checked as in
Proposition \ref{prop3}.

Similarly to prove that $R_0\in
S^v_{\kappa;\tau}\left(\bbb{B}(\sss{H}_\text{{\upshape
f}},\sss{D})\right)$ we need to show that
$$
\|R_0\ N^{\alpha}_{z}\|_{\bbb{B}(\sss{H}_\text{{\upshape
f}},\sss{D})}=\|(\num{1}_{\sss{H}_\text{f}}-\Delta_\theta)\ R_0\
N^{\alpha}_{z}\|_{\bbb{B}(\sss{H}_\text{{\upshape f}})}\leqslant
C_{\alpha}\ v'(\cdot)
$$ uniformly in $\lambda$. Since
$N^{\alpha}_{z}$ is bounded on $\sss{H}_\text{f}$ it is sufficient
to show that
$\|(\num{1}_{\sss{H}_\text{f}}-\Delta_\theta)R_0(\lambda;z)\|_{\bbb{B}(\sss{H}_\text{{\upshape
f}})}\leqslant C'_{\alpha}\ v'(z)$. Observe that
$\|(\num{1}_{\sss{H}_\text{f}}-\Delta_\theta)R_0(\lambda,[\kappa]-\gamma^\ast;\eta)\|_{\bbb{B}(\sss{H}_\text{{\upshape
f}})}=\|(\num{1}_{\sss{H}_\text{f}}-\Delta_\theta)\tau(\gamma^\ast)^{-1}R_0(\lambda,[\kappa];\eta)\|_{\bbb{B}(\sss{H}_\text{{\upshape
f}})}$. The commutation relation
$$
-\Delta_\theta\tau(\gamma^\ast)^{-1}=\tau(\gamma^\ast)^{-1}\left(|\gamma^\ast|^2+i2\gamma^\ast\cdot\nabla_\theta-\Delta_\theta\right)
$$
and  the straightforward bound
$$\|\left(|\gamma^\ast|^2+i2\gamma^\ast\cdot\nabla_\theta-\Delta_\theta\right)(\num{1}_{\sss{H}_\text{f}}-\Delta_\theta)^{-1}\|_{\bbb{B}(\sss{H}_\text{{\upshape
f}})}\leqslant C(1+|\gamma^\ast|^2)\leqslant
C'(1+|\kappa(k,\eta)|^2)
$$ imply
$$\|(\num{1}_{\sss{H}_\text{f}}-\Delta_\theta)R_0(\lambda,z)\|_{\bbb{B}(\sss{H}_\text{{\upshape
f}})}\leqslant C'_{\alpha}\
v'(z)\|(\num{1}_{\sss{H}_\text{f}}-\Delta_\theta)R_0(\lambda,[\kappa];\eta)\|_{\bbb{B}(\sss{H}_\text{{\upshape
f}})}$$ with $v':=v\circ \jmath_\kappa.$ Finally observe that
\begin{equation}\label{eqbound}
     \|(\num{1}_{\sss{H}_\text{f}}-\Delta_\theta)R_0(\lambda,[\kappa];\eta)\|_{\bbb{B}(\sss{H}_\text{{\upshape
f}})}\leqslant f([\kappa];\eta)\leqslant C''.
\end{equation}
The first inequality above
follows by an expansion on the Fourier basis, for fixed $[\kappa]$ and $\eta$;
the second follows from the fact that
$[\kappa]$ takes values on a compact set and the explicit dependence
on $\eta$ is through the bounded function $\phi$.
The bound (\ref{eqbound}) implies that  $R_0\in
S^v_{\kappa;\tau}\left(\bbb{B}(\sss{H}_\text{{\upshape
f}},\sss{D})\right)\cap
S^1_{\kappa;\tau}\left(\bbb{B}(\sss{H}_\text{{\upshape f}})\right)$
uniformly in $\lambda$.

To prove that $R_j \in S^1_{\kappa;\tau}\left(\bbb{B}(\sss{H}_\text{{\upshape f}})\right)$,
we observe that for any $\alpha \in \N^d$ one has
$$
\partial_z^\alpha
R_j(\lambda,z)=R_0(\lambda,z)\ M_{z;j}^\alpha(\lambda,z)
$$
where $M_{z;j}^\alpha$ is a linear combination of terms which
are product of $N_z^\beta$ and $\partial_z^\delta L_j$ with
$|\beta|,|\delta|\leqslant|\alpha|$. Thus it is sufficient to prove that
$L_j\in S^1_{\kappa;\tau}\left(\bbb{B}(\sss{H}_\text{{\upshape f}})\right)$
for every $j \in \N$. The latter claim is proved by induction on $j \in \N$.
Referring to (\ref{R induction}), one has trivially
that $L_1\in S^1_{\kappa;\tau}\left(\bbb{B}(\sss{H}_\text{{\upshape
f}})\right)$.  $L_{j+1}$ is a linear combination of products of
$N_z^\alpha$ (with $0\leqslant|\alpha|\leqslant j+1$) and
$M^\beta_{z,i}$ (with $|\beta|+i=j+1$ and $0\leqslant i\leqslant
j$). Then the induction hypothesis on $L_i$
for all $i=1,\ldots,j$ implies that $L_{j+1}$ is in
$S^1_{\kappa;\tau}\left(\bbb{B}(\sss{H}_\text{{\upshape
f}})\right)$.

Finally observing that
$\|\partial^\alpha_zR_j\|_{\bbb{B}(\sss{H}_\text{{\upshape
f}},\sss{D})}\leqslant\|
M_{z;j}^\alpha\|_{\bbb{B}(\sss{H}_\text{{\upshape f}})}\
\|R_0\|_{\bbb{B}(\sss{H}_\text{{\upshape f}},\sss{D})}$ and using
the fact that $R_0\in
S^v_{\kappa;\tau}\left(\bbb{B}(\sss{H}_\text{{\upshape
f}},\sss{D})\right)$ it follows that $R_j\in
S^v_{\kappa;\tau}\left(\bbb{B}(\sss{H}_\text{{\upshape
f}},\sss{D})\right)\cap
S^1_{\kappa;\tau}\left(\bbb{B}(\sss{H}_\text{{\upshape f}})\right)$
uniformly in $\lambda$, for all $j\in\N$.

As explained in Section \ref{appA2}, we can apply the result of
Proposition B.4 in \cite{stefan_book} to special  $\tau$-equivariant
symbols obtaining $H_0 \moy \pi\in
S^{v^2}_{\kappa;\tau}(\bbb{B}(\sss{H}_\text{f}))$. However the
$\tau$-equivariance of $H_0 \moy \pi$ and its derivatives implies
that the norms are bounded in $z$, hence $H_0 \moy \pi\in
S^1_{\kappa;\tau}(\bbb{B}(\sss{H}_\text{f}))$ which implies by
adjointness also $\pi \moy H_0 \in
S^1_{\kappa;\tau}(\bbb{B}(\sss{H}_\text{f}))$. By construction
$[H^\sss{Z};\text{Op}_\varepsilon(\pi)]=\text{Op}_\varepsilon(H_0
\moy \pi -\pi \moy H_0)=\sss{O}_0(\varepsilon^\infty)$ where the
remainder is bounded in the norm of $\bbb{B}(\sss{H}_\tau)$.

The operator $\text{Op}_\varepsilon(\pi)$ is only approximately a
projection, since
$\text{Op}_\varepsilon(\pi)^2=\text{Op}_\varepsilon(\pi \moy
\pi)=\text{Op}_\varepsilon(\pi)+\sss{O}_0(\varepsilon^\infty)$.
We obtain the super adiabatic projection $\Pi_\varepsilon$ by using
the trick in \cite{NeSo}. Indeed, one notices that, for $\varepsilon$ sufficiently small,
the spectrum of  $\text{Op}_\varepsilon(\pi)$ does not contain e.g. the points
$\{ 1/2 \}$ and $\{3/2\}$. Thus, the formula
\begin{equation}\label{eq_true_proj}
\Pi_{\varepsilon} = \frac{i}{2 \pi} \oint_{|z -1|=\nicefrac{1}{2}} (\text{Op}_\varepsilon(\pi) - z)^{-1}.
\end{equation}
yields an orthogonal projector such that $\Pi_{\varepsilon} = \text{Op}_\varepsilon(\pi) + \sss{O}_0(\varepsilon^\infty)$.

Finally, equation \eqref{eq_hof_stima1} follows by observing that
$[H^\sss{Z};\Pi_{\varepsilon}]=\sss{O}_0(\varepsilon^\infty)$
implies $$
[\expo{-i\frac{t}{\varepsilon}H^\sss{Z}};\Pi_{\varepsilon}]=\sss{O}_0(\varepsilon^\infty|t|)
$$
as proved in Corollary 3.3 in \cite{stefan_book}.

\subsubsection*{Step 2. Construction of the intertwining unitary}

The construction of the intertwining unitary follows as in the proof
of Proposition 2 of \cite{PST2}. Firstly one constructs a formal
symbol $u\asymp\sum_{j=0}^{\infty}\varepsilon^j u_j$ such that:
{\upshape(i)} $u^\dag \moy u=u \moy
u^\dag=\num{1}_{\sss{H}_\text{f}}$; {\upshape(ii)} $u \moy \pi \moy
u^\dag=\pi_\text{r}$.

The existence of such a symbol follows from a recursive procedure
starting from $u_0$ and using the expansion of
$\pi\asymp\sum_{j=0}^{\infty}\varepsilon^j \pi_j$ obtained above.
However, the symbol $u$ which comes out of this procedure is not
unique.

Since $u_0$ is \emph{right $\tau$-covariant} (see the end of Section
\ref{sec_rel_pat}) in $\kappa$, then one can prove by induction that
the same is also true for all the symbols $u_j$ and hence for the
full symbol $u$. Finally, since $u_0\in
S^1(\bbb{B}(\sss{H}_\text{f}))$ one deduces by induction also
$u_j\in S^1(\bbb{B}(\sss{H}_\text{f}))$ for all $j\in\N$. The
quantization of this symbol is an element of
$\bbb{B}(\sss{H}_\tau,\sss{H}_\text{ref})$  satisfying the following
properties:
\renewcommand{\labelenumi}{{\rm(\roman{enumi})}}
\begin{enumerate}
    \item
    $\text{Op}_\varepsilon(u)\text{Op}_\varepsilon(u)^\dag=\num{1}_{\sss{H}_\text{ref}}+\sss{O}_0(\varepsilon^\infty)$,
    \item
    $\text{Op}_\varepsilon(u)^\dag\text{Op}_\varepsilon(u)=\num{1}_{\sss{H}_\tau}+\sss{O}_0(\varepsilon^\infty)$,
    \item $\text{Op}_\varepsilon(u)\Pi_\varepsilon\text{Op}_\varepsilon(u)^\dag=\Pi_\text{r}+\sss{O}_0(\varepsilon^\infty)$.
\end{enumerate}

Nevertheless $\text{Op}_\varepsilon(u)$ can be modified by an
$\sss{O}_0(\varepsilon^\infty)$ term using the same technique of
Lemma 3.3 (Step II) in \cite{PST1} to obtain the true unitary
$U_\varepsilon$.

\subsubsection*{Step 3. Effective dynamics}

The last step of the proof is identical to the corresponding part
(Proposition 3) of \cite{PST2}.

\bigskip


\goodbreak
\subsection{Hofstadter-like Hamiltonians}

We now focus on the special case of a single isolated energy band
$E_\ast$, i.e. $m=1$, and we comment on the relation between the
effective Hamiltonian, the celebrated Peierls substitution and
Hofstadter-like Hamiltonians (see Section 1).

In this special case,
$\pi_0({\kappa})=\ketbra{\psi_\ast({\kappa})}{\psi_\ast({\kappa})}$
and $u_0({\kappa})=\ketbra{\chi}{\psi_\ast({\kappa})}+u_0^\bot$
where $\psi_\ast(k)$ is the eigenvector of $H_\text{{\upshape
per}}(k)$ corresponding to the eigenvalue $\sss{E}_\ast(k)$. Let
$h\in S^1(\bbb{B}(\sss{H}_\text{\upshape f}))$ be a resummation of
the formal symbol $u \moy \pi \moy H_0 \moy \pi \moy u^{-1}$. A
straightforward computation yields
$$
h_0 = u_0 \, \pi_0 \, H_0 \, \pi_0 \, u_0^\dag =
\ketbra{\chi}{\psi_\ast}\ \ketbra{\psi_\ast}{\psi_\ast}\ H_0\
\ketbra{\psi_\ast}{\psi_\ast}\ \ketbra{\psi_\ast}{\chi}= E_\ast \,
\pi_\text{r}.
$$
Since $\pi_\text{r}$ is one-dimensional, $h_0$ can be regarded as a
scalar-valued symbol with explicit expression
$$
h_0(k,\eta)= E_\ast(k,\eta)
=\sss{E}_\ast\left(k-A(\eta)\right)+\phi(\eta).
$$
By considering the quantization of the latter, the effective one-band Hamiltonian reads
\begin{equation}\label{eq75}
\text{{\upshape
Op}}_\varepsilon(h_0)=E_\ast({k},i\varepsilon{\nabla}_{{k}})=\sss{E}_\ast\left({k}-{A}(i\varepsilon{\nabla}_{{k}})\right)+\phi(i\varepsilon{\nabla}_{{k}}).
\end{equation}
The latter formula corresponds to the momentum-space reformulation
of the well-known \emph{\textbf{Peierls substitution}}
\cite{peierls,ashcroft}.

To illustrate this point, we specialize to the case of a uniform external magnetic field
and zero external electric field, setting $\phi=0$ and $A_0=0$ in
\eqref{eq69}. The leading order contribution to the dynamics in the
almost invariant subspace is therefore given by a bounded operator,
acting on the reference Hilbert space
$L^2({M_{\Gamma^\ast}},d^2\underline{k})$, defined as the
quantization (in the sense of Section \ref{sec_quan_pair})
of the function $\sss{E}_\ast\circ\jmath_\kappa: (k,
\eta) \mapsto \sss{E}_\ast(k - A(\eta))$, defined on $\mathbb{T}^d\times \mathbb{R}^d$.

Loosely speaking, the above procedure corresponds to the following
\virg{substitution rule}: one may think to quantize the smooth
function  $\sss{E}_\ast: \T^d \to\R$ by formally replacing the
variables $(k_1,k_2)$ with the operators $(\bbb{K}'_1,\bbb{K}'_2)$
defined by
\begin{equation}\label{eq76}
\bbb{K}'_1:=k_1+\dfrac{i}{2}(\iota_q\varepsilon)\dfrac{\partial}{\partial
k_2},\ \ \ \ \ \ \ \ \
\bbb{K}'_2:=k_2-\dfrac{i}{2}(\iota_q\varepsilon)\dfrac{\partial}{\partial
k_1},
\end{equation}
regarded as unbounded operators acting in
$L^2({M_{\Gamma^\ast}},d^2\underline{k})$. To make this procedure
rigorous, one can expand $\sss{E}_\ast$ in its Fourier series, i.e.
$\sss{E}_\ast(k)=\sum_{n,m\in\Z}c_{n,m}\ \expo{i2\pi(na+mb)\cdot k}$
and define the \emph{\textbf{Peierls' quantization}} of
$\sss{E}_\ast$ as the operator obtained by the same series expansion
with the phases $\expo{i2\pi(na+mb)\cdot k}$ replaced by the unitary
operators $\expo{i2\pi(na+mb)\cdot \bbb{K}'}$ (the series is
norm-convergent, in view of the regularity of $\sss{E}_\ast$). This
fixes uniquely the prescription for the quantization.

To streamline the notation, one introduces new coordinates
$\xi_1:=2\pi({a}\cdot{k})$ and $\xi_2:=2\pi({b}\cdot{k})$ such that
the function $\sss{E}'_\ast$,
$\sss{E}'_\ast(\xi_1,\xi_2):=\sss{E}_\ast({k}(\xi))$ becomes
 $(2\pi\Z)^2$-periodic.
The change of variables induces  a unitary map from
$L^2({M_{\Gamma^\ast}}^2,d^2\underline{k})$ to
$L^2(\num{T}^2,d^2{\xi})$ which intertwines the operators
\eqref{eq76} with the operators (recall
$\varepsilon=\nicefrac{2\pi}{h_B}$)
\begin{equation}\label{eq77}
\bbb{K}_1:=\xi_1+i\pi\left( \frac{\iota_q}{h_B}\right)\dfrac{\partial}{\partial
\xi_2},\ \ \ \ \ \ \ \ \ \bbb{K}_2:=\xi_2-i\pi\left( \frac{\iota_q}{h_B}\right)\dfrac{\partial}{\partial \xi_1},
\end{equation}
so that $2\pi(a\cdot \bbb{K}')\mapsto\bbb{K}_1$ and $2\pi(b\cdot
\bbb{K}')\mapsto\bbb{K}_2$.

Let $F: \num{T}^2 \rightarrow \num{C}$  be  sufficiently regular
that its Fourier series
$F(\xi_1,\xi_2)=\sum_{n,m\in\Z}f_{n,m}\expo{i(n\xi_1+m\xi_2)}$ is
uniformly-convergent. We define the \emph{Peierls quantization} of
$F$ as
$$
\widehat{F}:=\sum_{n,m\in\Z}f_{n,m} \expo{i(n\bbb{K}_1+m\bbb{K}_2)}.
$$ Let
$\ssss{U}_{0}=\expo{i\bbb{K}_1}$ and
$\ssss{V}_{0}=\expo{i\bbb{K}_2}$ (\emph{Hofstadter unitaries}), acting on $L^2(\num{T}^2,d^2{\xi})$ as
\begin{equation}\label{eq52}
 (\ssss{U}_{0}\psi)(\xi_1,\xi_2)=\expo{i\xi_1}\psi\left(\xi_1,\xi_2-\pi\frac{\iota_q}{h_B}\right),\ \ \ \ \ \
(\ssss{V}_{0}\psi)(\xi_1,\xi_2)= \expo{i\xi_2}\psi\left(\xi_1+\pi\frac{\iota_q}{h_B},\xi_2\right).
\end{equation}
We regard \eqref{eq52} as the definition of the two unitaries, so
there is no need to specify the domain of definition of the
generators \eqref{eq77}. Thus the Peierls quantization of the
function $F$ defines a bounded operator on $L^2(\num{T}^2,d^2{\xi})$
given, in terms of the Hofstadter unitaries, by
\begin{equation}\label{eq79}
 \widehat{F}(\ssss{U}_{0},\ssss{V}_{0})=\sum_{n,m=-\infty}^{+\infty}f_{n,m}\, \expo{i\pi nm\left(\frac{\iota_q}{h_B}\right)}\  \ssss{U}_{0}^n\ \ssss{V}_{0}^m,
\end{equation}
where the fundamental commutation relation $\ssss{U}_{0}
\ssss{V}_{0}=\expo{-i2\pi\left(\frac{\iota_q}{h_B}\right)}\
\ssss{V}_{0} \ssss{U}_{0}$ has been used. Formula \eqref{eq79}
defines a \emph{Hofstadter-like Hamiltonian} with {\emph{deformation
parameter}} $\nicefrac{1}{h_B}$. Indeed, the special case $
H_\text{Hof} = \ssss{U}_{0} + \ssss{U}_{0}^{-1} + \ssss{V}_{0} +
\ssss{V}_{0}^{-1} $ is (up to a unitary equivalence) the celebrated
Hofstadter Hamiltonian \cite{Hof}.

Summarizing, we draw the following
\begin{concl}\label{corol_hof} Under the assumption of Theorem \ref{teo1}, for every
$\ttt{V}_{\Gamma} \in L^2_{\rm loc}(\R^2,d^2r)$, in the Hofstadter
regime ($h_B\to \infty$), the dynamics generated by the Hamiltonian
$\ttb{H}_\text{{\upshape BL}}$ \eqref{Hamilt BL} in the subspace
related to a single isolated Bloch band, is
 approximated up to an error of order $\nicefrac{1}{h_B}$ (and up to a unitary transform)
 by the dynamics generated on the reference Hilbert space $L^2(\num{T}^2,d^2\xi)$ by a  Hofstadter-like Hamiltonian,
 i.e. by a power series in the Hofstadter unitaries $\ssss{U}_{0}$ and $\ssss{V}_{0}$ defined by \eqref{eq52}.
\end{concl}


\newpage
\section{Space-adiabatic theory for the Harper regime}\label{Sec_Har}

\subsection{Adiabatic parameter for strong magnetic fields}\label{Sec_Har_par}

We now consider the case of a strong external magnetic field. Since
we are interested in the limit $B\to+\infty$ we set $\ttb{A}_0=0$
and $\Phi=0$ in the Hamiltonian \eqref{eq2}. By exploiting the gauge
freedom,  we choose
 \begin{equation}\label{eeq0}
{\nabla}_r\cdot\ttb{A}_\Gamma=0,\ \ \ \ \ \ \ \ \ \ \ \ \ \int_{M_\Gamma}\ttb{A}_\Gamma(r)\ d^2r=0,
\end{equation}
this choice being always possible  \cite{sobo}. Let us denote  by
$Q_{r} =(Q_{r_1},Q_{r_2})$ the multiplication operators by $r_1$ and
$r_2$ and with $P_{r}=(P_{r_1},P_{r_2})=-i\hslash{\nabla_{r}}$.
Taking into account conditions \eqref{eeq0} and $A_0=0$, $\phi=0$,
the Hamiltonian \eqref{eq2}  is rewritten as
\begin{equation}\label{eeq1}
\ttb{H}_\text{BL}=\dfrac{1}{2m}\left[\left(P_{r_1}+\dfrac{qB}{2c}Q_{r_2}\right)^2+\left(P_{r_2}-\dfrac{qB}{2c}Q_{r_1}\right)^2\right]+\widetilde{\ttt{V}}_\Gamma(Q_r)+\widetilde{\ttt{W}}(Q_r)
\end{equation}
where
\begin{align}
&\widetilde{\ttt{V}}_\Gamma(Q_r) =\ttt{V}_\Gamma(Q_r)+\frac{q^2}{2mc^2}|\ttb{A}_\Gamma(Q_r)|^2\label{eeq1b}\\
&\widetilde{\ttt{W}}(Q_r,P_r) =-\frac{q}{mc}(\ttt{A}_\Gamma)_1(Q_r)\
\left[P_{r_1}+\dfrac{qB}{2c}Q_{r_2}\right]-\frac{q}{mc}(\ttt{A}_\Gamma)_2(Q_r)\
\left[P_{r_2}-\dfrac{qB}{2c}Q_{r_1}\right]\label{eeq1c}
\end{align}
with $(\ttt{A}_\Gamma)_1$ and $(\ttt{A}_\Gamma)_2$ the
$\Gamma$-periodic components of the vector potential
$\ttb{A}_\Gamma$. The first of \eqref{eeq0} assures that
$\widetilde{\ttt{W}}$ is a symmetric operator.


It is useful to define two new pairs of canonical dimensionless operators:
\begin{equation}\label{eeq2}
 (\text{fast})\ \left\{
\begin{aligned}
 &K_1:=-\frac{1}{2\delta}\ {b}^\ast\cdot{Q}_r-\iota_q\frac{\delta}{\hslash}\ {a}\cdot{P}_r\\
 &K_2:=\phantom{-}\frac{1}{2\delta}\ {a}^\ast\cdot{Q}_r-\iota_q\frac{\delta}{\hslash}\ {b}\cdot {P}_r
\end{aligned}
\right.\ \ \ \ \ \ \ \
(\text{slow})\ \left\{
\begin{aligned}
 &G_1:=\dfrac{1}{2}\ {{b}^\ast\cdot {Q}_r}-\iota_q\frac{\delta^2}{\hslash}\ {a}\cdot {P}_r\\
&G_2:=\dfrac{1}{2}\ {a}^\ast\cdot {Q}_r+\iota_q\frac{\delta^2}{\hslash}\  {b}\cdot {P}_r
\end{aligned}
\right.
\end{equation}
where $\delta:=\sqrt{\hslash_B}=\sqrt{\nicefrac{\Phi_0}{2\pi
Z\Phi_B}}$ according to the notation introduced in Section
\ref{sec_int}. Since $\delta^2\propto \nicefrac{1}{B}$, the limit of
strong magnetic field corresponds to $\delta\to0$. We consider
$\delta$ as the \emph{adiabatic parameter} in the Harper regime. A
direct computation shows that
\begin{equation}\label{eeq3}
[K_1;K_2]=i \iota_q\  \num{1}_{\sss{H}_\text{phy}},\ \ \ \ \ [G_1;G_2]=i \iota_q\ \delta^2\ \num{1}_{\sss{H}_\text{phy}},\ \ \ \ \ \ [K_j;G_k]=0,\ \ \ \ \ j,k=1,2.
\end{equation}

\goodbreak

These new variables  are important for three reasons:
\begin{enumerate}
 \item[(a)] they make evident a
separation of scales between the {slow degrees of freedom} related
to the the dynamics induced by periodic potential and the {fast
degrees of freedom} related to the  cyclotron motion induced by the
external magnetic field. Indeed, for  $\ttt{V}_\Gamma=0$, the
\emph{fast variables} $(K_1,K_2)$ (the \emph{kinetic momenta})
describe the kinetic energy of the cyclotron motion, while the
\emph{slow variables} $(G_1,G_2)$ correspond semiclassically to the
center of the cyclotron orbit and are conserved quantities.
 \item[(b)] The new variables are dimensionless.
According to the notation used in Section
\ref{Sec_Hofs_par} let
$H_\text{BL}:=\nicefrac{1}{\sss{E}_0}\ttb{H}_\text{BL}$ be the
dimensionless Bloch-Landau Hamiltonian with
$\sss{E}_0:=\nicefrac{\hslash^2}{m\Omega_\Gamma}$ the \emph{natural
unit of energy}.
\item[(c)] The use of the new variables simplifies the expression of the $\Gamma$-periodic functions
appearing in $\ttb{H}_\text{BL}$. Indeed,
${a}^\ast\cdot{Q}_r=G_2+\delta\ K_2$ and ${b}^\ast\cdot
{Q}_r=G_1-\delta\ K_1$, hence if $f_\Gamma$ is any $\Gamma$-periodic
function one has
\begin{equation}\label{Eq f_Gamma}
f_\Gamma(Q_r)=f(G_2+\delta\ K_2,G_1-\delta\ K_1)
\end{equation}
where $f$ is the $\Z^2$-periodic function related to $f_\Gamma$.
\end{enumerate}

\noindent In terms of the new variables \eqref{eeq2}, the
Hamiltonian $H_\text{BL}$ reads
\begin{equation}\label{eeq4} H_\text{BL}=\frac{1}{\delta^2}\
\Xi(K_1,K_2)+V\left(G_2+\delta\ K_2,G_1-\delta\
K_1\right)+\frac{1}{\delta}\ W(K_1,G_1,K_2,G_2)
\end{equation}
 where
\begin{equation}\label{eeq5}
\Xi(K_1,K_2):=\frac{1}{2\Omega_\Gamma}\left[|{a}|^2{K_2}^2+ |{b}|^2{K_1}^2-{a}\cdot{b}\ \{K_1;K_2\}\right]
\end{equation}
is a  quadratic function of the operators $K_1$ and $K_2$
($\{\cdot;\cdot\}$ denotes the anticommutator), $V$ is  the
$\Z^2$-periodic function related to the $\Gamma$-periodic function
$\nicefrac{1}{\sss{E}_0}\widetilde{\ttt{V}}_\Gamma$ and $W$ denotes
the  function $\nicefrac{1}{\sss{E}_0}\widetilde{W}$ with respect
the new canonical pairs, namely
\begin{equation}\label{eeq6}
W(K_1,G_1,K_2,G_2)=f_1\left(G_2+\delta\ K_2,G_1-\delta\
K_1\right)K_1-f_2\left(G_2+\delta\ K_2,G_1-\delta\ K_1\right)K_2
\end{equation}
where $f_1$ and $f_2$ are the $\Z^2$-periodic dimensionless
functions
$$
f_1({a}^\ast\cdot {r},{b}^\ast\cdot
{r}):=2\pi\frac{Z\Omega_\Gamma}{\Phi_0}({a}^\ast\cdot \ttb{A}_\Gamma)(r)
\qquad \mbox{and} \qquad  f_{2}({a}^\ast\cdot
{r},{b}^\ast\cdot{r}):=2\pi\frac{Z\Omega_\Gamma}{\Phi_0}(
{b}^\ast\cdot \ttb{A}_\Gamma)(r).
$$
An easy computation shows that the
first gauge condition of \eqref{eeq0} is equivalent to
\begin{equation}\label{eeq00}
\frac{\partial f_1}{\partial x_1}(x_1,x_2)+ \frac{\partial
f_2}{\partial x_2}(x_1,x_2)=0.
\end{equation}
Obviously $W$ is a symmetric operator, since $\widetilde{\ttt{W}}$
is symmetric.

The problem has a natural time-scale which is fixed by the
\emph{cyclotron frequency} $\omega_\text{c}=\frac{|q|B}{mc}$. With
respect to the (fast) \emph{ultramicroscopic time-scale}
$\tau:=\omega_\text{c}s$, equation \eqref{eq1} becomes
\begin{equation}\label{eeq7}
i\frac{1}{\delta^2}\dfrac{\partial}{\partial\tau}\psi=H_\text{BL}\psi,
\qquad \qquad \delta^2 =\frac{\sss{E}_0}{\hslash\omega_\text{c}}.
\end{equation}
Thus the physically relevant Hamiltonian is
\begin{equation}\label{eeq8}
H_\text{BL}^\delta:= \delta^2\ H_\text{BL} =\Xi(K_1,K_2)+\delta\ W(K_1,G_1,K_2,G_2)+\delta^2\ V\left(G_2+\delta\ K_2,G_1-\delta\ K_1\right).
\end{equation}

\goodbreak


\subsection{Separation of scales: the von Neumann unitary}\label{sec_neum_unit}
The commutation relations \eqref{eeq3} show that $(K_1,K_2)$ and $(G_1,G_2)$ are two pairs of canonical conjugate operators. The \emph{Stone-von Neumann uniqueness Theorem} (see \cite{bra-rob2} Corollary 5.2.15) assures the existence of a unitary map $\ssss{W}$ (called  \emph{von Neumann unitary})
\begin{equation}\label{eeq8'}
\ssss{W}:\sss{H}_\text{phy}{\longrightarrow}\sss{H}_\text{w}:=\sss{H}_{\text{s}}\otimes\sss{H}_{\text{f}}= L^2(\R,dx_\text{s})\otimes  L^2(\R,dx_\text{f})
\end{equation}
such that
\begin{align}
&\ssss{W}G_1\ssss{W}^{-1}= Q_{\text{s}}=\text{multiplication by}\ \ x_\text{s},&& \ssss{W}G_2\ssss{W}^{-1}= P_{\text{s}}=-i \iota_q \delta^2\dfrac{\partial}{\partial x_\text{s}}\label{eeq9a}\\
&\ssss{W}K_1\ssss{W}^{-1}= Q_{\text{f}}=\text{multiplication by}\
\ x_\text{f},&& \ssss{W}K_2\ssss{W}^{-1} =
P_{\text{f}}=-i \iota_q \ \dfrac{\partial}{\partial x_\text{f}}.\label{eeq9b}
\end{align}
 The explicit costruction of the von Neumann unitary $\ssss{W}$ is described in Appendix \ref{appB}.

Let $X_j:=G_j+(-1)^j\delta\ K_j$ with $j=1,2$. From \eqref{eeq9a}
and \eqref{eeq9b} it follows that
\begin{equation}\label{eeq10}
X'_1:=\ssss{W}X_1\ssss{W}^{-1}=Q_{\text{s}}-\delta\ Q_{\text{f}},\ \
\ \ \ \ \ \ \ X'_2:=\ssss{W}X_2\ssss{W}^{-1}=P_{\text{s}}+\delta\
P_{\text{f}}.
\end{equation}
Since $X_1$ and $X_2$ commute, one can use the spectral calculus to
define any measurable function of $X_1$ and $X_2$. For any $f\in
L^\infty(\R^2,d^2x)$ one defines $f(X_1,X_2):=\int_{\R^2}f(x_1,x_2)\
d\num{E}_{x_1}^{(1)}\ d\num{E}_{x_2}^{(2)}$ where $d\num{E}^{(j)}$
is the projection-valued measure corresponding to $X_j$. In view of
the unitarity of $\sss{W}$, and observing that
$d{\num{E}'}^{(j)}:=\ssss{W}d{\num{E}}^{(j)}\ssss{W}^{-1}$ is the
projection-valued measure of $X'_j$, one obtains that
$$
\ssss{W}f(X_1,X_2)\ssss{W}^{-1}=\int_{\R^2}f(x_1,x_2)\
d{\num{E}'_{x_1}}^{(1)}\ {d\num{E}'_{x_2}}^{(2)}=f(X'_1,X'_2).
$$
So the effect of the conjugation through $\ssss{W}$ on a function
$f$ of the operators $X_1$ and $X_2$ formally amounts to replace the
operators $X_j$ with $X'_j$ inside $f$. \hfill

In view of the above remark, one can easily rewrite
$H_\text{BL}^\delta$ making explicit the r\^ole of the fast and slow
variables, obtaining
\begin{equation}\label{eeq11}
H^\ssss{W}:=\ssss{W}H_\text{BL}^\delta\ssss{W}^{-1}=\num{1}_{\sss{H}_{\text{s}}}\ \otimes\  \Xi(Q_{\text{f}},P_{\text{f}})+\delta\ W(Q_{\text{f}},Q_{\text{s}},P_{\text{f}},P_{\text{s}})+\delta^2\ V\left(P_{\text{s}}+\delta\ P_{\text{f}},Q_{\text{s}}-\delta\ Q_{\text{f}}\right)
\end{equation}
where, according to \eqref{eeq6},
\begin{equation}\label{eeq12}
W(Q_{\text{f}},Q_{\text{s}},P_{\text{f}},P_{\text{s}})=f_1\left(P_{\text{s}}+\delta\
P_{\text{f}},Q_{\text{s}}-\delta\ Q_{\text{f}}\right)\
Q_{\text{f}}-f_2\left(P_{\text{s}}+\delta\
P_{\text{f}},Q_{\text{s}}-\delta\ Q_{\text{f}}\right)\ P_{\text{f}}.
\end{equation}


\subsection{Relevant part of the spectrum: the Landau bands}\label{sec_land_band}
The existence of a separation between fast and slow degrees of
freedom and the decomposition of the physical Hilbert space
$\sss{H}_\text{phy}$ into the product space
$\sss{H}_\text{w}=\sss{H}_{\text{s}}\otimes\sss{H}_{\text{f}}$  are
the first two ingredients to develop the SAPT. According to the
general scheme, we \virg{replace} the canonical operators
corresponding to the slow degrees of freedom with classical
variables which will be re-quantized \virg{a posteriori}.
Mathematically, we show that the Hamiltonian $H^\ssss{W}$ acting in
$\sss{H}_\text{w}$ is  the Weyl quantization of the operator-valued
function (symbol) $H_\delta$,
\begin{equation}\label{eeq13}
H_\delta(p_\text{s},x_\text{s}):=
\Xi(Q_{\text{f}},P_{\text{f}})+\delta\
\underbrace{W(Q_{\text{f}},x_{\text{s}},P_{\text{f}},p_{\text{s}})}_{=
W_\delta(p_\text{s},x_\text{s})}+\delta^2\
\underbrace{V\left(p_{\text{s}}+\delta\
P_{\text{f}},x_{\text{s}}-\delta\
Q_{\text{f}}\right)}_{=V_\delta(p_\text{s},x_\text{s})}.
\end{equation}
The quantization is defined (formally) by the rules
$$
x_{\text{s}}\longmapsto\text{Op}_\delta(x_{\text{s}}):=
Q_{\text{s}}\ {\otimes\ \num{1}_{\sss{H}_\text{f}}},\ \ \ \ \ \ \ \
p_{\text{s}}\longmapsto\text{Op}_\delta(p_{\text{s}}):=
P_{\text{s}}\ {\otimes\ \num{1}_{\sss{H}_\text{f}}}.
$$
For every $(p_\text{s},x_\text{s})\in \R^{2}$, equation
\eqref{eeq13} defines an unbounded operator
$H_\delta(p_\text{s},x_\text{s})$ which acts in the Hilbert space
$\sss{H}_\text{f}$. To make the quantization procedure rigorous, as
explained in Appendix A of \cite{PST1}, we need to consider
$H_\delta$ as function from $\R^2$ into some Banach space which is
also a domain of self-adjointness for
$H_\delta(p_\text{s},x_\text{s})$.  We take care of this details in
the Section \ref{Sec symbol class}.

To complete the list of ingredients needed for the SAPT, we need to
analize the spectrum of the principal part of the symbol
\eqref{eeq13} as $(p_\text{s},x_\text{s})$ varies in $\R^2$. The
principal part of the symbol, denoted by
$H_0(p_\text{s},x_\text{s})$, is given by \eqref{eeq13} when
$\delta=0$, so it reads:
\begin{equation}\label{eeq14}
H_0(p_\text{s},x_\text{s}):=\Xi(Q_{\text{f}},P_{\text{f}})=\dfrac{1}{2\Omega_\Gamma}\left[|{a}|^2{P_{\text{f}}}^2+
|{b}|^2{Q_{\text{f}}}^2-{a}\cdot{b}\
\left\{Q_{\text{f}};P_{\text{f}}\right\}\right].
\end{equation}
Since the principal symbol is constant on the phase space, i.e.
$H_0(p_\text{s},x_\text{s})=\Xi$ for all
$(p_\text{s},x_\text{s})\in\R^2$, we are reduced to compute the
spectrum of $\Xi$. As well-known {(see Remark \ref{rkdsa} below)},
the spectrum of $\Xi$ is pure point with
$\sigma(\Xi)=\{\lambda_n:=(n+\nicefrac{1}{2}) \, :\, n\in\N\}$. We
refer to the eigenvalue $\lambda_n$ as the $n$-th \emph{Landau
level}.

\begin{figure}[htbp]
\begin{center}
\fbox{
\includegraphics[width=12cm]{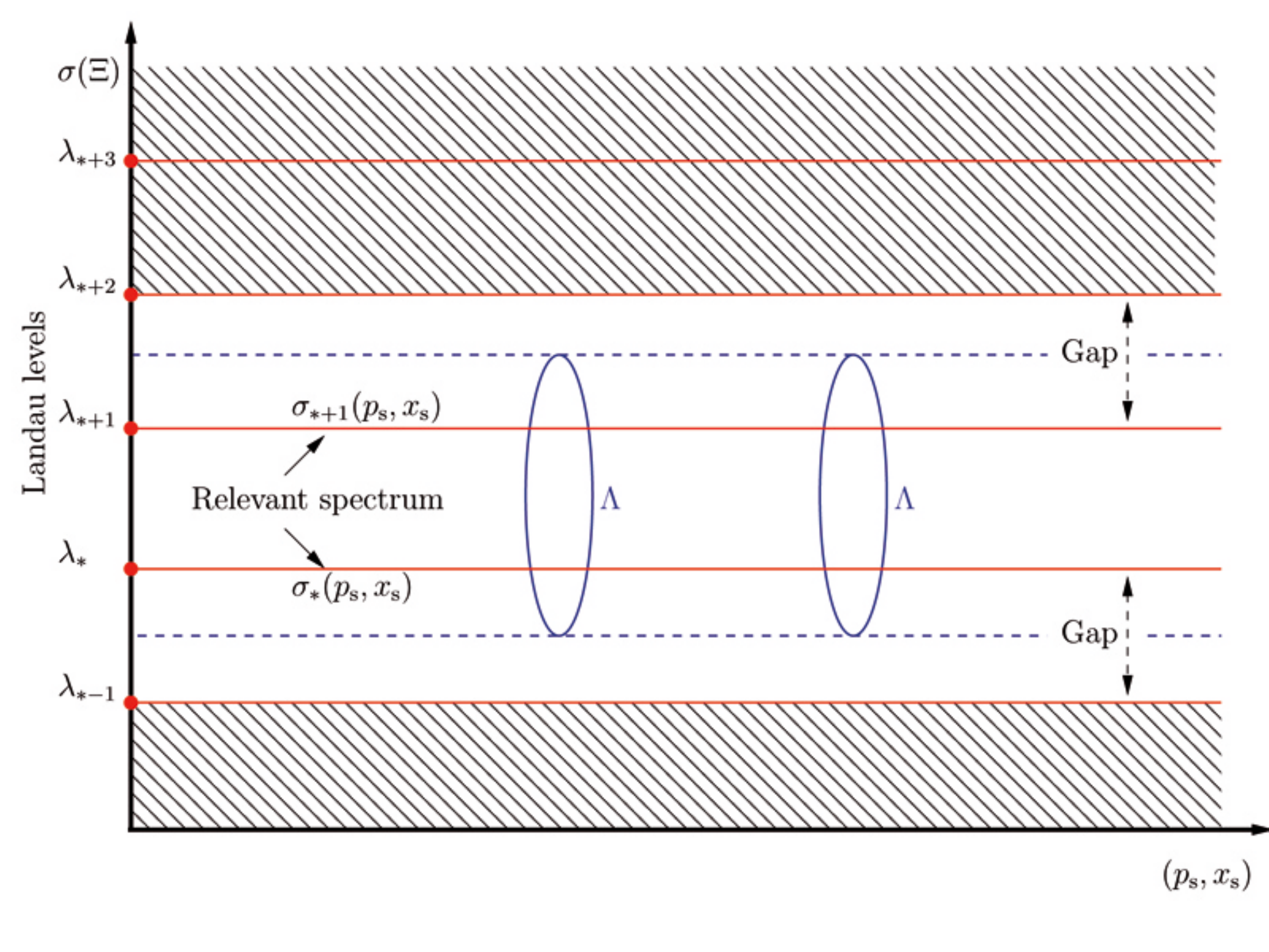}
}
\end{center}
\caption{{\footnotesize Structure of the spectrum of $H_0$. The
picture shows a \virg{relevant part of the spectrum} consisting of
two Landau bands of constant energy $\lambda_\ast$ and
$\lambda_{\ast+1}$. }}\label{fig02}
\end{figure}

The spectrum of the symbol $H_0$ consists of a collection of
constant functions $\sigma_n:\R^2\to\R$, $n\in\N$,
$\sigma_n(p_\text{s},x_\text{s}) \equiv \lambda_n$, which we call
\emph{Landau bands}. The band $\sigma_n$ is separated by the rest of
the spectrum by a constant gap. In the gap condition (analogous to
\eqref{eq64}) one can choose $C_\text{g}=1$. Therefore, each finite
family of contiguous Landau bands defines a relevant part of the
spectrun appropriate to develop the SAPT.

\begin{rk}[{The domain of self-adjointness}]\label{rkdsa}
 We describe explicitly the
domain of self-adjointness of $H_0(p_\text{s},x_\text{s})$.
Mimicking the standard theory of Landau levels, one introduces
operators
\begin{align}
&\rrr{a}\phantom{^\dag}:=\dfrac{i}{\sqrt{2}}\dfrac{\ell}{\Omega_\Gamma}\left[(a_1+ia_2)P_\text{f}-(b_1+ib_2)Q_\text{f}\right]=\dfrac{i}{\sqrt{2}}\left[\overline{z_{a}}\ P_\text{f}-\overline{z_{b}}\ Q_\text{f}\right]\label{eeq14a}\\
&
\rrr{a}^\dag:=\dfrac{-i}{\sqrt{2}}\dfrac{\ell}{\Omega_\Gamma}\left[(a_1-ia_2)P_\text{f}-(b_1-ib_2)Q_\text{f}\right]=\dfrac{-i}{\sqrt{2}}\left[z_{a}\
P_\text{f}-z_{b}\ Q_\text{f}\right]\label{eeq14b},
\end{align}
where $z_{a}:=\frac{1}{\ell}(a_1-ia_2)$ and
$z_{b}:=\frac{1}{\ell}(b_1-ib_2)$. It is  easy to check that
\begin{equation}\label{eeq15}
\rrr{a}\rrr{a}^\dag=\Xi(Q_\text{f},P_\text{f})+\iota_q\ \frac{1}{2}
\num{1}_{\sss{H}_\text{f}},\ \ \ \ \ \ \rrr{a}^\dag
\rrr{a}=\Xi(Q_\text{f},P_\text{f})-\iota_q\ \frac{1}{2}
\num{1}_{\sss{H}_\text{f}},\ \ \ \ \ \
[\rrr{a};\rrr{a}^\dag]=\iota_q\  \num{1}_{\sss{H}_\text{f}}.
\end{equation}


Without loss of generality, we suppose that $\iota_q=1$. Let
$\psi_0$ be the \emph{ground state} defined by $\rrr{a}\psi_0=0$. A
simple computation shows that
$\psi_0(x_\text{f})=C\expo{-(\beta-i\alpha) x_\text{f}^2}$, where
$C>0$ is a normalization constant, and $\alpha\in\R$,  $\beta>0$ are
related to the geometry of the lattice $\Gamma$ by
$\alpha:=\nicefrac{a\cdot b}{2|a|^2}$ and
$\beta:=\nicefrac{\Omega_\Gamma}{2|a|^2}$. Since $\psi_0$ is a fast
decreasing smooth function, the vectors
$\psi_n:=(n!)^{-\frac{1}{2}}(\rrr{a}^\dag)^n\psi_0$, with
$n=0,1,\ldots$, are well defined. From the algebraic relations
\eqref{eeq15} it follows straightforwardly that: (i)
$\rrr{a}\psi_n=\sqrt{n}\psi_{n-1}$; (ii) the family of vectors
$\{\psi_n\}_{n\in\N}$ is an orthonormal basis for $\sss{H}_\text{f}$
called the \emph{generalized Hermite basis}; (iii)
$\Xi\psi_n=\lambda_n\psi_n$; (iv) the spectrum of $\Xi$ is pure
point with $\sigma(\Xi)=\{\lambda_n\ :\ n\in\N\}$.

Let $\sss{L}\subset\sss{H}_\text{f}$ be the set of the finite linear
combinations of the elements of the basis $\{\psi_n\}_{n\in\N}$. The
unbounded operators $\rrr{a}$, $\rrr{a}^\dag$ and $\Xi$ are well
defined on $\sss{L}$ and on this domain  $\rrr{a}^\dag$ acts as the
adjoint of $\rrr{a}$ and $\Xi$ is symmetric. Both $\rrr{a}$ and
$\rrr{a}^\dag$ are closable and we will denote their closure by the
same symbols. The operator $\Xi$ is essentially selfadjoint on the
domain $\sss{L}$ (the  deficiency indices are both zero) and so
 its domain of selfadjointness
$\sss{F}:=\sss{D}(\Xi)$ is the closure of $\sss{L}$ with respect to
the \emph{graph norm}
$\|\psi\|_{\Xi}^2:=\|\psi\|^2_{\sss{H}_\text{f}}+\|\Xi\psi\|^2_{\sss{H}_\text{f}}$.
The graph norm is equivalent to the more simple \emph{regularized
norm}
$$
\|\psi\|_{\sss{F}}:=\|\Xi\psi\|_{\sss{H}_\text{f}}.
$$
The domain $\sss{F}$ has the structure of an Hilbert space with
hermitian structure provided by the \emph{regularized scalar
product}
$(\psi;\varphi)_{\sss{F}}:=(\Xi\psi;\Xi\varphi)_{\sss{H}_\text{f}}$.\hfill
$\blacklozenge\lozenge$
\end{rk}



\subsection{Symbol class and asymptotic expansion}
\label{Sec symbol class}

In this section, we firstly identify the Banach space in which  the
symbol $H_\delta$ defined by \eqref{eeq13} takes values and,
secondarily, we explain in which sense $H_\delta$ is a
\virg{semiclassical symbol} in a suitable H\"ormander symbol class.
The main results are contained in Proposition {\ref{prop_def_dom}}.
Readers who are not interested in technical details can jump
directly to the next section. For the definitions of the
H\"{o}rmander classes $S^1(\bbb{B}(\sss{H}_\text{f}))$ and $
S^1(\bbb{B}(\sss{F};\sss{H}_\text{{\upshape f}}))$  we refer to
Section \ref{appA2}.

\begin{propos}\label{prop_def_dom}
Assume that {\upshape Assumption ($\text{A}_\text{s}$)} holds true.
Then for all $(p_\text{{\upshape s}},x_\text{{\upshape s}})\in\R^2$
the operator $H_\delta(p_\text{{\upshape s}},x_\text{{\upshape s}})$
is essentially self-adjoint on the dense domain
$\sss{L}\subset\sss{H}_\text{{\upshape f}}$ consisting of finite
linear combinations of  generalized Hermite functions, and its
domain of self-adjointness is the domain $\sss{F}$ on which the
operator $H_0=\Xi$ is self-adjoint. Finally,  $H_\delta$ is in the
H\"{o}rmander class $S^1(\bbb{B}(\sss{F};\sss{H}_\text{{\upshape
f}}))$.
\end{propos}

In particular, $H_\delta(p_\text{{\upshape s}},x_\text{{\upshape
s}})$ is a bounded operator from the Hilbert space $\sss{F}$ to the
Hilbert space $\sss{H}_\text{f}$ for all $(p_\text{{\upshape
s}},x_\text{{\upshape s}})\in\R^2$. The proof of the Proposition
\ref{prop_def_dom} follows from the {\it Kato-Rellich Theorem}
showing that for any $(p_\text{{\upshape s}},x_\text{{\upshape
s}})\in\R^2$ the operator $H_\delta(p_\text{{\upshape
s}},x_\text{{\upshape s}})$ differs from $H_0$ by a relatively
bounded perturbation. The latter claim will be proved in
Lemmas \ref{lem_semi1} and \ref{lem_semi2} below.}\\

In view of {\upshape Assumption ($\text{A}_\text{s}$)}, $\widetilde{\ttt{V}}_\Gamma\in
C^\infty_{\text{b}}(\R^2,\R)$ so its Fourier series
$$
\widetilde{\ttt{V}}_\Gamma(r)=\sum_{n,m\in\Z}w_{n,m}\ \expo{i2\pi n\
{a}^\ast\cdot r}\expo{i2\pi m\ {b}^\ast\cdot r}
$$
converges uniformly and moreover
$\sum_{n,m=-\infty}^{+\infty}|m|^{\alpha_1}|n|^{\alpha_2}\left|w_{n,m}\right|\leqslant
C_\alpha$ for all $\alpha\in\N^2$. 

Let $V$ be the $\Z^2$-periodic function related to
$\nicefrac{1}{\sss{E}_0}\widetilde{\ttt{V}}_\Gamma$, as in Section
\ref{Sec_Har_par}. In view of (\ref{Eq f_Gamma}) one has
\begin{align}\label{eeq16}
{\ssss{W} \frac{\widetilde{\ttt{V}}_\Gamma}{\sss{E}_0}(X_1,X_2)
\ssss{W}^{-1}} =V(P_\text{s}+\delta P_\text{f},Q_\text{s}-\delta
P_\text{f}) &=\sum_{n,m=-\infty}^{+\infty}v_{n,m}\ \expo{i2\pi
(nP_\text{s}+mQ_\text{s})}\ \expo{i2\pi\delta (n P_\text{f}-m
Q_\text{f})}
\end{align}
with $v_{n,m}:=\nicefrac{1}{\sss{E}_0}w_{n,m}$ and where we used the
fact that fast and slow variables commute and
$[Q_\text{s};P_\text{s}]=\delta^2[Q_\text{f};P_\text{f}]$. The
operator \eqref{eeq16} can be seen as the Weyl quantization of the
operator-valued symbol
\begin{equation}\label{eeq17}
 V_\delta(p_\text{s},x_\text{s}):=\sum_{n,m=-\infty}^{+\infty}v_{n,m}\ \expo{i2\pi (np_\text{s}+mx_\text{s})}\ \expo{i2\pi\delta (n P_\text{f}-m Q_\text{f})}
\end{equation}
with   quantization rule
\begin{equation}\label{eeq_quan_rul}
\text{Op}_\delta\left(\expo{i2\pi
(np_\text{s}+mx_\text{s})}\right)= \expo{i2\pi
(nP_\text{s}+mQ_\text{s})}\otimes\num{1}_{\sss{H}_\text{f}}.
\end{equation}
\begin{lem}\label{lem_semi1}
Let {\upshape Assumption ($\text{A}_\text{s}$)} hold true. Then
$V_\delta\in S^1(\bbb{B}(\sss{H}_\text{f}))\cap
S^1(\bbb{B}(\sss{F};\sss{H}_\text{{\upshape f}}))$. In particular
$V_\delta(p_\text{{\upshape s}},x_\text{{\upshape s}})$ is a bounded
self-adjoint operator on $\sss{H}_\text{{\upshape f}}$ for all
$(p_\text{{\upshape s}},x_\text{{\upshape s}})\in\R^2$.
\end{lem}

\Proof It is sufficient to show that $V_\delta\in
S^1(\bbb{B}(\sss{H}_\text{f}))$ since
$$
\|\partial^\alpha
V_\delta\|_{\bbb{B}(\sss{F};\sss{H}_\text{f})}=\|(\partial^\alpha
V_\delta)\
\Xi^{-1}\|_{\bbb{B}(\sss{H}_\text{f})}\leqslant2\|\partial^\alpha
V_\delta\|_{\bbb{B}(\sss{H}_\text{f})}
$$ in view of $\|\Xi^{-1}\|_{\bbb{B}(\sss{H}_\text{f})}=2$. Let
$\alpha:=(\alpha_1,\alpha_2)\in\N^2$, then
$$
\left\|\partial_{p_\text{s}}^{\alpha_1}\partial_{x_\text{s}}^{\alpha_2}V_\delta(p_\text{s},x_\text{s})\right\|_{\bbb{B}(\sss{H}_\text{f})}\leqslant(2\pi)^{|\alpha|}\sum_{n,m+-\infty}^{+\infty}|n|^{\alpha_1}|m|^{\alpha_2}\left|v_{n,m}\right|\leqslant\frac{(2\pi)^{|\alpha|}}{\sss{E}_0}C_\alpha
$$ for all $(p_\text{s},x_\text{s})\in\R^2$, as a consequence of the
 unitarity of $\expo{i2\pi\delta (n P_\text{f}-m Q_\text{f})}$.
The self-adjointness follows by observing that $\{v_{n,m}\}$ are the
Fourier coefficients of a real function.\CVD

{\upshape Assumption ($\text{A}_\text{s}$)} implies that the
$\Gamma$-periodic functions $a^\ast\cdot \ttb{A}_\Gamma$ and
$b^\ast\cdot \ttb{A}_\Gamma$ are elements of
$C^\infty_{\text{b}}(\R^2,\R)$. By the same arguments above, one
proves that the operators $f_j(P_\text{s}+\delta
P_\text{f},Q_\text{s}-\delta P_\text{f})$, $j=1,2$, appearing in
\eqref{eeq12}, are the Weyl quantization of the operator-valued
functions
\begin{equation}\label{eeq18}
 f_\delta^{(j)}(p_\text{s},x_\text{s}):=\sum_{n,m=-\infty}^{+\infty}f^{(j)}_{n,m}\
 \expo{i2\pi (np_\text{s}+mx_\text{s})}\ \expo{i2\pi\delta (n P_\text{f}-m Q_\text{f})}\ \ \ \ \ \ \
 j=1,2
\end{equation}
according to  \eqref{eeq_quan_rul}. The coefficients
$\frac{1}{2\pi}\frac{\Phi_0}{Z\Omega_\Gamma}f^{(j)}_{n,m}$ are the
Fourier coefficients of $a^\ast\cdot \ttb{A}_\Gamma$ if $j=1$ and of
$b^\ast\cdot \ttb{A}_\Gamma$ if $j=2$. Thus, equation \eqref{eeq12}
shows that the operator
$W(Q_{\text{f}},Q_{\text{s}},P_{\text{f}},P_{\text{s}})$ coincides
with the Weyl quantization of the operator-valued symbol
\begin{equation}\label{eeq19}
 W_\delta(p_\text{s},x_\text{s}):=f_\delta^{(1)}(p_\text{s},x_\text{s})\ Q_\text{f}+f_\delta^{(2)}(p_\text{s},x_\text{s})\ P_\text{f},
\end{equation}
defined, initially, on the dense domain $\sss{L}$.

\begin{lem}\label{lem_semi2}
Let {\upshape Assumption ($\text{A}_\text{s}$)} hold true. Then
$f_\delta^{(j)}\in S^1(\bbb{B}(\sss{H}_\text{f}))\cap
S^1(\bbb{B}(\sss{F};\sss{H}_\text{{\upshape f}}))$, for $j=1,2$. For
all $(p_\text{{\upshape s}},x_\text{{\upshape s}})\in\R^2$, the
bounded operators $f_\delta^{(j)}(p_\text{{\upshape
s}},x_\text{{\upshape s}})$ are self-adjoint while
 $W_\delta(p_\text{{\upshape s}},x_\text{{\upshape s}})$ is
symmetric on the dense domain $\sss{L}$
and infinitesimally bounded
with respect to $\Xi$. Finally $W_\delta\in
S^1(\bbb{B}(\sss{F};\sss{H}_\text{f}))$.
\end{lem}

\Proof As  in the first part of the proof of Lemma \ref{lem_semi1},
one proves that $f_\delta^{(j)}\in
S^1(\bbb{B}(\sss{H}_\text{f}))\cap
S^1(\bbb{B}(\ssss{F};\sss{H}_\text{{\upshape f}}))$ and its
self-adjointness. The operator  $W_\delta(p_\text{{\upshape
s}},x_\text{{\upshape s}})$ is a linear combination of $Q_\text{f}$
and $P_\text{f}$, which are densely defined on $\sss{L}$, multiplied
by bounded operators. Using \eqref{eeq00} one checks by a direct
computation that $W_\delta(p_\text{{\upshape s}},x_\text{{\upshape
s}})$ acts as a symmetric operator on $\sss{L}$.
Since $Q_\text{f}$ and $P_\text{f}$  are infinitesimally bounded
with respect to $\Xi$, then the same holds true for
$W_\delta(p_\text{{\upshape s}},x_\text{{\upshape s}})$,
$(p_\text{{\upshape s}},x_\text{{\upshape s}})\in\R^2$. The last
claim follows by observing that
$$
\|\partial^\alpha (f^{(j)}_\delta
X_\text{f})\|_{\bbb{B}(\sss{F};\sss{H}_\text{f})}=\|(\partial^\alpha
f^{(j)}_\delta)\ X_\text{f} \,
\Xi^{-1}\|_{\bbb{B}(\sss{H}_\text{f})}\leqslant \|X_\text{f} \,
\Xi^{-1}\|_{\bbb{B}(\sss{H}_\text{f})}\ \|\partial^\alpha
f^{(j)}_\delta\|_{\bbb{B}(\sss{H}_\text{f})},
$$
with $j=1,2$ and $X_\text{f}=Q_\text{f}$ or $P_\text{f}$. Since
$\|X_\text{f} \, \Xi^{-1}\|_{\bbb{B}(\sss{H}_\text{f})}\leqslant C$
and $f_\delta^{(j)}\in S^1(\bbb{B}(\sss{H}_\text{f}))$, the claim
is proved. \CVD

Lemmas \ref{lem_semi1} and \ref{lem_semi2}, together with the fact
that $H_0 \equiv \Xi$ is clearly in
$S^1(\bbb{B}(\sss{F};\sss{H}_\text{{\upshape f}}))$ imply the last
part of Proposition \ref{prop_def_dom}.


\subsection{Semiclassics: the $\sss{O}(\delta^4)$-approximated symbol}

In this section we consider the asymptotic expansion for
the symbol $H_\delta$ in the parameter $\delta$. The Fourier
expansion \eqref{eeq17} for  $V_\delta$ and the similar expression
for $W_\delta$, namely
\begin{equation}\label{eeq21}
 W_\delta(p_\text{s},x_\text{s})=\sum_{n,m=-\infty}^{+\infty} \expo{i2\pi (np_\text{s}+mx_\text{s})}\ \expo{i2\pi\delta (n P_\text{f}-m Q_\text{f})}\left[f^{(1)}_{n,m}\ Q_\text{f}+f^{(2)}_{n,m}\
 P_\text{f}\right],
\end{equation}
suggest a way to expand the symbol $H_\delta$ in powers of $\delta$.
By inserting the expansion $\expo{i2\pi\delta
I_{n,m}}=\sum_{j=0}^{+\infty}\frac{(i2\pi\delta)^j}{j!}{I_{n,m}}^j$,
with $I_{n,m}:=n P_\text{f}-m Q_\text{f}$, in \eqref{eeq17} and
\eqref{eeq21} and by exchanging the order of the series one obtains
the \emph{formal expansions}
\begin{equation}\label{eeq22}
 V_\delta(p_\text{s},x_\text{s}) \simeq \sum_{j=0}^{+\infty}\delta^j\ V_{j+2}(p_\text{s},x_\text{s})\qquad
\qquad
W_\delta(p_\text{s},x_\text{s}) \simeq \sum_{j=0}^{+\infty}\delta^j\
W_{j+1}(p_\text{s},x_\text{s})
\end{equation}
where
\begin{align}
& V_{j+2}(p_\text{s},x_\text{s}):=\dfrac{(i2\pi)^j}{j!}\sum_{n,m=-\infty}^{+\infty}\expo{i2\pi (np_\text{s}+mx_\text{s})}\  {I_{n,m}}^j\ v_{n,m}\label{eeq23a}\\
&
W_{j+1}(p_\text{s},x_\text{s}):=\dfrac{(i2\pi)^j}{j!}\sum_{n,m=-\infty}^{+\infty}
\expo{i2\pi (np_\text{s}+mx_\text{s})}\ {I_{n,m}}^j\
\left[f^{(1)}_{n,m}\ Q_\text{f}+f^{(2)}_{n,m}\
P_\text{f}\right]\label{eeq23b}.
\end{align}
In view  of \eqref{eeq00}, one easily shows that the operators $W_j$ are (formally) symmetric.

The justification of the formal expansions above requires some cautions:
(i)  we need to specify the domains of definitions of the
unbounded operators ${I_{n,m}}^j$ and consequently the domains of
definitions of $V_{j}$ and $W_{j}$;
(ii) we need to justify the
exchange of the order of the series in
the equations $\eqref{eeq17}$ and $\eqref{eeq21}$.

As for (i), one notices that
\begin{equation}\label{eeq24}
 I_{n,m} =\alpha_{n,m}\ \rrr{a}+\overline{\alpha_{n,m}}\ \rrr{a}^\dag,\ \ \ \ \ \ \  \alpha_{n,m}:=\frac{nz_b-mz_a}{\sqrt{2}}.
\end{equation}
For all $(n,m)\in\Z^2$ the operators $I_{n,m}$ are essentially
self-adjoint on the invariant dense domain $\sss{L}$ (their
deficiency indices are both zero). The powers ${I_{n,m}}^j$ are also
well defined and essentially self-adjoint on $\sss{L}$, as
consequence of the \emph{Nelson Theorem} (Theorem X.39 in
\cite{red-sim2}) since the set $\{\psi_n\}_{n\in\N}$ of the
generalized Hermite functions is a \emph{total set of analytic
vectors} for every $I_{n,m}$ (see Example 2 Section X.6 of
\cite{red-sim2}). The domain  of self-adjointness for ${I_{n,m}}^j$
is the closure of $\sss{L}$ with respect the corresponding graph
norm.

The operator $V_{j}(p_\text{s},x_\text{s})$ defined by equation
\eqref{eeq23a} is an homogeneous polynomial of degree $j-2$ in
$\rrr{a}$ and $\rrr{a}^\dag$. It is symmetric (hence closable) and
essentially self-adjoint on the invariant dense domain $\sss{L}$.
Analogously, the operators
\begin{equation}\label{eeq25}
 M^j_{n,m}:={I_{n,m}}^j\left[f^{(1)}_{n,m}\ Q_\text{f}+f^{(2)}_{n,m}\ P_\text{f}\right]={I_{n,m}}^j\left[g_{n,m}\ \rrr{a}+\overline{g_{n,m}}\ \rrr{a}^\dag\right],\ \ \ \ \ g_{n,m}:=\frac{z_af^{(1)}_{n,m}+z_bf^{(2)}_{n,m}}{\sqrt{2}}
\end{equation}
which appear in the right-hand side of equation \eqref{eeq23b}, are
essentially self-adjoint on $\sss{L}$ {since} the set of the
generalized Hermite functions provides a total set of analytic
vectors. Thus we answered to point (i).

Since the generalized Hermite functions are a  total set of analytic
vectors for all $I_{n,m}$ then the series
$\sum_{j=0}^{+\infty}\frac{(i2\pi\delta)^j}{j!}{I_{n,m}}^j \psi$
converges in norm for every $\psi \in \sss{L}$. From this
observation, the fact that the series of coefficients $v_{n,m}$,
$f^{(1)}_{n,m}$ and $f^{(2)}_{n,m}$  are absolutely convergent and
that $Q_\text{f}$ end $P_\text{f}$ leave invariant the domain
$\sss{L}$, one argues that for all $\psi\in\sss{L}$ the double
series which defines $V_\delta(p_\text{s},x_\text{s})\psi$ and
$W_\delta(p_\text{s},x_\text{s})\psi$ are absolutely convergent,
hence the order of the sums can be exchanged. Thus the series
appearing on the r.h.s. of \eqref{eeq22} agrees with $V_\delta$
(respectively, with $W_\delta$) on the dense domain $\sss{L}$. By a
density argument, the equality in \eqref{eeq22} holds true on the
full domain of definition of $V_\delta$ (which is
$\sss{H}_\text{f}$) and $W_\delta$ respectively.

In view of the above, we write the \virg{semiclassical expansion} of the
symbol $H_\delta$ as:
\begin{align}\label{eeq26}
H_\delta(p_\text{s},x_\text{s})=\Xi+\sum_{j=1}^{+\infty}\delta^j\ H_{j}(p_\text{s},x_\text{s}),\ \ \ \ \ \ \ \ H_{j}(p_\text{s},x_\text{s}):=W_{j}(p_\text{s},x_\text{s})+V_{j}(p_\text{s},x_\text{s})
\end{align}
with $V_1=0$.

Proposition \ref{prop_def_dom} shows that the natural domain for the
full symbol $H_\delta(p_\text{s},x_\text{s})$ is the domain
$\sss{F}$ of self-adjointness of  $\Xi$.
However, if we want to truncate the series \eqref{eeq26} at the
$j$-th order, we must be careful in the determination of the domain
of definition of the single terms and to control the remainder.
Every term in the expansion \eqref{eeq26} is  essentially
self-adjoint  on $\sss{L}$. However, the $j$-th order term
$H_j$ is the sum of two  homogeneous polynomials in $Q_\text{f}$ and
$P_\text{f}$ (or equivalently in $\rrr{a}$ and $\rrr{a}^\dag$),
$W_j$ of degree $j$ and $V_j$ of degree $j-2$. Since $W_j=0$ if
$\ttb{A}_\Gamma=0$, one obtains
$$
\text{deg}\ H_j= \left\{
  \begin{array}{ll}
    j,       & \hbox{if $\ttb{A}_\Gamma\neq0$} \\
    j-2,     & \hbox{if $\ttb{A}_\Gamma=0$}
  \end{array}
\right.
$$
where $\text{deg}\ H_j$ means the degree of $H_j$ as a polynomial in
$Q_\text{f}$ and $P_\text{f}$. If $\text{deg}\ H_j>2$ then the
operator $H_j$ is not bounded by the principal symbol $\Xi$, and in this sense it cannot be considered as a
\virg{small perturbation} in the sense of Kato.
Moreover, some other problems appear (see Remark \ref{rk2}).
In order to avoid these problems,
we truncate the expansion \eqref{eeq26} up to the
polynomial term of degree $2$, i.e. up to order $\delta^2$ if
$\ttb{A}_\Gamma=0$ and up to order $\delta^4$ if $\ttb{A}_\Gamma\neq0$.

Hereafter let $\natural$ be the \emph{indicator function} of the periodic
vector potential, defined as
$$
\natural = \left\{
  \begin{array}{ll}
    0,       & \hbox{if $A_\Gamma\neq0$} \\
    1,     & \hbox{if $A_\Gamma=0$.}
  \end{array}
\right.
$$
Let
$\widetilde{H}^\natural_\delta(p_\text{s},x_\text{s}):=\Xi+\sum_{j=1}^{2(1+\natural)}\delta^jH_{j}(p_\text{s},x_\text{s})$,
namely
\begin{align}
 &\widetilde{H}^1_\delta(p_\text{s},x_\text{s})=\Xi+\delta^2\sum_{n,m=-\infty}^{+\infty}v_{n,m}\ \expo{i2\pi (np_\text{s}+mx_\text{s})}\left(\num{1}_{\sss{H}_\text{f}}+i2\pi\delta I_{n,m}+\dfrac{1}{2}(i2\pi)^2\delta^2{I_{n,m}}^2\right)\label{eeq27a}\\
&\widetilde{H}^0_\delta(p_\text{s},x_\text{s})=\Xi+\delta\sum_{n,m=-\infty}^{+\infty}\
\expo{i2\pi (np_\text{s}+mx_\text{s})}\left[M^0_{n,m}+\delta(i2\pi
M^1_{n,m}+v_{n,m}\num{1}_{\sss{H}_\text{f}})\right].\label{eeq27b}
\end{align}
We call  $\widetilde{H}^\natural_\delta$
the \emph{approximated symbol} up to  order $\delta^{2(1+\natural)}$.
As a consequence of the Kato-Rellich theorem we have the following result:

\begin{propos}\label{prop_def_dom_aprx}
Under {\upshape Assumption ($\text{A}_\text{s}$)}  there exists a constant
$\delta_0$ such that for every $\delta<\delta_0$ and for every
$(p_\text{s},x_\text{s})\in\R^2$ the operator
$\widetilde{H}^\natural_\delta(p_\text{s},x_\text{s})$ (both for
$\natural=0$ or 1) is self-adjoint on the domain $\sss{F}$ and
bounded from below. Moreover $\widetilde{H}^\natural_\delta\in
S^1(\bbb{B}(\sss{F};\sss{H}_\text{{\upshape f}}))$.
\end{propos}

\Proof As proved in  Lemma \ref{appA3lem2}, $\rrr{a}$ and
$\rrr{a}^\dag$ are infinitesimally bounded with respect to $\Xi$.
This fact and {\upshape Assumption ($\text{A}_\text{s}$)}, which assures the fast decay of
the coefficients $v_{n,m}$ and $g_{n,m}$ (see \eqref{eeq25}), imply
that the operators
\begin{align*}
 &\sum_{n,m=-\infty}^{+\infty}v_{n,m}\ \expo{i2\pi (np_\text{s}+mx_\text{s})}\left(\num{1}_{\sss{H}_\text{f}}+i2\pi\delta I_{n,m}\right),&&
\sum_{n,m=-\infty}^{+\infty}\ \expo{i2\pi (np_\text{s}+mx_\text{s})}\left[M^0_{n,m}+\delta v_{n,m}\num{1}_{\sss{H}_\text{f}}\right]
\end{align*}
are infinitesimally bounded with respect to $\Xi$ and are elements
of $S^1(\bbb{B}(\sss{F};\sss{H}_\text{{\upshape f}}))$.

The operators ${I_{n,m}}^2$ and $M^1_{n,m}$ are only bounded (and
not infinitesimally bounded) with respect to $\Xi$. First of all it
is easy to check that
$\|{\rrr{a}^\sharp}^2\psi\|_{\sss{H}_\text{f}}\leqslant
3\|\Xi\psi\|_{\sss{H}_\text{f}}$ for every $\psi\in\sss{F}$, where
$\rrr{a}^\sharp$ means $\rrr{a}$ or $\rrr{a}^\dag$.
Then, for every $\psi\in\sss{F}$,
\begin{equation}\label{eeq28}
\|{I_{n,m}}^2\psi\|^2_{\sss{H}_\text{f}}\leqslant|\alpha_{n,m}|^4\left(\|{\rrr{a}}^2\psi\|_{\sss{H}_\text{f}}+\|\{\rrr{a};\rrr{a}^\dag\}^2\psi\|_{\sss{H}_\text{f}}+\|{\rrr{a}^\dag}^2\psi\|_{\sss{H}_\text{f}}\right)^2\leqslant27\frac{d^4}{\Omega^2}(n^2+m^2)^2\
\|\Xi\psi\|_{\sss{H}_\text{f}}^2
\end{equation}
where we used the inequality
$(\alpha+\beta+\gamma)^2\leqslant3(\alpha^2+\beta^2+\gamma^2)$, the
identity $\{\rrr{a};\rrr{a}^\dag\}=2\Xi$ and the bound
$|\alpha_{n,m}|^2\leqslant\nicefrac{d^2}{\Omega}(n^2-m^2)$ with
$d^2:=\text{max}\{|a|,|b|\}$. Assumption ($\text{A}_\text{s}$) assures that
the operator $\delta^42\pi^2\sum_{n,m=-\infty}^{+\infty}v_{n,m}\
\expo{i2\pi (np_\text{s}+mx_\text{s})}{I_{n,m}}^2$, which appears in
\eqref{eeq27a}, is bounded by $\Xi$ by a constant $\delta^4 C$, with
$C\propto\sum_{n,m=-\infty}^{+\infty}v_{n,m}(n^2+m^2)$, and is in
$S^1(\bbb{B}(\sss{F};\sss{H}_\text{{\upshape f}}))$. The claim for
$\widetilde{H}^1_\delta$ follows from the Kato-Rellich theorem
fixing $\delta_0:=C^{-\frac{1}{4}}$.

The claim for $\widetilde{H}^0_\delta$ follows in the same way
proving an inequality of the type \eqref{eeq28} for
$M^1_{n,m}=c_{n,m}\rrr{a}^2+\overline{c_{n,m}}{\rrr{a}^\dag}^2+2\Re(d_{n,m})\Xi+\iota_q\Im(d_{n,m})$
where $c_{n,m}:=\alpha_{n,m}g_{n,m}$ and
$d_{n,m}:=\alpha_{n,m}\overline{g_{n,m}}$. Observe that the series
of coefficients $g_{n,m}$ decays rapidly, then also the serie
$c_{n,m}$ and $d_{n,m}$ have a fast decay and in particular are
bounded. This implies that in the inequality of type \eqref{eeq28}
we can find a global constant which does not depend on $n$ and $m$.
\CVD

It is useful to have  explicit expressions of the first terms $H_j$,
in terms of $\rrr{a}$ and $\rrr{a}^\dag$. From equations
\eqref{eeq26}, \eqref{eeq27a} and \eqref{eeq27b}, using the Fourier
expansion of the derivatives of $V$ and
$g:=\nicefrac{1}{\sqrt{2}}\left(z_af_1+z_bf_2\right)$, it is easy to
check the following:\\
\\
{\bf - Case 1: $\ttb{A}_\Gamma=0$ - }\ In this situation
$$\widetilde{H}^1_\delta=\Xi+\delta^2 H_2+\delta^3 H_3+\delta^4 H_4$$
with
\begin{align}
 H_{2}(p_\text{s},x_\text{s})&=V(p_\text{s},x_\text{s})\ \num{1}_{\sss{H}_\text{f}}\label{eeq29a}\\
H_{3}(p_\text{s},x_\text{s})&=-\dfrac{1}{\sqrt{2}}\left[D_z(V)\ \rrr{a}+D_{\overline{z}}(V)\ \rrr{a}^\dag\right]\label{eeq29b}\\
H_{4}(p_\text{s},x_\text{s})&=\dfrac{1}{4}
\left[|D_z|^2(V)\
2\Xi
+{D}^2_z(V)\ \rrr{a}^2+D^2_{\overline{z}}(V)\ {\rrr{a}^\dag}^2\right]
\label{eeq29c}
\end{align}
where $D_z$ is the  differential operator defined by
$D_z:=\left(z_a\frac{\partial }{\partial
x_\text{s}}-z_b\frac{\partial }{\partial p_\text{s}}\right)$ and
$D_{\overline{z}}$ is obtained by replacing $z_a$ and $z_b$ with
$\overline{z}_a$ and $\overline{z}_b$. Since $V$ is real,
$D_{\overline{z}}(V)=\overline{D_z(V)}$, which shows that $H_3$ is
symmetric. The explicit expression of the second order differential
operator $|D_z|^2:=D_z\circ D_{\overline{z}}$ is
\begin{equation}\label{eeq30}
|D_z|^2=\dfrac{1}{\Omega_\Gamma}\left(|a|^2\ \dfrac{\partial^2
}{\partial x_\text{s}^2}-2a\cdot b\ \dfrac{\partial^2 }{\partial
x_\text{s}\partial p_\text{s}}+|b|^2\ \dfrac{\partial^2 }{\partial
p_\text{s}^2}\right).
\end{equation}
\\
\\
{\bf - Case 2: $\ttb{A}_\Gamma\neq0$ -}\  In this situation
$$\widetilde{H}^0_\delta=\Xi+\delta H_1+\delta^2 H_2$$ with
\begin{align}
H_{1}(p_\text{s},x_\text{s})&=g(p_\text{s},x_\text{s})\ \rrr{a}+\overline{g}(p_\text{s},x_\text{s})\ \rrr{a}^\dag\label{eeq30a}\\
H_{2}(p_\text{s},x_\text{s})&=V\
\num{1}_{\sss{H}_\text{f}}-\sqrt{2}D_z(\overline{g})\ \Xi
-\frac{1}{\sqrt{2}}\left[D_z(g)\
\rrr{a}^2+D_{\overline{z}}(\overline{g})\
{\rrr{a}^\dag}^2\right]\label{eeq30b}.
\end{align}
In the computation of \eqref{eeq30b} we used the first of the gauge
conditions \eqref{eeq0} which assures that
$$
D_z(\overline{g})=\frac{1}{\sqrt{2}\Omega_\Gamma}\left[|a|^2\ \dfrac{\partial f_1 }{\partial x_\text{s}}+a\cdot b\left( \dfrac{\partial f_2 }{\partial x_\text{s}}- \dfrac{\partial f_1 }{\partial p_\text{s}}\right)-|b|^2\ \dfrac{\partial f_2}{\partial p_\text{s}}\right]
$$
is a real function. From the definition of $g,f_1$ and $f_2$ it follows that
\begin{equation}\label{eeq31}
g(a^\ast\cdot r,b^\ast\cdot r)
=\pi \sqrt{2} \frac{Z\ell}{\Phi_0}
\left[(\ttt{A}_\Gamma)_1-i( \ttt{A}_\Gamma)_2\right](r),
\end{equation}
namely $g$ is the dimensionless $\Z^2$-periodic function related to
the $\Gamma$-periodic function $(\ttt{A}_\Gamma)_1-i( \ttt{A}_\Gamma)_2$, up to a
 multiplicative constant.


\goodbreak

\subsection{Main result: effective dynamics for strong magnetic fields}
\label{Sec_Har_ham}

\subsubsection*{Preliminary estimates on the remainder}

The difference $\rrr{R}^\natural_\delta :=H_\delta-\widetilde{H}^\natural_\delta$ is
a self-adjoint element of $S^1(\bbb{B}(\sss{F};\sss{H}_\text{{\upshape f}}))$, which we call
 the \emph{remainder symbol}. To develop the SAPT for the Harper regime
we need to  estimate the order of the remainder symbol. The next result shows essentially
that

\begin{equation}\label{eeq35}
H_\delta(p_\text{s},x_\text{s})=\widetilde{H}^\natural_\delta(p_\text{s},x_\text{s})+\sss{O}(\delta^{2(\natural+1)}),\
\ \ \ \ \ \ \ \pi_\text{r}\ H_\delta(p_\text{s},x_\text{s})\
\pi_\text{r}=\pi_\text{r}\
\widetilde{H}^\natural_\delta(p_\text{s},x_\text{s})\
\pi_\text{r}+\sss{O}(\delta^{2(\natural+1)+1}),
\end{equation}
where
\begin{equation}\label{eeq35proj}
\pi_\text{\upshape r}:=\sum_{i=1}^m\ketbra{\psi_{k_i}}{\psi_{k_i}}
\end{equation}
is the projection on the subspace spanned by the finite family of
generalized Hermite functions $\{\psi_{k_i}\}_{i=1}^m$. In other
words, the error done by replacing the true symbol $H_\delta$ with
the approximated symbol $\widetilde{H}^\natural_\delta$ (which has
order $2(\natural+1)$ in $\delta$) is of the same order of the
approximated symbol, so in this sense
$\widetilde{H}^\natural_\delta$ is not a good approximation for
$H_\delta$. On the other side, what we need to develop the SAPT is
to control the operator $\pi_\text{r}\ H_\delta\ \pi_\text{r}$,
which is well approximated by $\pi_\text{r}\
\widetilde{H}^\natural_\delta\ \pi_\text{r}$ up to an error of order
$2(\natural+1) +1$ in $\delta$.

\begin{propos}\label{prop_ord_rem}
Let {\upshape Assumption ($\text{A}_\text{s}$)} hold true. Then
$\rrr{R}^\natural_\delta$ has  order
$\sss{O}(\delta^{2(\natural+1)})$, \ie there exist a constant $C$
such that $
\|\rrr{R}^\natural_\delta(p_\text{s},x_\text{s})\|_{\bbb{B}(\sss{D},\sss{H}_\text{{\upshape
f}})}\leqslant C\delta^{2(\natural+1)} $ for all
$(p_\text{s},x_\text{s})\in\R^2$.
Moreover 
$\|\rrr{R}^\natural_\delta\ \pi_\text{r}\|_{\bbb{B}(\sss{H}_\text{{\upshape f}})}=\|\pi_\text{r}\ \rrr{R}^\natural_\delta\|_{\bbb{B}(\sss{H}_\text{{\upshape f}})}\leqslant C\delta^{2(\natural+1)+1},
$ for all $(p_\text{s},x_\text{s})\in\R^2$, i.e.   $\rrr{R}^\natural_\delta\ \pi_\text{r}$, $\pi_\text{r}\ \rrr{R}^\natural_\delta$ and $[\rrr{R}^\natural_\delta;\pi_\text{r}]$ are $\bbb{B}(\sss{H}_\text{{\upshape f}})$-valued symbols of order $\sss{O}(\delta^{2(\natural+1)+1})$.
\end{propos}

\Proof \textbf{(Case $\natural=1$)} The explicit expression  of the
remainder symbol  is
\begin{equation}\label{eeq32}
 \rrr{R}^1_\delta(p_\text{s},x_\text{s})=\delta^2\sum_{n,m=-\infty}^{+\infty}v_{n,m}\ \expo{i2\pi (np_\text{s}+mx_\text{s})}\left[\expo{i2\pi\delta I_{n,m}}-\left(\num{1}_{\sss{H}_\text{f}}+i2\pi\delta I_{n,m}+\dfrac{1}{2}(i2\pi)^2\delta^2{I_{n,m}}^2\right)\right].
\end{equation}
and from \eqref{eeq32} it follows that
$\|\rrr{R}^1_\delta(p_\text{s},x_\text{s})\|_{\bbb{B}(\sss{D},\sss{H}_\text{f})}\leqslant\delta^2\sum_{n,m=-\infty}^{+\infty}|v_{n,m}|\Lambda_{n,m}$ with
 \begin{align}\label{eeq33}
\Lambda_{n,m}:=\sup_{\psi\in\sss{H}_\text{f}} \setminus \{0\}
\dfrac{\left\|\left[\expo{i2\pi\delta
I_{n,m}}-\left(\num{1}_{\sss{H}_\text{f}}+i2\pi\delta
I_{n,m}+\dfrac{1}{2}(i2\pi)^2\delta^2{I_{n,m}}^2\right)\right]\Xi^{-1}\psi\right\|_{\sss{H}_\text{f}}}{\|\psi\|_{\sss{H}_\text{f}}}
\end{align}
since  $\|\psi\|_{\sss{F}}:=\|\Xi\psi\|_{\sss{H}_\text{f}}$ and
$\sss{F}=\Xi^{-1}\sss{H}_\text{f}$. The operators $I_{n,m}$ are
essentially self-adjoint on $\sss{L}$ and we denote their closure
with the same symbol. Since the operators ${I_{n,m}}^2$ are
positive, we can consider the resolvent operators
$R_{n,m}:=({I_{n,m}}^2+\num{1}_{\sss{H}_\text{f}})^{-1}$. Let
suppose that
\begin{equation}\label{eeq34}
\zeta_{n,m}(\delta):= \left\|\left[\expo{i2\pi\delta I_{n,m}}-\left(\num{1}_{\sss{H}_\text{f}}+i2\pi\delta I_{n,m}+\dfrac{1}{2}(i2\pi)^2\delta^2{I_{n,m}}^2\right)\right]R_{n,m}\right\|_{\bbb{B}(\sss{H}_\text{f})}\leqslant \zeta(\delta),
\end{equation}
for all $n,m\in\Z$, with $\sup_\delta \zeta(\delta) < + \infty$. Then equation \eqref{eeq33} would imply
$$\Lambda_{n,m}\leqslant
\zeta(\delta)\left\|({I_{n,m}}^2+\num{1}_{\sss{H}_\text{f}})\,
\Xi^{-1}\right\|_{\bbb{B}(\sss{H}_\text{f})}.$$
Noticing that
${I_{n,m}}^2={\alpha_{n,m}}^2\ \rrr{a}^2+\overline{\alpha_{n,m}}^2\
{\rrr{a}^\dag}^2+2|\alpha_{n,m}|^2\ \Xi$ and observing that
$\|\Xi^{-1}\|_{\bbb{B}(\sss{H}_\text{f})}=2$,
$\|{\rrr{a}}^2\Xi^{-1}\|_{\bbb{B}(\sss{H}_\text{f})}=1$ and
$\|{\rrr{a}^\dag}^2\Xi^{-1}\|_{\bbb{B}(\sss{H}_\text{f})}=2\sqrt{2}$,
 one deduces from inequality \eqref{eeq34} that
$$
\|\rrr{R}^1_\delta(p_\text{s},x_\text{s})\|_{\bbb{B}(\sss{D},\sss{H}_\text{f})}\leqslant C_1\left(\sum_{n,m=-\infty}^{+\infty}|v_{n,m}|(|n|+|m|)^2\right)\delta^2\zeta(\delta)\leqslant C_2 \delta^2\zeta(\delta)
$$
for suitable positive constants $C_1$ and $C_2$.

It remains to prove the inequality \eqref{eeq34} and the estimate on
$\zeta(\delta)$. By spectral calculus one has that
$\zeta_{n,m}(\delta)=\sup_{t\in\sigma(I_{n,m})}|Z_\delta(t)|\leqslant\sup_{t\in\R}|Z_\delta(t)|=:\zeta(\delta)$
where
$$
Z_\delta(t):=4\pi^2\delta^2\ \dfrac{\expo{i2\pi\delta t}-\left(1+i2\pi\delta t-\dfrac{1}{2}(2\pi\delta t)^2\right)}{(2\pi\delta t)^2+4\pi^2\delta^2}.
$$
After some manipulations and the change of variable $\tau:=2\pi\delta t$ one has that
$$
G_\delta(\tau):=\dfrac{1}{4\pi^4\delta^4}\left|Z_\delta\left(\dfrac{\tau}{2\pi\delta}\right)\right|^2\leqslant \dfrac{\tau^4+4\tau^2\cos(\tau)-8\tau\sin(\tau)-8\cos(\tau)+8}{\tau^4}<C_3.
$$
Thus $ \zeta(\delta)^2=4\pi^4\delta^4\
\sup_{\tau\in\R}G_\delta(\tau)\leqslant 4\pi^4C_3\delta^4$, hence
$\|\rrr{R}^1_\delta(p_\text{s},x_\text{s})\|_{\bbb{B}(\sss{D},\sss{H}_\text{f})}\leqslant
C\delta^4$. This concludes the first part of the proof.

\medskip

Since
$\|\rrr{R}^1_\delta\pi_\text{r}\|_{\bbb{B}(\sss{H}_\text{f})}\leqslant\sum_{i=1}^m\|\rrr{R}^1_\delta\ketbra{\psi_{k_i}}{\psi_{k_i}}\|_{\bbb{B}(\sss{H}_\text{f})}$,
then it is enough to show that for any Hermite vector $\psi_k$ the
inequality
$\|\rrr{R}^1_\delta\ketbra{\psi_{k}}{\psi_{k}}\|_{\bbb{B}(\sss{H}_\text{f})}\leqslant
C_k\delta^5$ holds true. Observing that
$\|\rrr{R}^1_\delta\ketbra{\psi_{k}}{\psi_{k}}\|_{\bbb{B}(\sss{H}_\text{f})}=\|\rrr{R}^1_\delta\psi_k\|_{\sss{H}_\text{f}}$,
one deduces
\begin{align*}
 &\lim_{\delta\to0}\delta^{-5}\|\rrr{R}^1_\delta\ketbra{\psi_{k}}{\psi_{k}}\|_{\bbb{B}(\sss{H}_\text{f})}=\lim_{\delta\to0}\left\|\sum_{n,m=-\infty}^{+\infty}v_{n,m}\expo{i2\pi (np_\text{s}+mx_\text{s})}\left(\sum_{j=3}^{+\infty}\dfrac{(i2\pi)^j\delta^{j-3}}{j!}{I_{n,m}}^j\right)\psi_k\right\|_{\sss{H}_\text{f}}\\
&\leqslant\dfrac{4}{3}\pi^3\sum_{n,m=-\infty}^{+\infty}|v_{n,m}|\ \|{I_{n,m}}^3\psi_k \|_{\sss{H}_\text{f}}\leqslant\dfrac{32}{3}\pi^3 C'  \|{A^\dag}^3\psi_k \|_{\sss{H}_\text{f}}=\dfrac{32}{3}\sqrt{(k+3)!}\ \pi^3 C'=:C_k
\end{align*}
where $C':=\sum_{n,m=-\infty}^{+\infty}|v_{n,m}|\ |\alpha_{n,m}|^3$
is finite in view of Assumption ($\text{A}_\text{s}$).

This shows that for all $\delta\in[0,\delta_0)$ (for a suitable
$\delta_0>0$) the norm
$\|\rrr{R}^1_\delta\ketbra{\psi_{k}}{\psi_{k}}\|_{\bbb{B}(\sss{H}_\text{f})}$
is bounded by  $C_k\delta^5$ and so it follows that
$\|\rrr{R}^1_\delta\pi_\text{r}\|_{\bbb{B}(\sss{H}_\text{f})}\leqslant
m C\delta^5$ with $C:=\max_{1,\ldots,m}\{C_{k_i}\}$. Finally
$\|\pi_\text{r}\rrr{R}^1_\delta\|_{\bbb{B}(\sss{H}_\text{f})}
=\|(\rrr{R}^1_\delta\pi_\text{r})^\dag\|_{\bbb{B}(\sss{H}_\text{f})}
=\|\rrr{R}^1_\delta\pi_\text{r}\|_{\bbb{B}(\sss{H}_\text{f})}$.

\textbf{(Case $\natural=0$)} The proof proceeds as in the previous
case. Divide the remainder symbol in two terms
$\rrr{R}^0_\delta=\rrr{R}^0_0+ \rrr{R}^0_1$ where:
\begin{align*}
\rrr{R}^0_0(p_\text{s},x_\text{s})&:=\delta\sum_{n,m=-\infty}^{+\infty}\expo{i2\pi (np_\text{s}+mx_\text{s})}\left(\expo{i2\pi\delta I_{n,m}}-\num{1}_{\sss{H}_\text{f}}-i2\pi\delta I_{n,m}\right)M^0_{n,m}\\
\rrr{R}^0_1(p_\text{s},x_\text{s})&:=\delta^2\sum_{n,m=-\infty}^{+\infty}v_{n,m}\ \expo{i2\pi (np_\text{s}+mx_\text{s})}\left(\expo{i2\pi\delta I_{n,m}}-\num{1}_{\sss{H}_\text{f}}\right).
\end{align*}
The control of $\rrr{R}^0_1$ is easy, indeed
$\|\rrr{R}^0_1\|_{\bbb{B}(\sss{D},\sss{H}_\text{f})}\leqslant2\|\rrr{R}^0_1\|_{\bbb{B}(\sss{H}_\text{f})}\leqslant
4C\delta^2$ where $C:=\sum_{n,m=-\infty}^{+\infty}|v_{n,m}|$.
Moreover (with the same technique used for the case $\natural=1$),
one can check that for any Hermite vector $\psi_k$ the function
$t_1(\delta):=\frac{1}{\delta^3}\|\rrr{R}^0_1(p_\text{s},x_\text{s})\psi_k\|_{\sss{H}_\text{f}}$
is bounded by a constant $C_k>0$ in a suitable interval
$[0,\delta_0)$. This assures that $\|\rrr{R}^0_1\
\pi_\text{r}\|_{\bbb{B}(\sss{H}_\text{f})}$ is of order
$\sss{O}(\delta^3)$.

To control $\rrr{R}^0_0$ we need to estimate
$\Sigma_{n,m}:=\|\left(\expo{i2\pi\delta
I_{n,m}}-\num{1}_{\sss{H}_\text{f}}-i2\pi\delta
I_{n,m}\right)M^0_{n,m}\Xi^{-1}\|_{\bbb{B}(\sss{H}_\text{f})}$. Let
$R'_{n,m}$ be the resolvent
$(I_{n,m}+i\num{1}_{\sss{H}_\text{f}})^{-1}$. It is easy to check
that
$\|(I_{n,m}+i\num{1}_{\sss{H}_\text{f}})M^0_{n,m}\Xi^{-1}\|_{\bbb{B}(\sss{H}_\text{f})}$
is bounded by a linear expression in $|n|$ and $|m|$. Indeed, as
proved in Proposition \ref{prop_def_dom_aprx}, both $M^0_{n,m}$ and
$M^1_{n,m}$ are bounded by $\Xi$. By spectral calculus
$\|\left(\expo{i2\pi\delta
I_{n,m}}-\num{1}_{\sss{H}_\text{f}}-i2\pi\delta
I_{n,m}\right)R'_{n,m}\|^2_{\bbb{B}(\sss{H}_\text{f})}$ is bounded
by the maximum in $\tau$ of the function $F_\delta(\tau):=
4\pi^2\delta^2\ \frac{\tau^2-2\tau\sin\tau-2\cos\tau+2}{\tau^2}$.

The last part follows  observing that $M^0_{n,m}$ is a  linear
combinations of $\rrr{a}$ and $\rrr{a}^\dag$ and so they act
splitting a Hermite vector $\psi_k$ as
$c^k_{n,m}\psi_{k-1}+d^k_{n,m}\psi_{k+1}$ where for a fixed $k$ the
coefficients depend on $f^{(j)}_{n,m}$.  To conclude the proof it is
sufficient to notice that
$t_0(\delta):=\frac{1}{\delta^2}\|\left(\expo{i2\pi\delta
I_{n,m}}-\num{1}_{\sss{H}_\text{f}}-i2\pi\delta
I_{n,m}\right)\psi_k\|_{\sss{H}_\text{f}}$ is bounded by a
constant $C_k>0$ in a suitable interval $[0,\delta_0)$. \CVD

\subsubsection*{Derivation of the adiabatically decoupled effective dynamics}

We recall that the Weyl quantization of the symbol $H_\delta$ is the
Hamiltonian \eqref{eeq11}, namely
$\text{Op}_\delta(H_\delta)=H^\ssss{W}$. As for the approximated
symbol $\widetilde{H}^\natural_\delta$, we pose
$\widetilde{H}^\natural:=\text{Op}_\delta(\widetilde{H}^\natural_\delta)$.
Both $H^\ssss{W}$ and $\widetilde{H}^\natural$ are bounded operators
from $L^2(\R,dx_\text{{\upshape s}})\otimes \sss{F}$ to
$\sss{H}_\text{w}:=L^2(\R,dx_\text{{\upshape s}})\otimes
\sss{H}_\text{{\upshape f}}$.

\begin{teo}\label{teo1}
Let {\upshape Assumption ($\text{A}_\text{s}$)} be satisfied. Let
$\{\sigma_n(\cdot)\}_{n\in\sss{I}}$, with $\sss{I} = \{n, \ldots,
n+m-1 \}$,  be a family of Landau bands for $\Xi$ and let
$\pi_{\text{\upshape r}}: =
\sum_{n\in\sss{I}}\ketbra{\psi_{n}}{\psi_{n}}$ be the spectral
projector of $H_0=\Xi$ corresponding to the  set
$\{\sigma_n(p_\text{{\upshape s}},x_\text{{\upshape
s}})\}_{n\in\sss{I}}$. Then:

\medskip

\noindent \textbf{1. Almost-invariant subspace:} there exists an
orthogonal projection
$\Pi^\natural_\delta\in\bbb{B}(\sss{H}_\text{{\upshape w}})$, with
$\Pi^\natural_\delta=\text{{\upshape
Op}}_\delta(\pi)+\sss{O}_0(\delta^\infty)$,  $\pi(p_\text{{\upshape
s}},x_\text{{\upshape s}})\asymp\sum_{j=0}^\infty \delta^j\
\pi_j(p_\text{{\upshape s}},x_\text{{\upshape s}})$,  and
$\pi_0(p_\text{{\upshape s}},x_\text{{\upshape s}}) \equiv
\pi_{\text{\upshape r}}$ , such that
\begin{equation}\label{eeq_ord_sym}
[\widetilde{H}^\natural;\Pi^\natural_\delta]=\sss{O}_0(\delta^\infty), \qquad  [H^\ssss{W};\Pi^\natural_\delta]=\sss{O}_0(\delta^{2(\natural+1)+1}).
\end{equation}

\medskip

\noindent \textbf{2. Effective dynamics:} let $\Pi_\text{{\upshape r}}:=\num{1}_{\sss{H}_{\text{\upshape s}}}\otimes\pi_\text{{\upshape r}}\in\bbb{B}(\sss{H}_\text{{\upshape w}})$  and $\sss{K}:=\text{{\upshape Ran}}\ \Pi_\text{{\upshape r}}\simeq L^2(\R,dx_\text{s})\otimes \C^m$. Then there exists a
unitary operator
$
U^\natural_\delta\in\bbb{B}(\sss{H}_\text{{\upshape w}})
$ such that
\renewcommand{\labelenumi}{{\rm(\roman{enumi})}}
\begin{enumerate}
    \item $U^\natural_\delta=\text{{\upshape Op}}_\delta(u)+\sss{O}_0(\delta^\infty)$,
    where the symbol $u\asymp\sum_{j=0}^{\infty}\delta^j u_j$
    has principal part $u_0 \equiv \num{1}_{\sss{H}_\text{{\upshape f}}}$;
    \item $\Pi_\text{{\upshape r}}=U^\natural_\delta\ \Pi^\natural_\delta\ {U^\natural_\delta}^{-1};$
    \item Let $h^\natural$ in $S^1(\bbb{B}(\sss{H}_\text{{\upshape
f}}))$ be a resummation of the formal symbol
$u \moy \pi \moy \widetilde{H}^\natural_\delta \moy \pi \moy
u^{\dag}$ and define the {\upshape effective Hamiltonian} by
$H^\delta_{\text{\upshape eff}}:=\text{{\upshape Op}}_\delta(
h^\natural)$. Since  $[H^\delta_{\text{\upshape
eff}};\Pi_\text{{\upshape r}}]=0$,  $H^\delta_{\text{\upshape eff}}$
is a  bounded operator on $\sss{K}$. Then
\begin{equation}\label{eq_eff_dyn}
U^\natural_\delta \, \Pi^\natural_\delta \, H^{\sss{W}} \, \Pi^\natural_\delta \, {U^\natural_\delta}^{-1} =  H^\delta_{\text{\upshape eff}}+\sss{O}_0(\delta^{2(\natural+1)+1})\in\bbb{B}(\sss{K}).
\end{equation}
\end{enumerate}

\noindent \textbf{2'. Effective dynamics for a single Landau band
when $\ttb{A}_{\Gamma} = 0$:} Consider a single Landau band
$\sigma_\ast(\cdot)=\lambda_\ast$, so that $\pi_\text{{\upshape
r}}=\ketbra{\psi_\ast}{\psi_\ast}$. Then, up to the order
$\delta^4$, one has that
\begin{equation}\label{eeq36}
H^\delta_{\text{\upshape eff}}=\lambda_\ast\num{1}_{\sss{H}_\text{{\upshape s}}}+\delta^2\ V(P_\text{{\upshape s}},Q_\text{{\upshape s}})+\delta^4\ \dfrac{\lambda_\ast}{2}\ Y(P_\text{{\upshape s}},Q_\text{{\upshape s}})+\sss{O}_0\left(\delta^{5}\right)
\end{equation}
where  $V(P_\text{{\upshape s}},Q_\text{{\upshape
s}}):=\text{{\upshape Op}}_\delta(V)$ is the Weyl quantization of
the $\Z^2$-periodic function $V(p_\text{{\upshape
s}},x_\text{{\upshape s}})$ related to the $\Gamma$-periodic
potential $\ttt{V}_\Gamma$, while $Y(P_\text{{\upshape
s}},Q_\text{{\upshape s}}):=\text{{\upshape Op}}_\delta(|D_z|^2(V))$
is the Weyl quantization of the function
$|D_z|^2(V)(p_\text{{\upshape s}},x_\text{{\upshape s}})$ defined
through the differential operator \eqref{eeq30}.
\end{teo}

 The derivation of the effective dynamics when $\ttb{A}_{\Gamma}
\neq 0$ will be considered in Section \ref{Sec_Har_per}.

\goodbreak

\subsection*{Proof of Theorem \ref{teo1}}
\label{Sec_proof Har}

\subsubsection*{Step 1. {Almost-invariant subspace}}

As explained in the first part of proof of the Theorem \ref{teo2}
 one constructs a
formal symbol $\pi$ (the \emph{Moyal projection}) such that:
{\upshape(i)} $\pi\moy\pi\asymp\pi$;
{\upshape(ii)} $\pi^\dag=\pi$; {\upshape(iii)}
$\widetilde{H}^\natural_\delta\moy\pi\asymp\pi\moy
\widetilde{H}^\natural_\delta$. Such a symbol
$\pi\asymp\sum_{j=0}^{\infty}\delta^j\pi_j$ is constructed recursively  order by order starting from
$\pi_0=\pi_\text{r}$ and $\widetilde{H}^\natural_\delta$ and it is unique (see Lemma 2.3. in \cite{PST1}).
The recursive relations are
\begin{equation}\label{eq_rec_p1}
\pi_n:=\pi^{\text{D}}_n+\pi^{\text{OD}}_n
\end{equation}
where the \emph{diagonal part} is $\pi^{\text{D}}_n:=\pi_\text{r}G_n\pi_\text{r}+(\num{1}_{\sss{H}_\text{f}}-\pi_\text{r})G_n(\num{1}_{\sss{H}_\text{f}}-\pi_\text{r})$ with
\begin{equation}\label{eq_rec_p2}
G_n:=\left[\left(\sum_{j=0}^{n-1}\delta^j \pi_j\right)\sharp\left(\sum_{j=0}^{n-1}\delta^j \pi_j\right)-\left(\sum_{j=0}^{n-1}\delta^j \pi_j\right)\right]_n.
\end{equation}
The \emph{off-diagonal part} is defined by the implicit relation
$[H_0;\pi^{\text{OD}}_n]=-F_n$ where
\begin{equation}\label{eq_rec_p3}
F_n:=\left[\widetilde{H}^\natural_\delta \, \sharp \,
\left(\sum_{j=0}^{n-1}\delta^j \pi_j+\delta^n
\pi^{\text{D}}_n\right)-\left(\sum_{j=0}^{n-1}\delta^j
\pi_j+\delta^n \pi^{\text{D}}_n\right)\, \sharp \,
\widetilde{H}^\natural_\delta\right]_n.
\end{equation}

The uniqueness  allows us to construct $\pi$ only locally and this
local construction is explained in the
 second part of Lemma 2.3 in \cite{PST1}.
 In our case we can choose a $(p_\text{s},x_\text{s})$-independent
positively oriented complex circle $\Lambda\subset\C$, symmetric
with respect to the real axis, which encloses the family of
(constant) spectral bands
$\{\sigma_n(\cdot)=\lambda_{n}\}_{n\in\sss{I}}$ and such that
$\text{dist}(\Lambda,\sigma(H_0))\geqslant\frac{1}{2}$ (see Figure
\ref{fig02}). For all $\lambda\in\Lambda$
we construct recursively
the \emph{Moyal resolvent} (or \emph{parametrix})
$R^\natural(\lambda;\cdot):=\sum_{j=0}^\infty\delta^jR^\natural_j(\lambda;\cdot)$
 of $\widetilde{H}^\natural_\delta$, following the same technique explained during the proof of Theorem \ref{teo2}.
The approximants of the symbol $\pi$ are related to the approximants
of the Moyal resolvent by the usual Riesz formula
$\pi_j(z):=\dfrac{i}{2\pi}\oint_{\Lambda}d\lambda\
R^\natural_j(\lambda;z)$ where $z:=(p_\text{s},x_\text{s})\in\R^2$.
Some care is required to show (iii) since, by construction,
$\widetilde{H}^\natural_\delta\moy \pi$ takes values in
$\bbb{B}(\sss{H}_\text{f})$ while $\pi\moy
\widetilde{H}^\natural_\delta$ takes values in $\bbb{B}(\sss{F})$.
To solve this problem one can use the same argument proposed in
Lemma 7 of \cite{PST2}.

The technical and new part of the proof consist in  showing that
$\pi\in S^1(\bbb{B}(\sss{H}_\text{f}))\cap
S^1(\bbb{B}(\sss{H}_\text{f},\sss{F}))$. The Riesz formula
 implies
$\|(\partial^\alpha_z\pi_j)(z)\|_\flat\leqslant 2\pi\
\sup_{\lambda\in\Lambda}\|\partial^\alpha_z
R^\natural_j(\lambda;z)\|_\flat$  for all $\alpha\in\N^2$  ($\flat$
means either $\bbb{B}(\sss{H}_\text{f})$ or
$\bbb{B}(\sss{H}_\text{f};\sss{F})$ and
$\partial^{\alpha}_z:=\partial^{\alpha_1}_{p_\text{s}}\partial^{\alpha_2}_{x_\text{s}}$)
since $\Lambda$ does not depend on $z$. Then we need only to show
that  $R_j^\natural(\lambda;\cdot)\in
S^1(\bbb{B}(\sss{H}_\text{f}))\cap
S^1(\bbb{B}(\sss{H}_\text{f},\sss{F}))$. The choice of $\Lambda$
assures
$\|R^\natural_0(\lambda;z)\|_{\bbb{B}(\sss{H}_\text{f})}=\|(\Xi-\lambda\num{1}_{\sss{H}_\text{f}})^{-1}\|_{\bbb{B}(\sss{H}_\text{f})}\leqslant2$.
Moreover $\partial^\alpha_zR_0^\natural(\lambda;z)=0$ for all
$\alpha\neq0$ and this implies that $R^\natural_0\in
S^1(\bbb{B}(\sss{H}_\text{f}))$ uniformly in $\lambda$. Since $
\|R^\natural_0(\lambda;z)\|_{\bbb{B}(\sss{H}_\text{f},\sss{F})}=\|\Xi
(\Xi-\lambda\num{1}_{\sss{H}_\text{f}})^{-1}\|_{\bbb{B}(\sss{H}_\text{f})}\leqslant
\infty$  one concludes that $R^\natural_0\in
S^1(\bbb{B}(\sss{H}_\text{f},\sss{F}))$ uniformly in $\lambda$.

By means of equation \eqref{R induction}, one has
$R_j^\natural=-R_0^\natural L^\natural_j$ where $L^\natural_j$ is
the \emph{$j$-th order obstruction} for $R_0^\natural$ to be the
Moyal resolvent. In view of this recursive relation, the proof of
$R^\natural_j\in S^1(\bbb{B}(\sss{H}_\text{f}))$ for all $j\in\N$ is
reduced to show that $L^\natural_j\in
S^1(\bbb{B}(\sss{H}_\text{f}))$ for all $j\in\N$.

The first order obstruction,  computed by means of
\eqref{L_obstruction}, is
\begin{align*}
L^\natural_1(\lambda;z)&=\delta^{-1}[(\widetilde{H}^\natural_\delta(z)-\lambda\num{1}_{\sss{H}_\text{f}})\moy
R^\natural_0(\lambda;z)-\num{1}_{\sss{H}_\text{f}}]_1=H_1(z)\
R^\natural_0(\lambda;z)-\dfrac{i}{2}\{\Xi;R^\natural_0(\lambda;z)\}_{p_\text{s},x_\text{s}}.
\end{align*}
Since $\Xi$ and $R^\natural_0$ do not depend on $z\in\R^2$ it
follows that $L^\natural_1=H_1R^\natural_0$. The operator $H_1$ is
linear in $\rrr{a}$ and $\rrr{a}^\dag$ (with all its derivative) if
$\natural=0$ or $H_1=0$ if $\natural=1$. In both cases $H_1$ (with
its derivatives) is infinitesimally bounded with respect to $\Xi$
(see Lemma \ref{appA3lem2}). This shows that $L^\natural_1\in
S^1(\bbb{B}(\sss{H}_\text{f}))$ (but not in
$S^1(\bbb{B}(\sss{H}_\text{f},\sss{F}))$ if $\natural=0$).

We proceed by induction assuming that $L^\natural_j\in
S^1(\bbb{B}(\sss{H}_\text{f}))$ for all $j\leqslant m\in\N$. The
$(m+1)$-th order obstruction $L^\natural_{m+1}$ can be computed by
means of equation \eqref{L_obstruction} and the
 Moyal  formula for the expansion of $\sharp$ (see equation (A.9) in \cite{stefan_book}).
After some manipulations, one gets
\begin{align*}
L^\natural_{m+1}(\lambda;z)&=\frac{1}{(2i)^{m+1}}\sum_{\substack{\alpha_1+\alpha_2+r+l=m+1\\
\\
0\leqslant l\leqslant m,\ 0\leqslant r\leqslant
2(\natural+1)}}\dfrac{(-1)^{|\alpha|+1}}{\alpha_1!\alpha_2!}\left(\partial^{\alpha_1}_{x_\text{s}}\partial^{\alpha_2}_{p_\text{s}}
H_{r}R_0^\natural\right)(\lambda;z)\
\left(\partial^{\alpha_1}_{p_\text{s}}\partial^{\alpha_2}_{x_\text{s}}L^\natural_l\right)(\lambda;z).
\end{align*}
 Since $H_{r}R_0^\natural\in S^1(\bbb{B}(\sss{H}_\text{f}))$ uniformly
in $\lambda$ (see Proposition \ref{prop_ord_rem}) then
$L^\natural_{m+1}\in S^1(\bbb{B}(\sss{H}_\text{f}))$, and this
concludes the inductive argument.

Finally to prove $R_j^\natural\in
S^1(\bbb{B}(\sss{H}_\text{f},\sss{F}))$, observe that $
\|\partial^\alpha_zR^\natural_j\|_{\bbb{B}(\sss{H}_\text{f},\sss{F})}=\|\Xi
\, R^\natural_0 \,
(\partial^\alpha_zL^\natural_j)\|_{\bbb{B}(\sss{H}_\text{f})}\leqslant\
C_\alpha \|\Xi \,
R^\natural_0\|_{\bbb{B}(\sss{H}_\text{f})}\leqslant+\infty $ for all
$j,\alpha\in\N$.
\begin{rk}\label{rk2}
It clearly emerges from the proof that the order
$\delta^{2(\natural+1)}$ is the best approximation which can be
obtained with this technique. The obstruction  is the condition
$H_rR_0^\natural\in S^1(\bbb{B}(\sss{H}_\text{f}))$, which can be
satisfied by the resolvent
$R_0^\natural:=(\Xi-\zeta\num{1}_\text{f})^{-1}$ only for
$0\leqslant r\leqslant2(\natural+1)$. \hfill $\blacklozenge\lozenge$
\end{rk}
Proposition A.9 of \cite{stefan_book} assures that
$\widetilde{H}^\natural_\delta\moy \pi\in
S^1(\bbb{B}(\sss{H}_\text{f}))$  and, by adjointness, also $\pi\moy
\widetilde{H}^\natural_\delta \in S^1(\bbb{B}(\sss{H}_\text{f}))$.
By construction
$[\widetilde{H}^\natural;\text{Op}_\delta(\pi)]=\text{Op}_\delta([\widetilde{H}^\natural_\delta;\pi]_\sharp
)=\sss{O}_0(\delta^\infty)$ where $
[\widetilde{H}^\natural_\delta;\pi]_\sharp:=
\widetilde{H}^\natural_\delta\moy \pi -\pi\moy
\widetilde{H}^\natural_\delta=\sss{O}(\delta^\infty)$ denotes the
\emph{Moyal commutator}. Observing that $
[H_\delta;\pi]_\sharp=[\widetilde{H}^\natural_\delta+\rrr{R}^\natural_\delta;\pi]_\sharp=[\rrr{R}^\natural_\delta;\pi]_\sharp+\sss{O}(\delta^\infty)$
and since Proposition \ref{prop_ord_rem} implies
$[\rrr{R}^\natural_\delta;\pi]_\sharp=[\rrr{R}^\natural_\delta;\pi_\text{r}]+\sss{O}(\delta^{2(\natural+1)+1})=\sss{O}(\delta^{2(\natural+1)+1})$,
it follows $[H_\delta;\pi]_\sharp=\sss{O}(\delta^{2(\natural+1)+1})$
which implies after the quantization
$[H^\ssss{W};\text{Op}_\delta(\pi)]=\sss{O}_0(\delta^{2(\natural+1)+1})$.

The last step is to obtain the true projection $\Pi_\delta^\natural$
(the \emph{super adiabatic projection}) from $\text{Op}_\delta(\pi)$
by means of the formula \eqref{eq_true_proj}. Since
$\Pi_\delta^\natural-\text{Op}_\delta(\pi)=\sss{O}_0(\delta^\infty)$,
one recovers the estimates \eqref{eeq_ord_sym}.

\subsubsection*{Step 2. {Construction of the intertwining unitary}}

The construction of the intertwining unitary follows as in the proof
of Theorem 3.1 of \cite{PST1}. Firstly one constructs a formal
symbol $u\asymp\sum_{j=0}^{\infty}\delta^j u_j$ such that:
{\upshape(i)} $u^\dag \moy u=u \moy
u^\dag=\num{1}_{\sss{H}_\text{f}}$; {\upshape(ii)} $u \moy \pi \moy
u^\dag=\pi_\text{r}$.

The existence of such a symbol follows from a recursive procedure
starting from $u_0$  (which  can be fixed to be
$\num{1}_{\sss{H}_\text{f}}$ in our specific case) and using the
expansion of $\pi\asymp\sum_{j=0}^{\infty}\delta^j \pi_j$ obtained
above. However, the symbol $u$ which comes out of this procedure is
not unique. The recursive relations are
\begin{equation}\label{eq_rec_u1}
u_n:=a_n+b_n\hspace{1cm}\text{with}\hspace{1cm}a_n:=-\frac{1}{2}A_n,\ \ \ \ \ \ \ b_n:=[\pi_\text{r};B_n]
\end{equation}
where
\begin{equation}\label{eq_rec_u2}
A_n:=\left[\left(\sum_{j=0}^{n-1}\delta^j u_j\right)\sharp\left(\sum_{j=0}^{n-1}\delta^j u_j\right)^\dag-\num{1}_{\sss{H}_\text{f}}\right]_n
\end{equation}
and
\begin{equation}\label{eq_rec_u3}
B_n:=\left[\left(\sum_{j=0}^{n-1}\delta^j u_j+\delta^n
a_n\right)\moy \pi \moy \left(\sum_{j=0}^{n-1}\delta^j u_j+\delta^n
a_n\right)^\dag-\pi_\text{r}\right]_n
\end{equation}
Since $u_0=\num{1}_{\sss{H}_\text{f}}\in
S^1(\bbb{B}(\sss{H}_\text{f}))$, then it follows by induction that
$u_j\in S^1(\bbb{B}(\sss{H}_\text{f}))$ for all $j\in\N$.

The quantization of $u$ is an element of $\bbb{B}(\sss{H}_\text{w})$
but it is not a true unitary. Nevertheless $\text{Op}_\delta(u)$ can
be modified by an $\sss{O}_0(\delta^\infty)$ term using the same
technique of Lemma 3.3 (Step II) in \cite{PST1} to obtain the true
unitary $U_\delta^\natural$.

\subsubsection*{Step 3. {Effective dynamics}}

By construction
$[H^\delta_\text{eff};\Pi_\text{r}]=\text{Op}_\delta([h^\natural;\pi_\text{r}]_\sharp)=
[U_\delta^\natural \, \Pi^\natural_\delta \,
\widetilde{H}^\natural_\delta \, \Pi^\natural_\delta \,
{U_\delta^\natural}^{-1}; \Pi_\text{r}]=0$ since
$\Pi_\text{r}={U_\delta^\natural}\, \Pi^\natural_\delta \,
{U_\delta^\natural}^{-1}$. Moreover equation \eqref{eq_eff_dyn}
follows observing that $U^\natural_\delta \, \Pi^\natural_\delta \,
H^{\sss{W}} \, \Pi^\natural_\delta \, {U^\natural_\delta}^{-1} -
H^\delta_{\text{\upshape eff}}$ coincides with the quantization of
$u\moy\pi\moy \rrr{R}^\natural_\delta\moy\pi \moy u^{\dag}$ which is
a symbol of order $\sss{O}(\delta^{2(\natural+1)+1})$.

\subsubsection*{Step 4. {The case of a single Landau band when $\ttb{A}_{\Gamma} = 0$}}

We need to expand the Moyal product $h^{\natural=1}=u\moy \pi\moy
\widetilde{H}^1_\delta\moy\pi\moy u^{\dag}=\pi_\text{r}\moy u\moy
\widetilde{H}^1_\delta\moy
u^{\dag}\moy\pi_\text{r}+\sss{O}(\delta^\infty)$ up to the order
$\delta^4$. To compute the various terms of the expansion
$h^{\natural=1}\asymp\sum_{j=0}^\infty\delta^jh_j$ it is useful to
define $\chi_j:=[u\moy \widetilde{H}^1_\delta\moy u^{\dag}]_j$, so
that $h_j=\pi_\text{r}\chi_j\pi_\text{r}$. Observing that
\begin{align*}
u\moy \widetilde{H}^1_\delta-\left(\sum_{j=0}^{m-1}\delta^j\
\chi_j\right)\moy u&=\left(u\moy \widetilde{H}^1_\delta\moy
u^{\dag}-\sum_{j=0}^{m-1}\delta^j\chi_j\right)\moy
u+\sss{O}(\delta^\infty)=\delta^{m}\chi_{m} +\sss{O}(\delta^{m+1})
\end{align*}
one obtains the useful formula
\begin{equation}\label{eeq40}
\chi_{m}=\left[u\moy
\widetilde{H}^1_\delta-\left(\sum_{j=0}^{m-1}\delta^j\
\chi_j\right)\moy u\right]_{m}.
\end{equation}

At the zeroth order  one finds $ h_0= \pi_0 \, u_0 \, H_0 \,
u_0^{\dag}\,\pi_0 = \pi_\text{r}\, \Xi \, \pi_\text{r}=\lambda_\ast
\, \pi_\text{r} $ since $u_0=\num{1}_{\sss{H}_\text{f}}$ and
$\pi_0=\pi_\text{r}$. Its quantization  is the operator
$\text{Op}_\delta({h}_0)=\lambda_\ast\num{1}_{\sss{H}_\text{s}}$
acting on $\sss{K}=L^2(\R,dx_\text{s})$.

As for the first order ($m=1$),  $
\chi_1=u_1H_0+u_0H_1-\chi_0u_1+[u_0\moy H_0]_1-[\chi_0\moy
u_0]_1=[u_1;\Xi] $ since $\chi_0=u_0H_0 u_0^{-1}=\Xi$  and $H_1=0$.
Then $
{h}_1=\pi_\text{r}[u_1;\Xi]\pi_\text{r}=\lambda_\ast(\pi_\text{r}u_1\pi_\text{r}-\pi_\text{r}u_1\pi_\text{r})=0
$, hence $\text{Op}_\delta(h_1)=0$.

At the second order ($m=2$), one obtains after some manipulations
$\chi_2 = H_2+u_2 \, \Xi-\Xi \, u_2- \chi_1 \, u_1$ which implies $
h_2=\pi_\text{r}H_2\pi_\text{r}-\pi_\text{r}\chi_1u_1\pi_\text{r}$.
We need to compute $u_1$. Using equations \eqref{eq_rec_u1},
\eqref{eq_rec_u2} and \eqref{eq_rec_u3} one obtains that
$-2a_{1}:=\left[u_0\moy{u_0}^\dag-\num{1}_{\sss{H}_\text{f}}\right]_{1}=0$
and $b_1:=[\pi_\text{r};B_1]$ with $B_1=[u_0\moy\pi \moy
{u_0}^\ast-\pi_\text{r}]_1=\pi_1$ since $a_1=0$. To compute $\pi_1$
we use equations \eqref{eq_rec_p1}, \eqref{eq_rec_p2} and
\eqref{eq_rec_p3}. Since
$G_1=[\pi_\text{r}\moy\pi_\text{r}-\pi_\text{r}]_1=0$ it follows
that $\pi^\text{D}_1=0$. In the case of a single energy band in the
relevant part of the spectrum, the implicit relation which defines
$\pi^\text{OD}_n$ can be solved, obtaining the useful equation
\begin{equation}\label{eq_rec_p4}
 \pi^\text{OD}_n=\pi_\text{r}F_n(\Xi-\lambda_\ast\num{1}_{\sss{H}_\text{f}})^{-1}(\num{1}_{\sss{H}_\text{f}}-\pi_\text{r})-
(\num{1}_{\sss{H}_\text{f}}-\pi_\text{r})(\Xi-\lambda_\ast\num{1}_{\sss{H}_\text{f}})^{-1}F_n\pi_\text{r}.
\end{equation}
Since
$F_1=[\widetilde{H}^1_\delta\moy\pi_\text{r}-\pi_\text{r}\moy\widetilde{H}^1_\delta]_1=H_1\pi_\text{r}-\pi_\text{r}H_1=0$,
being $H_1=0$, it follows $B_1=\pi_1= \pi^\text{OD}_1=0$ and
consecutively $u_1=b_1=0$. Then $ h_2=\pi_\text{r} \, H_2 \,
\pi_\text{r}=V \, \pi_\text{r}$, according to \eqref{eeq29a}, and
its quantization defines on $\sss{K}$ the operator
$\text{Op}_\delta({h}_2)=V(P_\text{s},X_\text{s})$.

Considering \eqref{eeq40} at the third order (m=3) and using
$u_1=0$, one obtains after some computations
$\chi_3=H_3+u_3\,\Xi-\Xi \, u_3-\chi_1 \, u_2$ which implies $
h_3=\pi_\text{r} \, H_3 \, \pi_\text{r}-\pi_\text{r} \, \chi_1 \,
u_2 \, \pi_\text{r}$. Thus we need to compute $u_2$. Since $u_1=0$,
it follows
$-2a_2=[u_0\moy{u_0}^\dag-\num{1}_{\sss{H}_\text{f}}]_2=0$,
$B_2=[u_0\moy\pi \moy {u_0}^\dag-\pi_\text{r}]_2=\pi_2$ and
$b_2=[\pi_\text{r};\pi_2]$.  Since  $\pi_1=0$, one has that
$G_2=[\pi_\text{r}\moy\pi_\text{r}-\pi_\text{r}]_2=0$ which  implies
$\pi_2^\text{D}=0$. To compute $\pi_2^\text{OD}$ we need
 $F_2=[\widetilde{H}^1_\delta\moy\pi_\text{r}-\pi_\text{r}\moy\widetilde{H}^1_\delta]_2=[H_2;\pi_\text{r}]
 =[\num{1}_{\sss{H}_\text{f}};\pi_\text{r}]=0$, where $H_2=V\ \num{1}_{\sss{H}_\text{f}}$ has been used.
Then $B_2=\pi_2= \pi^\text{OD}_2=0$ and consequently $u_2=b_2=0$.
Therefore ${h}_3=\pi_\text{r}H_3\pi_\text{r}$, and equation
\eqref{eeq29b} implies that  $\pi_\text{r}H_3\pi_\text{r}=0$ in view
of
$\pi_\text{r}\rrr{a}\pi_\text{r}=\bra{\psi_\ast}\rrr{a}\ket{\psi_\ast}\pi_\text{r}=0$
and
 similarly for $\rrr{a}^\dag$. Then $\text{Op}_\delta({h}_3)=0$.

To compute the fourth order, we do not need to compute $u_3$ and
$\pi_3$. Indeed, by computing \eqref{eeq40} at the fourth order
(m=4) one finds $ \chi_4=H_4+u_4 \, \Xi-\Xi \,
u_4+u_3\,H_1-\chi_3\,u_1=H_4+u_4\,\Xi-\Xi \, u_4$ since
$H_1=u_1=u_2=0$. Then ${h}_4=\pi_\text{r} H_4
\pi_\text{r}=\frac{\lambda_\ast}{2}|D_z(V)|^2\ \pi_\text{r}$,
according to equation \eqref{eeq29c}, and its quantization yields
$\text{Op}_\delta({h}_4)=\frac{\lambda_\ast}{2}Y(P_\text{s},Q_\text{s})$.

\subsection{Harper-like Hamiltonians}

The first term in \eqref{eeq36} is a multiple of the identity, and
therefore does not contribute to the dynamics as far as the
expectation values of the observables are concerned. The leading
term, providing a non-trivial contribution to the dynamics at the
original microscopic time scale $s\propto\delta^2\tau$, is the
bounded operator $V(P_\text{{\upshape s}},Q_\text{{\upshape s}})$
acting on the reference Hilbert space $L^2(\num{R},dx_\text{s})$.
This operator is the Weyl quantization of the $\Z^2$-periodic smooth
function $V$ defined on the classical phase space $\R^2$. Hereafter
we write $x_\text{s} \equiv x$ to simplify the notation.

The quantization procedure can be reformulated by introducing the
unitary operators {$\ssss{U}_{\infty}:=\expo{-i2\pi Q_\text{s}}$}
and {$\ssss{V}_{\infty}:=\expo{-i2\pi P_\text{s}}$} (\emph{Harper
unitaries}), acting on $L^2(\num{R},dx)$ as (recall
$\delta^2=\nicefrac{h_B}{2\pi}$)
\begin{equation}\label{eeq_har_unit}
(\ssss{U}_{\infty}\psi)(x)=\expo{- i 2\pi x}\psi(x), \qquad  \qquad
(\ssss{V}_{\infty}\psi)(x)=\psi(x- \iota_qh_B) .
\end{equation}

For any $\Z^2$-periodic function
$F(p,x)=\sum_{n,m=-\infty}^{+\infty}f_{n,m}\expo{-i2\pi(np+mx)}$
whose Fourier series is uniformly convergent, the $h_B$-Weyl
quantization of $F$ is given by the formula
\begin{equation}\label{eeq37}
 \widehat{F}(\ssss{U}_{\infty},
\ssss{V}_{\infty})= \sum_{n,m=-\infty}^{+\infty}f_{n,m}
\expo{-i\pi nm(\iota_q h_B)}\ \ssss{V}_{\infty}^n
\ssss{U}_{\infty}^m.
\end{equation}
where the fundamental commutation relation $\ssss{U}_{\infty}
\ssss{V}_{\infty}=\expo{-i2\pi\left({\iota_q}{h_B}\right)}\
\ssss{V}_{\infty} \ssss{U}_{\infty}$ has been used. Formula
\eqref{eeq37} defines a \emph{Harper-like Hamiltonian} with
deformation parameter $h_B$. Indeed, the special case
$H_\text{Har}=\sss{U}_{\infty}+\sss{U}_{\infty}^{-1}+\sss{V}_{\infty}+\sss{V}_{\infty}^{-1}$
is the celebrated Harper Hamiltonian, namely the operator acting on
$L^2(\R,dx)$ as
\begin{equation}\label{eq54}
(H_\text{Har}\psi)(x)=\psi(x + h_B)+\psi(x - h_B) + 2\cos( 2\pi
x)\psi(x).
\end{equation}

In analogy with Section 3, we summarize the discussion in the
following conclusion

\begin{concl}\label{corol_har}
Under the assumptions of Theorem \ref{teo1}, for every
$\ttt{V}_{\Gamma} \in C^\infty_{\text{\upshape b}}(\R^2,\R)$, in the
Harper regime ($h_B\to 0$), the dynamics generated by the
Hamiltonian $\ttb{H}_\text{{\upshape BL}}$ \eqref{Hamilt BL}
restricted to the spectral subspace corresponding to a Landau level
$\lambda_*$ is approximated up to an error of order ${h_B}$ (and up
to a unitary transform and an energy rescaling)
 by the dynamics generated on the reference Hilbert space $L^2(\num{R},dx)$ by a  Harper-like Hamiltonian,
 i.e. by a power series in  the Harper unitaries $\ssss{U}_{\infty}$ and $\ssss{V}_{\infty}$, defined by \eqref{eeq_har_unit}.
\end{concl}


\subsection{Coupling of Landau bands in a periodic magnetic potential}
\label{Sec_Har_per}

According to Theorem \ref{teo1}, the first non-trivial term which
describes the effective  dynamics in the almost invariant subspace
related to a single Landau band $\lambda_\ast$ is of order
$\delta^2\propto h_B$. An important ingredient in the proof is that
$\ttb{A}_\Gamma=0$ implies $H_1=0$. Moreover, the second non-trivial
correction appears  at order $\delta^4\propto{h_B}^2$ although
$H_3\neq0$. Indeed, the correction at order $\delta^3$ vanishes
since $H_3$, defined by \eqref{eeq29b}, is linear in $\rrr{a}$ and
$\rrr{a}^\dag$, hence $\bra{\psi_\ast}H_3\ket{\psi_\ast}=0$. This
observation suggests that for a family of Landau bands which
contains two contiguous bands $\{ \lambda_{\ast},\lambda_{\ast+1}
\}$ one has, in general, a second non-trivial correction of order
$\delta^3\propto{h_B}^{\frac{3}{2}}$ for the effective dynamics.
Indeed, in this case
one has $\pi_\text{r}H_3\pi_\text{r}\neq0$ since
$\bra{\psi_\ast}H_3\ket{\psi_{\ast+1}}$ is generally non zero.
Nevertheless, also in this case, the first non-trivial correction is
of order $\delta^2$.

Is there any mechanism to produce a non-trivial correction in the
effective dynamics with leading order $\delta\propto
{h_{B}}^{\frac{1}{2}}$\,? An affirmative answer  requires
$H_1\neq0$, and the latter condition is satisfied if we include in
the Hamiltonian $\ttb{H}_\text{BL}$ the effect of a
$\Gamma$-periodic vector potential $\ttb{A}_\Gamma$ (i.e.
$\natural=0$). Since in this situation $H_1$ is linear in $\rrr{a}$
and $\rrr{a}^\dag$,  to obtain a non-trivial effect we need to
consider a spectral subspace which contains at least two contiguous
Landau bands.

Our goal is to derive the (non-trivial) leading order for the
effective Hamiltonian in this framework. According to the notation
of Theorem \ref{teo1}, we need to expand the Moyal product
$h^{\natural=0}= u\moy \pi\moy \widetilde{H}^0_\delta\moy\pi\moy
u^{\dag}=\pi_\text{r}\moy u\moy \widetilde{H}^0_\delta\moy
u^{\dag}\moy\pi_\text{r}+\sss{O}(\delta^\infty)$ up to the first
order $\delta$. The symbols $\pi=\pi_\text{r}+\sss{O}(\delta)$ and
$u=\num{1}_{\sss{H}_\text{f}}+\sss{O}(\delta)$ are derived as in the
general construction showed in the proof of Theorem \ref{teo1}. Now
$\sss{K}:=\text{Ran}\ \Pi_\text{r}\simeq
L^2(\R,dx_\text{s})\otimes\C^2$.

Expanding at  zero order  one finds $h_0= \pi_0 \, u_0 \, H_0
u_0^{-1} \,\pi_0= \pi_\text{r} \, \Xi \, \pi_\text{r} =\pi_\text{r}
\,\Xi = \Xi \, \pi_\text{r}$ and its quantization is the operator
on $\sss{K}$ defined by
\begin{equation}\label{eeq42}
\text{Op}_\delta(h_0)=\left(\begin{array}{cc}
 \left(n_\ast+\frac{3}{2}\right)\num{1}_{\sss{H}_\text{s}}&  0\\
 0& \left(n_\ast+\frac{1}{2}\right)\num{1}_{\sss{H}_\text{s}}
\end{array}\right)=(n_\ast+1)\num{1}_{\sss{K}}+\left(\begin{array}{cc}
\frac{1}{2}\num{1}_{\sss{H}_\text{s}}&  0\\
 0& -\frac{1}{2}\num{1}_{\sss{H}_\text{s}}
\end{array}\right)
\end{equation}

As for the next order, from equation \eqref{eeq40} it follows
 $\chi_1=H_1+u_1H_0-H_0u_1$ (we use $\chi_0=H_0$)
which implies ${h}_1=\pi_\text{r}\chi_1\pi_\text{r}=\pi_\text{r} \,
H_1 \, \pi_\text{r}+\pi_\text{r}\,[u_1;\Xi]\,\pi_\text{r}$. To
conclude the computation we need  $u_1$ and $\pi_1$. Using the
recursive formulas \eqref{eq_rec_p1}, \eqref{eq_rec_p2},
\eqref{eq_rec_p3}, \eqref{eq_rec_u1}, \eqref{eq_rec_u2} and
\eqref{eq_rec_u3}, one obtains
$-2a_{1}:=\left[u_0\moy{u_0}^\dag-\num{1}_{\sss{H}_\text{f}}\right]_{1}=0$,
$b_1:=[\pi_\text{r};B_1]$ and $B_1=[u_0\moy\pi \moy
{u_0}^\ast-\pi_\text{r}]_1=\pi_1$ since  $a_1=0$. Observing that
$G_1=[\pi_\text{r}\moy\pi_\text{r}-\pi_\text{r}]_1=0$, it follows
that $\pi^\text{D}_1=0$ and so
 $u_1=[\pi_\text{r};\pi_1]=[\pi_\text{r};\pi_1^\text{OD}]$ which implies $\pi_\text{r}u_1\pi_\text{r}=0$. Finally
$\pi_\text{r}[u_1;\Xi]\pi_\text{r}=\pi_\text{r}[u_1;\pi_\text{r}\Xi\pi_\text{r}]\pi_\text{r}=0$ and so ${h}_1=\pi_\text{r}H_1\pi_\text{r}$. According to \eqref{eeq30a} the quantization of ${h}_1$ is an operator which acts on $\sss{K}$ as
\begin{equation}\label{eeq43}
\text{Op}_\delta({h}_1)=\sqrt{n_\ast+1}\left(\begin{array}{cc}
0&  \ \ssss{G}(P_\text{s},Q_\text{s})\\
\\
 \ssss{G}(P_\text{s},Q_\text{s})^\dag& 0
\end{array}\right)
\end{equation}
where the operator $\ssss{G}(P_\text{s},Q_\text{s})$ is defined on
$L^2(\R,dx_\text{s})$ as the Weyl quantization of the
$\Z^2$-periodic function $g$ defined by equation \eqref{eeq31}.
Summarizing, we obtained the following result:

\begin{teo}[Effective Hamiltonian with a periodic magnetic potential]
\label{Th 2bands} Under the assumptions of Theorem \ref{teo1}, in
the case $\ttb{A}_\Gamma\neq0$  the dynamics
 in the spectral subspace related to a family of two contiguous Landau bands
  $\{\sigma_{\ast+j}(\cdot)=\lambda_{\ast+j}\ |\ j=0,1\}$  is
   approximated by the effective Hamiltonian $H^\delta_\text{{\upshape eff}}:=\text{Op}_\delta({h}^{(\natural=0)})$ on the reference space $\sss{K}=L^2(\R,dx_\text{s})\otimes
   \C^2$ which is given, up to errors of order $\delta^2$, by
\begin{equation}\label{eeq44}
H^\delta_\text{{\upshape eff}}=(n_\ast+1)\num{1}_{\sss{K}}+\sqrt{n_\ast+1}\left(\begin{array}{cc}
\frac{1}{2\sqrt{n_\ast+1}}\num{1}_{\sss{H}_\text{s}}&  \ \delta\ \ssss{G}(P_\text{s},Q_\text{s})\\
\\
 \delta\ \ssss{G}(P_\text{s},Q_\text{s})^\dag& -\frac{1}{2\sqrt{n_\ast+1}}\num{1}_{\sss{H}_\text{s}}
\end{array}\right)+\sss{O}_0\left(\delta^2\right),
\end{equation}
according to the notation introduced in \eqref{eeq42} and
\eqref{eeq43}.
 \end{teo}

Equation \eqref{eeq31} shows that
$g(p_\text{s},x_\text{s})=g_1(p_\text{s},x_\text{s})-ig_2(p_\text{s},x_\text{s})$
where the function $g_1$ and $g_2$ are related to the component
$(\ttt{A}_\Gamma)_1$ and $(\ttt{A}_\Gamma)_2$ of the
$\Gamma$-periodic vector potential by the relation $g_j(a^\ast\cdot
r,b^\ast\cdot
r)=\pi\sqrt{2}\frac{Z\ell}{\Phi_0}(\ttt{A}_\Gamma)_j(r)$, $j=1,2$.
Let $\ssss{G}_j(P_\text{s},Q_\text{s})$ be the Weyl quantization of
$g_j$. By introducing the Pauli matrices
\begin{equation}\label{eeq46}
\sigma_1= \left(
\begin{array}{cc}
0 & 1 \\
1 &0
\end{array}\right),\ \ \ \ \ \ \sigma_2= \left(
\begin{array}{cc}
0 & -i \\
i& 0
\end{array}\right),\ \ \ \ \ \ \sigma_\bot= \left(
\begin{array}{cc}
1 & 0 \\
0& -1
\end{array}\right)
\end{equation}
one can rewrite the effective Hamiltonian \eqref{eeq44} in the form
\begin{equation}\label{eeq47}
H^\delta_\text{{\upshape eff}}=\left((n_\ast+1)\num{1}_{\C^2}+
\frac{1}{2}\sigma_\bot\right)\otimes\num{1}_{\sss{H}_\text{s}}+\delta\sqrt{n_\ast+1}\
\ \sum_{j=1}^2 {\sigma}_j \otimes \ssss{G}_j(P_\text{s},Q_\text{s})
+ \sss{O}_0\left(\delta^2\right).
\end{equation}

Clearly, the operator $\ssss{G}_j(P_\text{s},Q_\text{s})$ are
Harper-like Hamiltonians and can be represented as a power series of
the Harper unitaries $\ssss{U}_{\infty}$ and $\ssss{V}_{\infty}$ of
type \eqref{eeq37}. In this case the coefficients in the expansion
are (up to a multiplicative constant) the Fourier coefficients of
the components $(\ttt{A}_\Gamma)_j$ of the $\Gamma$-periodic vector
potential.

The determination of the spectrum of $H^{\delta}_\text{{\upshape
eff}}$ can be reduced to the (generally simpler) problem of the
computation of the spectrum of $\ssss{G}\ssss{G}^\dag$.

\begin{propos}\label{prop_sim_spec}
Let $H^{\delta=1}_\text{{\upshape eff}}$ be the first order approximation of the effective Hamiltonian \eqref{eeq44} (or \eqref{eeq47}). Then
$$
\sigma(H^{\delta=1}_\text{{\upshape eff}})=(n_\ast+1)+S_+\cup S_-,\ \ \ \ S_\pm:=\{\pm\sqrt{\nicefrac{1}{4}+\delta^2(n_\ast+1)\ \lambda}\ :\ \lambda\in\overline{\sigma}(\ssss{G}\ssss{G}^\dag)\}
$$
where $\overline{\sigma}(\ssss{G}\ssss{G}^\dag)=\sigma(\ssss{G}\ssss{G}^\dag)\cup\{0\}$ if $\{0\}\in\sigma(\ssss{G}^\dag\ssss{G})\setminus\sigma(\ssss{G}\ssss{G}^\dag)$ and $\overline{\sigma}(\ssss{G}\ssss{G}^\dag)=\sigma(\ssss{G}\ssss{G}^\dag)$ otherwise.
\end{propos}
\Proof We give only a sketch of the proof.
The term $(n_\ast+1)\num{1}_{\sss{K}}$ shifts the spectrum of a constant value $(n_\ast+1)$, then we can consider only the spectrum of $\ssss{B}:=H^{\delta=1}_\text{{\upshape eff}}-(n_\ast+1)\num{1}_{\sss{K}}$.  A simple computation shows that
$$
\ssss{B}^2=\left(
\begin{array}{cc}
\frac{1}{2}\num{1}_{\sss{H}_\text{s}} & \delta\sqrt{n_\ast+1}\ \ssss{G} \\
\delta\sqrt{n_\ast+1}\ \ssss{G}^\dag& -\frac{1}{2}\num{1}_{\sss{H}_\text{s}}
\end{array}\right)^2=\frac{1}{4}\num{1}_\sss{K}+\delta^2(n_\ast+1)\left(
\begin{array}{cc}
\ssss{G}\ssss{G}^\dag & 0  \\
0&\ssss{G}^\dag\ssss{G}
\end{array}\right)
$$
which implies that
$\sigma(\ssss{B}^2)=\{\nicefrac{1}{4}+\delta^2(n_\ast+1)\lambda\ :\
\lambda\in\sigma(\ssss{G}\ssss{G}^\dag)\cup\sigma(\ssss{G}^\dag\ssss{G})\}$.
The operators $\ssss{G}\ssss{G}^\dag$, $\ssss{G}^\dag\ssss{G}$ and
$\ssss{B}$ are bounded and self-adjoint. To show that
$\sigma(\ssss{G}\ssss{G}^\dag)\setminus\{0\}=\sigma(\ssss{G}^\dag\ssss{G})\setminus\{0\}$,
let $\lambda\in\sigma(\ssss{G}\ssss{G}^\dag)$ with $\lambda\neq 0$
and $\{\psi_n\}_{n\in\N}\subset\sss{H}_\text{s}\setminus
\text{Ker}(\ssss{G}^\dag)$ be a sequence of non zero vectors such
that
$\|(\ssss{G}\ssss{G}^\dag-\lambda)\psi_n\|_{\sss{H}_\text{s}}\to 0$
(\emph{Weyl's criterion}), then
$\|(\ssss{G}^\dag\ssss{G}-\lambda)\ssss{G}^\dag\psi_n\|_{\sss{H}_\text{s}}\leqslant\|\ssss{G}^\dag\|_{\bbb{B}(\sss{H}_\text{s})}\|(\ssss{G}\ssss{G}^\dag-\lambda)\psi_n\|_{\sss{H}_\text{s}}\to
0$. This implies that
$\sigma(\ssss{G}\ssss{G}^\dag)\cup\sigma(\ssss{G}^\dag\ssss{G})=\overline{\sigma}(\ssss{G}\ssss{G}^\dag)$.
Now let
$\varepsilon_\pm(\lambda):=\pm\sqrt{\nicefrac{1}{4}+\delta^2(n_\ast+1)\
\lambda}$ with $\lambda\in \overline{\sigma}(\ssss{G}\ssss{G}^\dag)$
and $\{\psi_n\}_{n\in\N}$ a sequence of generalized eigenvectors for
$\ssss{G}\ssss{G}^\dag$ relative to $\lambda$. Then
$\Psi_n^{(\pm)}:=((\nicefrac{1}{2}+\varepsilon_\pm)\psi_n,\delta\sqrt{n_\ast+1}\ssss{G}^\dag\psi_n)\in\sss{H}_\text{s}\otimes\C^2$
is a sequence of generalized eigenvectors for $\ssss{B}$ relative to
$\varepsilon_\pm$. \CVD

\appendix
\section{Some technical results}
Since we include in our analysis a periodic vector potential
$A_\Gamma$ (as a new ingredient with respect to the standard
literature) we include a short discussion of the self-adjointness
and the spectral properties of the operators $H_\text{{\upshape
BL}}$ and $H_\text{{\upshape per}}$.
\subsection{Self-adjointness of $H_\text{BL}$ and $H_\text{per}$}\label{appA1}
The \emph{second Sobolev space} $\ssss{H}^2(\R^2)$ is defined to be
the set of all $\psi\in L^2(\R^2)$ such that
$\partial^{n_1}_{x_1}\partial^{n_2}_{x_2}\psi\in L^2(\R^2)$ in the
sense of distributions for all $n:=(n_1,n_2)\in\N^2$ with
$|n|:=n_1+n_2\leqslant2$. One can proves that $\ssss{H}^2(\R^2)$ is
the closure of $C^\infty_\text{c}(\R^2,\C)$ with respect to the
\emph{Sobolev norm}
$\|\cdot\|_{\ssss{H}^2}:=\|(\num{1}-\Delta_x)\cdot\|_{L^2}$ and has
a Hilbert space structure.
 Similarly the \emph{second magnetic-Sobolev space} $\ssss{H}_{\text{\upshape M}}^2(\R^2)$ is defined to be the set of all $\psi\in L^2(\R^2)$ such that $D_1^{n_1} D_2^{n_2}\psi\in L^2(\R^2)$ in the sense of distributions for all $n\in\N^2$ with $|\alpha|\leqslant2$, where $D_1:=(\partial_{x_1}+\frac{i}{2}x_2)$ and $D_2:=(\partial_{x_2}-\frac{i}{2}x_1)$.
One can prove that
$\ssss{H}_\text{M}^2(\R^2)$ is the closure of $C^\infty_\text{c}(\R^2,\C)$ with respect to the \emph{magnetic-Sobolev norm} $\|\cdot\|_{\ssss{H}_\text{M}^2}:=\|(\num{1}-\Delta_M)\cdot\|_{L^2}$, where $\Delta_M:={D_1}^{2}+ {D_2}^2$ is the \emph{magnetic-Laplacian}. Moreover, $\ssss{H}_\text{M}^2(\R^2)$  has a natural Hilbert space structure. For further details see Section IX.6 and IX.7 of \cite{red-sim2} and Chapter 7 of \cite{lieb-loss}.\\
\\
{\bf Proof of Proposition \ref{prop1}.}\\
We prove the claim for the dimensionless operators, namely we fix all the physical constants equal to 1 in \eqref{eq2} and \eqref{eq588}.\\
\emph{- Step 1.} First of all we  prove that $H_\text{{\upshape per}}$ is essentially self-adjoint on
$C^\infty_\text{c}(\R^2,\C)$ and self-adjoint on $\ssss{H}^2(\R^2)$. Notice that
\begin{equation}\label{eq1001}
H_\text{per}=\dfrac{1}{2}\left[-i\nabla_x-{A}_\Gamma(x)\right]^2+V_\Gamma(x)=-\frac{1}{2}\Delta_x+T_1+ \dfrac{1}{2}T_2
\end{equation}
with $T_1:=i{A}_\Gamma\cdot\nabla_{{x}}$ and
$T_2:=i(\nabla_{{x}}\cdot{A}_\Gamma)+|{A}_\Gamma|^2+2V_\Gamma$. The
free Hamiltonian  $-\nicefrac{1}{2}\Delta_x$  is a self-adjoint
operator with domain $\ssss{H}^2(\R^2)$, essentially self-adjoint on
$C^\infty_\text{c}(\R^2,\C)$  and from Assumption
$(\text{A}_\text{w})$ it follows that $T_2$ is an   infinitesimally
bounded  with respect to $-\nicefrac{1}{2}\Delta_x$ (notice that
$T_2-2V_\Gamma$ is bounded). The symmetric operator $T_1$ is
unbounded with domain $\sss{D}(T_1)\supset \ssss{H}^2(\R^2)$. Let
$\psi\in \ssss{H}^2(\R^2)$, then
$$
\left\|(A_\Gamma)_j \partial_{x_j}\psi\right\|^2_{L^2}\leqslant\|(A_\Gamma)_j\|^2_\infty\ \int_{\R^2}\lambda^2_j\ |\widehat{\psi}({\lambda})|^2\ d^2\lambda
$$
with $\widehat{\psi}(\lambda)$ the Fourier transform of $\psi(x)$. For every $a>0$, if $b=\frac{1}{2a}$ then $\lambda^2_j\leqslant(a|{\lambda}|^2+b)^2$. It follows that $\left\|(A_\Gamma)_j \partial_{x_j}\psi\right\|^2_{L^2}\leqslant C\|(a|{\lambda}|^2+b)\ \widehat{\psi}\|^2_{L^2}$
which implies that for all $a'>0$ (arbitrary small) there exists a $b'$ (depending on $a'$) such that
$$
\left\|(A_\Gamma)_j \partial_{x_j}\psi\right\|^2_{L^2}\leqslant a'\ \||{\lambda}|^2\ \widehat{\psi}\|_{L^2}^2+b'\ \|\widehat{\psi}\|^2_{L^2}=a'\ \|\Delta_{{x}}\psi\|^2_{L^2}+b'\ \|\psi\|^2_{L^2}.
$$
This inequality implies that $T_1$ is infinitesimally bounded with respect to $-\nicefrac{1}{2}\Delta_x$ and  the thesis follows form the {\it Kato-Rellich Theorem} (see \cite{red-sim2} Theorem X.12).\\
\emph{- Step 2.} The  Bloch-Landau Hamiltonian is
\begin{equation}
H_\text{BL}:=\dfrac{1}{2}\left[-i\nabla_x- A_\Gamma\left(x\right)-A\left(x\right)\right]^2+V_\Gamma\left(x\right)+ \phi
\left(x\right).
\end{equation}
Assumptions $(\text{A}_\text{w})$ and (B) imply that
$({A}_\Gamma+{A})_j\in C^1(\R^2,\R)$, $j=1,2$, and $V_\Gamma+\phi\in
L^2_\text{loc}(\R^2)$ and this  assures that  $H_\text{BL}$ is
essentially self-adjoint on $C^\infty_\text{c}(\R^2,\C)$ (see
\cite{red-sim2} Theorem X.34).  Let $A=A_0+A_B$ be the decomposition
of the external vector potential with $A_0$ smooth and bounded and
$A_B=\frac{1}{2}(-x_2,x_1)$. By posing ${D}:=\nabla_{{x}}-i{A}_B$,
$\Delta_M:=|{D}|^2$ and ${A}_\text{b}:={A}_\Gamma+{A}_0$, the
Hamiltonian $H_\text{{\upshape BL}}$ reads
$$
H_\text{{\upshape BL}}=-\dfrac{1}{2}\Delta_M-{A}_\text{b}\cdot {D}+\frac{1}{2}T.
$$
where $T:=i(\nabla_{{x}}\cdot {A}_\text{b})+|{A}_\text{b}|^2+2(V_\Gamma+\phi)$. The operator $T-2V_\Gamma$ is bounded
and the observation that ${A}_\text{b}\cdot {D}$ is infinitesimally bounded with respect to $-\nicefrac{1}{2}\Delta_M$ is an immediate consequence of Lemma \ref{appA3lem2}. the assumption
$\int_{M_\Gamma}|{V}_\Gamma(x)|^2\ d^2x<+\infty$ implies that ${V}_\Gamma$   is \emph{uniformly locally} $L^2$ and hence, infinitesimally bounded with respect to $-\Delta_x$ (see Theorem XIII.96 in \cite{red-sim4}). As proved in \cite{avr_sim_mag} (Theorem 2.4)  this is enough to claim that
 $V_\Gamma$ is aslo infinitesimally bounded with respect to $-\Delta_M$. Therefore, by the {\it Kato-Rellich Theorem} it follows that the domain of self-adjointness of $H_\text{{\upshape BL}}$ coincides with the domain of self-adjointness of  the magnetic-Laplacian, which is $\ssss{H}_\text{M}^2(\R^2)$.
\hspace{\stretch{1}}$\blacksquare$
\subsection{Band spectrum of $H_\text{per}$}\label{appA1bis2}
We describe the spectral properties of the periodic Hamiltonian. The Bloch-Floquet transform maps unitarily $H_\text{per}$ in $H_\text{per}^\sss{Z}:=\int_{M_\Gamma^\ast}^\oplus H_\text{per}(k)\ d^2\underline{k}$. Then to have information about the spectrum of $H_\text{per}$ we need to study the spectra of the family of Hamiltonians
$$
H_\text{per}(k)=\dfrac{1}{2}\left[-i\nabla_{{\theta}}+{k}-{A}_\Gamma({\theta})\right]^2+V_\Gamma({\theta})=-\frac{1}{2}\Delta_\theta+T_1({k})+\frac{1}{2} T_2({k})
$$
where $T_1({k}):=i({A}_\Gamma-\frac{k}{2})\cdot\nabla_{\theta}$ and $T_2({k}):=i(\nabla_{\theta}\cdot{A}_\Gamma)+|{k}|^2+|{A}_\Gamma|^2+2V_\Gamma$ are operators acting on the Hilbert space $\sss{H}_\text{f}:=L^2(\num{V},d^2\theta)$ with $\num{V}:=\R^2/\Gamma$ (\emph{Voronoi torus}).\\
\\
{\bf Proof of Proposition \ref{prop2}.}\\
- (i) The operator $-\nicefrac{1}{2}\Delta_{\theta}$ on the Hilbert
space $\sss{H}_\text{f}$, is  essentially selfadjoint on
$C^\infty(\num{V})$, has domain of self-adjointness
$\sss{D}:=\ssss{H}^2(\num{V})$  and its spectrum is pure point with
$\{\expo{i\theta\cdot{\gamma}^\ast}\}_{{\gamma^\ast\in\Gamma^\ast}}$
a complete orthogonal system of eigenvectors. If Assumption
($\text{A}_\text{w}$) holds true then $T_2(k)$ infinitesimally
bounded with respect to $-\nicefrac{1}{2}\Delta_{\theta}$, indeed
$T-2V_\Gamma$ is bounded and $V_\Gamma$ is infinitesimally bounded
(see \cite{red-sim4} Theorem XIII.97).
With a Fourier estimate similar to those in the proof of Proposition \ref{prop1}  one can also show that $T_1({k})$ is infinitesimally bounded with respect to $-\nicefrac{1}{2}\Delta_{\theta}$, hence  the \emph{Kato-Rellich Theorem} implies that $H_\text{per}({k})$ is  essentially self-adjoint on $C^\infty(\num{T}^2)$ and self-adjoint on the domain $\sss{D}$. Moreover since $-\nicefrac{1}{2}\Delta_{\theta}$ is bounded below then also $H_\text{per}({k})$ is bounded below.\\
- (ii)
For all $\zeta$ in the resolvent set of $-\nicefrac{1}{2}\Delta_{\theta}$ the \emph{resolvent operator} $r_0(\zeta):=(-\nicefrac{1}{2}\Delta_{\theta}-\zeta\num{1}_{\sss{H}_\text{f}})^{-1}$ is a compact operator. Since $T_1(k)+\frac{1}{2}T_2(k)$ is a bounded perturbation of $-\nicefrac{1}{2}\Delta_{\theta}$ it follows that $H_\text{per}(k)$ has compact resolvent (see \cite{red-sim4} Theorem XIII.68)
and moreover it has a purely discrete spectrum with eigenvalues $\sss{E}_n({k})\to+\infty$ as $n\to+\infty$ (see \cite{red-sim4} Theorem XIII.64).\\
- (iii)  The continuity of the function $\sss{E}_n(\cdot)$  follows from the \emph{perturbation theory of discrete spectrum} (see \cite{red-sim4} Theorem XII.13). Indeed, as discussed in Remark \ref{rk_analyt},
 $H_\text{per}(\cdot)$ is an {\it analytic family (of type A) in the sense of Kato}. Finally, since
$H_\text{per}({k}-{\gamma}^\ast)=\tau({\gamma}^\ast)H_\text{per}({k})\tau({\gamma}^\ast)^{-1}$, with $\tau({\gamma}^\ast)$ a unitary operator, then $\sss{E}_n(\cdot)$ are $\Gamma^\ast$-periodic. \hfill$\blacksquare$
\subsection{The Landau Hamiltonian $H_\text{L}$}\label{appA1bis}
The \emph{Landau Hamiltonian} is the operator
\begin{equation}\label{eq1appA1b}
H_\text{L}:=-\frac{1}{2}\Delta_M=\frac{1}{2}\left(K_1^2+K_2^2\right)=\frac{1}{2}\left[\left(-i\frac{\partial}{\partial x_1}+\frac{1}{2}x_2\right)^2+\left(-i\frac{\partial}{\partial x_2}-\frac{1}{2}x_1\right)^2\right]
\end{equation}
where $K_j:=-iD_j$, with $j=1,2$, are the \emph{kinetic momenta}. The Landau Hamiltonian $H_\text{L}$ is essentially self-adjoint on on $C^\infty_\text{c}(\R^2;\C)\subset L^2(\R^2)$ (see \cite{red-sim2} Theorem X.34) and its domain of self-adjointness is exactly the second magnetic-Sobolev space $\ssss{H}_{\text{\upshape M}}^2(\R^2)$ defined in Section \ref{appA1}. To describe the spectrum of $H_\text{L}$ is helpful to introduce another pair of operators:
$G_1:=-i\partial_{x_1}-\frac{1}{2}x_2$ and $G_2:=i\partial_{x_2}-\frac{1}{2}x_1$. The operators $K_1,K_2,G_1,G_2$ are all essentially self-adjoint on $C^\infty_\text{c}(\R^2;\C)$ (they have \emph{deficiency indices} equal to  zero) and on this domain the following commutation relations hold true
\begin{equation}\label{eq2appA1b}
[K_1;K_2]=[G_1;G_2]=i\num{1},\ \ \ [G_j;K_i]=0.
\end{equation}
The last of the \eqref{eq1appA1b} implies $[G_j;H_\text{L}]=0$ hence the operators $G_1$ and $G_2$ are cause for the degeneration of the spectral eigenspaces of $H_\text{L}$. It is a common lore to introduce the \emph{annihilation operator} $\rrr{a}:=\nicefrac{i}{\sqrt{2}}(K_2-iK_1)$ (its adjoint $\rrr{a}^\dag$ is called \emph{creation operator})
and the \emph{degeneration operator} $\rrr{g}:=\nicefrac{i}{\sqrt{2}}(G_2-iG_1)$. They fulfill the following (bosonic) commutation relation
\begin{equation}\label{eq3appA1b}
[\rrr{a};\rrr{a}^\dag]=[\rrr{g};\rrr{g}^\dag]=\num{1},\ \ \ [\rrr{g};H_{\text{L}}]=[\rrr{g}^\dag;H_{\text{L}}]=0,\ \ \ [\rrr{a};H_{\text{L}}]=\rrr{a},\ \ \ [\rrr{a}^\dag;H_{\text{L}}]=-\rrr{a}^\dag.
\end{equation}
The last two relations follow from the equality $H_\text{L}=\rrr{a}\rrr{a}^\dag-\nicefrac{1}{2}\num{1}=\rrr{a}^\dag \rrr{a}+\nicefrac{1}{2}\num{1}$. Define the \emph{ground state} $\psi_0\in L^2(\R^2)$ as the normalized solution of
$\rrr{g}\psi_0=0=\rrr{a}\psi_0$, i.e. $\psi_0(x)=C\expo{-\frac{1}{4}|x|^2}$. The \emph{generalized Hermite function} of order $(n,m)$ is defined to be $\psi_{n,m}:=\frac{1}{\sqrt{n!m!}}\left(\rrr{g}^\dag\right)^m\left(\rrr{a}^\dag\right)^n\psi_0$. We will denote by $\sss{F}\subset L^2(\R^2)$ the set of the finite linear combinations of the vectors $\psi_{n,m}$ and we will call it the \emph{Hermite domain}. Clearly $\sss{F}\subset S(\R^2)$ (the Schwartz space).
\begin{lem}\label{appA3lem1}
 With the notation above:
\item[{\upshape (i)}] the set $\{\psi_{n,m} \ :\ n,m=0,1,2,\ldots\}$ is a complete orthonormal basis for $ L^2(\R^2)$ and so $\sss{F}$ is a dense domain;
\item[{\upshape (ii)}] the spectrum of $H_\text{L}$ is pure point and is given by $\{\lambda_n:=(n+\nicefrac{1}{2})\ :\ n=0,1,2,\ldots\}$, moreover $H_\text{L}\ \psi_{n,m}=\lambda_n\ \psi_{n,m}$ for every $m=0,1,2,\ldots$ (degeneration index);
\item[{\upshape (iii)}] $H_\text{L}$ is essentially self-adjoint on $\sss{F}$ and the closure of $\sss{F}$ with respect to the magnetic-Sobolev norm coincides with the magnetic-Sobolev space $\ssss{H}_{\text{\upshape M}}^2(\R^2)$.
\end{lem}
\Proof
- (i) Let $\ssss{W}:L^2(\R^2,d^2x)\to L^2(\R,du)\otimes L^2(\R,dv)$ be the unitary map such that the conjugate pairs $(K_1,K_2)$ and $(G_1,G_2)$ are transformed by $\ssss{W}\ldots\ssss{W}^{-1}$ into the canonical pairs $(u,-i\partial_u)$ and $(v,-i\partial_v)$. The existence of such a unitary $\ssss{W}$ will be discussed in Appendix \ref{appB}. Obviously $\rrr{a}\mapsto\widetilde{\rrr{a}}=\nicefrac{1}{\sqrt{2}}(u+\partial_u)$, $\rrr{g}\mapsto\widetilde{\rrr{g}}=\nicefrac{1}{\sqrt{2}}(v+\partial_v)$ and $\widetilde{\psi}_0:=\ssss{W}\psi_0$
is the solution of $\widetilde{\rrr{g}}\widetilde{\psi}_0=\widetilde{\rrr{a}}\widetilde{\psi}_0=0$, namely
$\widetilde{\psi}_0(u,v)=h_0(u)\otimes h_0(v)$ where $h_0(t):=\pi^{-\frac{1}{4}}\expo{-\frac{1}{2}t^2}$ is the $0$-th Hermite function. Then $\widetilde{\psi}_{n,m}(u,v):=(\ssss{W}{\psi}_{n,m})(u,v)=h_n(u)\otimes h_m(v)$ which shows that the functions $\widetilde{\psi}_{n,m}$ define an orthonormal basis for $L^2(\R,du)\otimes L^2(\R,dv)$ since the Hermite functions $h_n$ are an orthonormal system for $L^2(\R)$. The claim follows since $\ssss{W}$ is a unitary map.\\
- (ii) Clearly $H_\text{L}\psi_0=(\rrr{a}^\dag\rrr{a}+\nicefrac{1}{2}\num{1})\psi_0=\nicefrac{1}{2}\psi_0$ and from relations \eqref{eq3appA1b} it follows that $H_\text{L}\psi_{n,m}=\frac{1}{\sqrt{n!m!}}\left(\rrr{g}^\dag\right)^m H_\text{L}\left(\rrr{a}^\dag\right)^n\psi_0=\nicefrac{1}{2}\psi_{n,m}+\frac{1}{\sqrt{n!m!}}\left(\rrr{g}^\dag\right)^m (\rrr{a}^\dag\rrr{a})\left(\rrr{a}^\dag\right)^n\psi_0=\lambda_{n}\ \psi_{n,m}$. Then the generalized Hermite functions $\psi_{n,m}$ are a complete set of orthonormal eigenvectors for $H_\text{L}$. This proves that the spectrum of $H_\text{L}$ is pure point.\\
- (iii) The operator $H_\text{L}$ is essentially self-adjoint in $\sss{F}$ since the deficiency indices are both  zero. This implies that last part of the claim.
\CVD
\begin{lem}\label{appA3lem2}
The operators $K_1$, $K_2$, $\rrr{a}$ and $\rrr{a}^\dag$ are
infinitesimally bounded with respect to $H_\text{L}$.
\end{lem}
\Proof Since $K_1=\nicefrac{1}{\sqrt{2}}(\rrr{a}+\rrr{a}^\dag)$ and $K_2=\nicefrac{1}{i\sqrt{2}}(\rrr{a}-\rrr{a}^\dag)$
it is enough to prove the claim for $\rrr{a}$ and $\rrr{a}^\dag$.
Let $\psi:=\sum_{n,m=0}^{+\infty}c_{n,m}\ \psi_{n,m}\in\ssss{H}_{\text{\upshape M}}^2(\R^2)$.  An easy computation shows that
\begin{align*}
&\|\rrr{a}\psi\|^2_{L^2}=\sum_{n,m=0}^{+\infty}|c_{n,m}|^2\ n,&
&\|a^\dag\psi\|^2_{L^2}=\sum_{n,m=0}^{+\infty}|c_{n,m}|^2\ \left(n+1\right).
\end{align*}
Since  $n\leqslant n+1\leqslant2\left(n+\frac{1}{2}\right)\leqslant a\left(n+\frac{1}{2}\right)^2+\frac{1}{a}$ holds true for any $a>0$ (arbitrarily small), then
$$
\|\rrr{a}^\sharp \psi\|^2_{L^2}\leqslant a
\sum_{n,m=0}^{+\infty}|c_{n,m}|^2\
\left(n+\frac{1}{2}\right)^2+b\sum_{n,m=0}^{+\infty}|c_{n,m}|^2=a\|H_\text{L}\psi\|^2_{L^2}+b\|\psi\|^2_{L^2}
$$
with $b:=\frac{1}{a}+\frac{1}{2}$ where $\rrr{a}^\sharp$ denotes
either $\rrr{a}$ or $\rrr{a}^\dag$. \CVD

\section{Canonical transform for fast and slow variables}\label{appB}
This appendix is devoted to the concrete realization of the \emph{von Neumann unitary} $\ssss{W}$ introduced (in abstract way) in Section \ref{sec_neum_unit}. The unitary $\ssss{W}$  maps the fast and slow variables, which satisfy canonical commutation relation, into a set of canonical Schr\"{o}dinger operators.
In Section \ref{appB1} we derive a general version of the transform $\ssss{W}$ \virg{by hand}, as a composition of three sequential transforms.  In Section \ref{appB2} we compute the integral kernel of $\ssss{W}$.
\subsection{The transform $\ssss{W}$ built \virg{by hand}}\label{appB1}
Let $\sss{H}:=L^2(\R^2,d^2r)$ be the initial Hilbert space, with $r:=(r_1,r_2)$. Let $Q_r:=(Q_{r_1},Q_{r_2})$ where $Q_{r_j}$ is the multiplication operator by $r_j$  and $P_r:=(P_{r_1},P_{r_2})$ where $P_{r_1}:=-i\hslash\partial_{r_j}$, with $j=1,2$.
Consider the \emph{fast} and \emph{slow} operators
\begin{equation}\label{eeqB1}
 (\text{fast})\ \left\{
\begin{aligned}
 &K_1:=-\frac{\alpha}{2\beta}\ v\cdot{Q}_r-\frac{\alpha\beta}{\hslash}\  w^\ast\cdot{P}_r\\
 &K_2:=\phantom{-}\frac{\alpha}{2\beta}\ w\cdot{Q}_r-\frac{\alpha\beta}{\hslash}\  v^\ast\cdot {P}_r
\end{aligned}
\right.\ \ \ \ \ \ \ \
(\text{slow})\ \left\{
\begin{aligned}
 &G_1:=\frac{1}{2}\ v\cdot {Q}_r-\frac{\beta^2}{\hslash}\ w^\ast\cdot {P}_r\\
&G_2:=\frac{1}{2}\ w\cdot {Q}_r+\frac{\beta^2}{\hslash}\ v^\ast\cdot {P}_r
\end{aligned}
\right.
\end{equation}
with $\alpha,\beta\in\C$ and $v,w,v^\ast,w^\ast\in\R^2$ such that $v\cdot v^\ast=w\cdot w^\ast=1$, $v^\ast\cdot w=v\cdot w^\ast=0$ and $|v\wedge w|=\ell^2>0$.
\begin{rk} The choice $v=b^\ast$, $w=a^\ast$, $\alpha=\sqrt{\iota_q}$ and $\beta=\sqrt{\iota_q}\ \delta$ defines the operators \eqref{eeq2}, while the choice $v=v^\ast=(0,-1)$, $w=w^\ast=(-1,0)$, $\alpha=\beta=1$ defines the
\emph{kinetic momenta} and the related conjugate operators introduced in Section \ref{appA1bis}.
\hfill $\blacklozenge\lozenge$
\end{rk}
 Observing that $[a\cdot{Q}_r+b\cdot{P}_r;c\cdot{Q}_r+d\cdot{P}_r]=i\hslash(a\cdot d-b\cdot c)\num{1}_\sss{H}$ one deduce that the operators \eqref{eeqB1} verify the following \emph{canonical commutation relations} (CCR)
\begin{equation}\label{eebB2}
[K_1,K_2]=i\alpha^2\num{1}_\sss{H},\ \ \ \ \ \ [G_1,G_2]=i\beta^2\num{1}_\sss{H},\ \ \ \ \ \ [K_i,G_j]=0,\ \ i,j=1,2.
\end{equation}
The \emph{Stone-von Neumann uniqueness theorem} (see \cite{bra-rob2} Corollary 5.2.15) assures the existence of a unitary map $\ssss{W}$ (\emph{von Neumann unitary})
\begin{equation}\label{eeqB3}
\ssss{W}:\sss{H}{\longrightarrow}\sss{H}_\text{w}:=\sss{H}_{\text{s}}\otimes\sss{H}_{\text{f}}:= L^2(\R,dx_\text{s})\otimes  L^2(\R,dx_\text{f})
\end{equation}
such that

\begin{align}
&\ssss{W}G_1\ssss{W}^{-1}:=Q_{\text{s}}=\text{multiplication by}\ \ x_\text{s},&& \ssss{W}G_2\ssss{W}^{-1}:=P_{\text{s}}=-i {\beta}^2\dfrac{\partial}{\partial x_\text{s}}\label{eeqB4}\\
&\ssss{W}K_1\ssss{W}^{-1}:=Q_{\text{f}}=\text{multiplication by}\ \ x_\text{f},&& \ssss{W}K_2\ssss{W}^{-1}:=P_{\text{f}}=-i \alpha^2\dfrac{\partial}{\partial x_\text{f}}.\label{eeqB5}
\end{align}

In other words, $(Q_{\text{s}},P_{\text{s}})$ is a pair of operators which defines a Schr\"odinger representation on the Hilbert space $\sss{H}_{\text{s}}:=L^2(\R,dx_\text{s})$
while the pair $(Q_{\text{f}},P_{\text{f}})$ defines a Schr\"odinger representation  on the Hilbert space $\sss{H}_{\text{f}}:=L^2(\R,dx_\text{f})$.
Our purpose is to give an explicit construction for $\ssss{W}$.
First of all consider the change of coordinates $(r_1,r_2)\mapsto(k_1:=\frac{v\cdot r}{\ell},k_2:=\frac{w\cdot r}{\ell})$.  The inverse transforms are defined by $r_1(k)=\frac{1}{\ell}(w_2k_1-v_2k_2)$ and $r_2(k)=\frac{1}{\ell}(v_1k_2-w_1k_1)$. The map $\sss{J}:L^2(\R^2,d^2r)\to L^2(\R^2,d^2k)$ defined by $(\sss{J}\psi)(k):=\psi(r(k))$ if $\psi\in L^2(\R^2,dr)$ is  unitary since the change of coordinates is invertible and isometric. Moreover $\sss{J} Q_{r_j}\sss{J}^{-1}$ acts on $ L^2(\R^2,dk)$ as the multiplication by $r_j(k)$, while
$\sss{J} P_{r_j}\sss{J}^{-1}=\frac{v_j}{\ell}P_{k_1}+\frac{w_j}{\ell}P_{k_2}$ where $P_{k_j}:=-i\hslash\partial_{k_j}$, with $j=1,2$. Then
\begin{align}
&\sss{J}G_1\sss{J}^{-1}:=\phantom{-\frac{\alpha}{\beta}}\frac{\ell}{2}\ Q_{k_1}-\frac{\beta^2}{\hslash}\frac{1}{\ell}\ P_{k_2},&& \sss{J}G_2\sss{J}^{-1}:=\phantom{\frac{\alpha}{\beta}}\frac{\ell}{2}\ Q_{k_2}+\frac{\beta^2}{\hslash}\frac{1}{\ell}\ P_{k_1}\label{eeqB6}\\
&\sss{J}K_1\sss{J}^{-1}:=-\frac{\alpha}{\beta}\frac{\ell}{2}\ Q_{k_1}-\frac{\alpha\beta}{\hslash}\frac{1}{\ell}\  P_{k_2},&& \sss{J}K_2\sss{J}^{-1}:=\frac{\alpha}{\beta}\frac{\ell}{2}\ Q_{k_2}-\frac{\alpha\beta}{\hslash}\frac{1}{\ell}\  P_{k_1}.\label{eeqB7}
\end{align}
Let $\sss{F}_{2,\mu}:L^2(\R,dk_2)\to L^2(\R,d\zeta_2)$ be the \emph{$k_2$-Fourier transfor} of weight $\mu$,
defined by $(\sss{F}_{2,\mu}\psi)(\zeta_2):=\sqrt{\frac{|\mu|}{2\pi}}\int_\R\expo{-i\mu \zeta_2 k_2}\psi(k_2)\ dk_2$ and let
$\Pi_1:L^2(\R,dk_1)\to L^2(\R,d\zeta_1)$ be the \emph{$k_1$-parity operator} defined by $(\Pi_2\psi)(\zeta_1):=\psi(-\zeta_1)$ (namely by the change of coordinates $k_1\mapsto \zeta_1$). Let $\num{I}$ the unitary map which identifies $L^2(\R^2,d^2k)$ with
$L^2(\R,dk_1)\otimes L^2(\R,dk_2)$. Let
 $Q_\zeta:=(Q_{\zeta_1},Q_{\zeta_2})$ whit $Q_{\zeta_j}$ the multiplication operator by $\zeta_j$  and $P_\zeta:=(P_{\zeta_1},P_{\zeta_2})$ where $P_{\zeta_1}:=-i\hslash\partial_{\zeta_j}$, with $j=1,2$. One can check that
\begin{equation}
 \sss{F}_{2,\mu}Q_{k_2}\sss{F}_{2,\mu}^{-1}=-\frac{1}{\mu\hslash}P_{\zeta_2},\ \ \ \sss{F}_{2,\mu}P_{k_2}\sss{F}_{2,\mu}^{-1}=\mu\hslash Q_{\zeta_2},\ \ \ \Pi_1Q_{k_1}\Pi_1^{-1}=-Q_{\zeta_1},\ \ \ \Pi_1P_{k_1}\Pi_1^{-1}=-P_{\zeta_1}.
\end{equation}
Fix $\mu:=-\frac{\ell^2}{2\beta^2}$, then  the unitary map $\sss{L}:=(\Pi_1\otimes\sss{F}_{2,\mu})\circ\num{I}\circ\sss{J}:\sss{H}\to L^2(\R,d\zeta_1)\otimes  L^2(\R,d\zeta_2)$ acts on the operators \eqref{eeqB1} in the following way
\begin{align}
&\sss{L}G_1\sss{L}^{-1}:=-\frac{\ell}{2}\left( Q_{\zeta_1}- Q_{\zeta_2}\right),&& \sss{L}G_2\sss{J}^{-1}:=-\frac{1}{\ell}\frac{\beta^2}{\hslash}\left( P_{\zeta_1}- P_{\zeta_2}\right)\label{eeqB8}\\
&\sss{L}K_1\sss{L}^{-1}:=\frac{\alpha}{\beta}\frac{\ell}{2}\left( Q_{\zeta_1}+Q_{\zeta_2}\right),&& \sss{L}K_2\sss{L}^{-1}:=\phantom{-}\frac{1}{\ell}\frac{\alpha\beta}{\hslash}\left( P_{\zeta_1}+ P_{\zeta_2}\right).\label{eeqB9}
\end{align}
Now we can consider the change of coordinates $(\zeta_1,\zeta_2)\mapsto(x_\text{s},x_\text{f})$ defined by
$$
\left\{
\begin{aligned}
 &x_\text{s}=-\dfrac{\ell}{{2}}(\zeta_1-\zeta_2)\\
&x_\text{f}=\phantom{-}\dfrac{{\alpha}}{{\beta}}\frac{\ell}{2}(
\zeta_1+\zeta_2)
\end{aligned}
\right.
\ \ \ \ \ \ \ \ \ \ \
 \left\{
\begin{aligned}
 &\zeta_1=-\frac{1}{\ell}{\left( x_\text{s}-\frac{\beta}{\alpha}\ x_\text{f}\right)}\\
&\zeta_2=\phantom{-}\frac{1}{\ell}{\left(
x_\text{s}+\frac{\beta}{\alpha}\ x_\text{f}\right)}
\end{aligned}
\right.
$$
The jacobian of this transformation is
$|\partial(\zeta_1,\zeta_2)/\partial(x_\text{s},x_\text{f})|={\frac{2}{\ell^2}\left|\frac{\beta}{\alpha}\right|}=:C$,
then the map $(\sss{R}\psi)(x_\text{s},x_\text{f}):=\sqrt{C}\
\psi(\zeta(x_\text{s},x_\text{f}))$ defines a unitary map
$\sss{R}:L^2(\R^2,d^2\zeta)\to L^2(\R^2,dx_\text{s}\ dx_\text{f})$.
With a direct computation one can check that $\sss{R}
Q_{\zeta_j}\sss{R}^{-1}$ acts on $ L^2(\R^2,dx_\text{s}\
dx_\text{f})$ as the multiplication by
$\zeta_j(x_\text{s},x_\text{f})$, while $\sss{R}
P_{\zeta_j}\sss{R}^{-1}=\frac{\ell\hslash}{2\alpha\beta}\left(\frac{(-1)^j\alpha}{\beta}P_\text{s}+P_\text{f}\right)$,
with $j=1,2$. This shows that the unitary map
$\ssss{W}:=\num{I}\circ\sss{R}\circ\num{I}^{-1}\circ\sss{L}:=\num{I}\circ\sss{R}\circ\num{I}^{-1}\circ(\Pi_1\otimes\sss{F}_{2,\mu})\circ\num{I}\circ\sss{J}$
is the von Neumann unitary that verifies the relations \eqref{eeqB4}
and \eqref{eeqB5}.
\subsection{The integral kernel of $\ssss{W}$}\label{appB2}
The unitary operators $\sss{J}:L^2(\R^2,d^2k)\to L^2(\R^2,d^2r)$ and $\sss{R}:L^2(\R^2,d^2\zeta)\to L^2(\R^2,dx_\text{s}dx_\text{f})$ related to the change of coordinates $(r_1,r_2)\mapsto(k_1,k_2)$ and
$(\zeta_1,\zeta_2)\mapsto(x_\text{s},x_\text{f})$ can be written
an integral operators
$$
(\sss{J}\psi)(k)=\int_{\R^2}J(r;k)\ \psi(r)\ d^2r,\ \ \ \ \ \ (\sss{R}\varphi)(x_\text{s},x_\text{f})=\int_{\R^2}R(\zeta;x_\text{s},x_\text{f})\ \varphi(\zeta)\ d^2\zeta
$$
with distributional integral kernels
\begin{align*}
&J(r_1,r_2;k_1,k_2):=\delta\left(r_1-\frac{1}{\ell}(w_2k_1-v_2k_2)\right)\ \delta\left(r_2+\frac{1}{\ell}(w_1k_1-v_1k_2)\right)\\
&R(\zeta_1,\zeta_2;x_\text{s},x_\text{f}):=\sqrt{C}\
\delta\left(\zeta_1+\frac{1}{\ell}{\left(
x_\text{s}-\frac{\beta}{\alpha}\ x_\text{f}\right)}\right)\
\delta\left(\zeta_2-\frac{1}{\ell}{\left(
x_\text{s}+\frac{\beta}{\alpha}\
x_\text{f}\right)}\right)\end{align*} with
$C={\frac{2}{\ell^2}\left|\frac{\beta}{\alpha}\right|}$. The
$k_1$-parity operator $\Pi_1$ can be written as an integral operator
with the distributional kernel $\delta(r_1+k_1)$ while the integral
kernel of the $k_2$-Fourier transform
 $\sss{F}_{2,\mu}$, with $\mu:=-\frac{\ell^2}{2\beta^2}$, is $\frac{\ell}{2|\beta|\sqrt{\pi}}\expo{i\frac{\ell^2}{2\beta^2}\zeta_2k_2}$. Then the unitary map $\num{I}^{-1}\circ(\Pi_1\otimes \sss{F}_{2,\mu})\circ\num{I}:L^2(\R^2,d^2k)\to L^2(\R^2,d^2\zeta)$
as the integral distributional kernel
$$
L(k_1,k_2;\zeta_1,\zeta_2):=\delta(\zeta_1+k_1)\ \frac{\ell}{2|\beta|\sqrt{\pi}}\expo{i\frac{\ell^2}{2\beta^2}\zeta_2k_2}.
$$
Summarizing the total transform $\ssss{W}:L^2(\R^2,d^2r)\to L^2(\R^2,dx_\text{s}\ dx_\text{f})$ (up to the obvious identification $\num{I}$) can be expressed as an integral operator
$$
(\ssss{W}\psi)(x_\text{s},x_\text{f})=\int_{\R^2}W(r;x_\text{s},x_\text{f})\ \Psi(r)\ d^2r
$$
with a (total) integral distributional kernel
\begin{align*}
W(r_1,r_2;x_\text{s},x_\text{f})&:={
\frac{\ell}{\sqrt{2\pi|\alpha\beta|}}}\ \delta\left({v\cdot
r}-{\left( x_\text{s}-\frac{\beta}{\alpha}\
x_\text{f}\right)}\right)\ \expo{i\frac{ w\cdot r }{2\beta^2}{\left(
x_\text{s}+\frac{\beta}{\alpha}\ x_\text{f}\right)}}.
\end{align*}

\newpage

\bibliographystyle{alpha}
\bibliography{biblio(Adiabatic_theory)}
\end{document}